\documentclass[%
 aip,
 amsmath,amssymb,
 reprint,%
]{revtex4-1}
\usepackage{graphicx}
\usepackage{subfigure}
\usepackage{dcolumn}
\usepackage{bm}
\usepackage{color}
\usepackage[T1]{fontenc}
\usepackage{mathptmx}
\usepackage{multirow}
\usepackage{array}
\usepackage{float}
\usepackage{etoolbox}
\usepackage{booktabs}
\usepackage{url}
\usepackage{CJKutf8}
\usepackage{pifont}
\usepackage{epstopdf, epsfig}
\usepackage{graphicx}
\usepackage{longtable}
\usepackage[
    colorlinks=true,
    linkcolor=blue,
    urlcolor=blue,
    citecolor=blue
]{hyperref}

\usepackage{titlesec}
\titleclass{\subsubsubsection}{straight}[\subsubsection]
\newcounter{subsubsubsection}[subsubsection]
\renewcommand\thesubsubsubsection{\thesubsubsection.\arabic{subsubsubsection}}
\titleformat{\subsubsubsection}
  {\normalfont\normalsize\bfseries}{\thesubsubsubsection}{1em}{}
\titlespacing*{\subsubsubsection}
  {0pt}{1.5ex plus .1ex minus .2ex}{1ex plus .2ex}

\begin{document}
\preprint{AIP/123-QED}

\title{Multiscale thermodynamic nonequilibrium effects in Kelvin-Helmholtz instability and their relative importance}




\author{Zhongyi He \begin{CJK*}{UTF8}{gbsn} (何忠奕) \end{CJK*}}
\affiliation{School of Mathematics and Statistics, Key Laboratory of Analytical Mathematics and Applications (Ministry of Education), Fujian Key Laboratory of Analytical Mathematics and Applications (FJKLAMA), Center for Applied Mathematics of Fujian Province (FJNU), Fujian Normal University, 350117 Fuzhou, China}
\author{Yanbiao Gan \begin{CJK*}{UTF8}{gbsn} (甘延标) \end{CJK*}}
 \thanks{Corresponding author: gan@nciae.edu.cn}
  \affiliation{Hebei Key Laboratory of Trans-Media Aerial Underwater Vehicle,
North China Institute of Aerospace Engineering, Langfang 065000, China}
\author{Bin Yang \begin{CJK*}{UTF8}{gbsn} (杨斌) \end{CJK*}}
  \affiliation{School of Energy and Safety Engineering, Tianjin Chengjian University, Tianjin 300384, China}
\author{Demei Li \begin{CJK*}{UTF8}{gbsn} (李德梅)\end{CJK*}}
\affiliation{School of Mathematics and Statistics, Key Laboratory of Analytical Mathematics and Applications (Ministry of Education), Fujian Key Laboratory of Analytical Mathematics and Applications (FJKLAMA), Center for Applied Mathematics of Fujian Province (FJNU), Fujian Normal University, 350117 Fuzhou, China}
\author{Huilin Lai \begin{CJK*}{UTF8}{gbsn} (赖惠林)\end{CJK*}}
\thanks{Corresponding author: hllai@fjnu.edu.cn}
\affiliation{School of Mathematics and Statistics, Key Laboratory of Analytical Mathematics and Applications (Ministry of Education), Fujian Key Laboratory of Analytical Mathematics and Applications (FJKLAMA), Center for Applied Mathematics of Fujian Province (FJNU), Fujian Normal University, 350117 Fuzhou, China}
\author{Aiguo Xu \begin{CJK*}{UTF8}{gbsn} (许爱国) \end{CJK*}}
\affiliation{National Key Laboratory of Computational Physics, Institute of Applied Physics and Computational Mathematics, P. O. Box 8009-26, Beijing 100088, P.R.China}
\affiliation{National Key Laboratory of Shock Wave and Detonation Physics, Mianyang 621999, China}
\affiliation{HEDPS, Center for Applied Physics and Technology, and College of Engineering, Peking University, Beijing 100871, China}
\affiliation{State Key Laboratory of Explosion Science and Safety Protection, Beijing Institute of Technology, Beijing 100081, China}

\date{\today}

\begin{abstract}	
This study investigates the complex kinetics of thermodynamic nonequilibrium effects (TNEs) and their relative importance during the development of the Kelvin-Helmholtz instability (KHI) using high-order discrete Boltzmann models (DBMs). First, the capabilities and differences among various discrete velocity sets in capturing TNEs and distribution functions are assessed. Based on this analysis, practical guidelines for constructing discrete velocity stencils are proposed to enhance phase-space discretization and improve the robustness of high-order DBM simulation.
At different stages of KHI and under varying initial conditions, multiscale TNEs, such as viscous stresses of different orders, emerge
with distinct dominant roles. Specifically, three scenarios are identified:
(i) regimes dominated by first-order TNEs,
(ii) alternation between first- and second-order TNEs, and
(iii) states where second-order TNEs govern the system's behavior.
To quantitatively capture these transitions, criteria for TNE dominance at different orders in KHI evolution are established based on the relative thermodynamic nonequilibrium intensity (\(R_{\text{TNE}}\)).
In scenarios dominated by second-order TNEs, differences between first-order and second-order models are compared in terms of macroscopic quantities, nonequilibrium effects, and kinetic moments, revealing the physical limitations of low-order models in capturing TNEs.
Furthermore, the effectiveness, extensibility, and limitations of a representative high-order model are examined under second-order TNE-dominated conditions.
To encapsulate these findings, a nonequilibrium phase diagram that visually maps the multiscale characteristics of KHI is constructed. This diagram not only provides intuitive insights into the dynamic interplay of different nonequilibrium effects but also serves as a kinetic roadmap for selecting suitable models under diverse nonequilibrium conditions.
\end{abstract}

\maketitle

\section{\label{sec:level1} Introduction}

Kelvin-Helmholtz instability (KHI) is a fluid instability induced by shear flows. It typically arises at the interface between two fluids with different tangential velocities and densities or within shear layers of a single fluid\cite{chandrasekhar2013CC,drazin2004CUP}.
When shear stress at the fluid interface surpasses a critical threshold, surface fluctuations emerge, triggering mass and momentum exchange between adjacent fluid layers. Over time, this interaction evolves through both linear and nonlinear stages, ultimately forming vortices and driving turbulent mixing\cite{gan2011PRE,li2018IP}.
Besides velocity-induced interactions, KHI frequently couples with other instabilities, such as Rayleigh-Taylor instability (RTI) and Richtmyer-Meshkov instability (RMI)\cite{wang2017SCP,akula2017JOFM}. This coupling results in more intricate flow structures, further intensifies mixing at the fluid interface, and enhances turbulence generation\cite{zhou2017rayleigh1,zhou2017rayleigh2,zhou2019turbulent}.

KHI, as a fundamental mechanism driving turbulent mixing, is widely observed in both natural phenomena and engineering applications across various fields. Examples include shear layers in atmospheric and oceanic flows\cite{smyth2012ocean}, mixing layers in high-speed flows\cite{martens1994AIAA}, and turbulent boundary layers\cite{jiang2021impact}.
In astrophysics, KHI shapes some of the most dynamic and visually striking cosmic phenomena. It sculpts the turbulent flow structures of the solar corona\cite{howson2017AA}, fuels the ever-changing battlefront between the solar wind and Earth's magnetosphere\cite{hasegawa2004NAT}, and churns the colossal, swirling storms of Jupiter's Great Red Spot.
In the violent aftermath of supernova explosions, KHI acts as a cosmic mixer, intensifying the interaction between ejected stellar material and the surrounding interstellar medium\cite{nomoto1997type,gamezo2003SCI}. This turbulent interplay accelerates high-energy particles and fosters the diffusion of matter across space, seeding the next generation of stars and planetary systems\cite{perucho2007PRE,yang2018TAJ}.

In inertial confinement fusion (ICF), KHI arises at various stages of the implosion process\cite{zhou2017rayleigh1,zhou2017rayleigh2,zhou2019turbulent,sadler2022AIP}. During the early stage of shell expansion, laser heating induces a velocity shear layer between the rapidly expanding outer layers and the inner layers, triggering KHI.
During fuel compression, sustained laser or particle beam compression amplifies KHI due to pressure and velocity gradients. During shockwave propagation, laser-driven shockwaves travel inward through the shell, forming new velocity shear layers at the shell-fuel interface and further triggering KHI\cite{sadler2022AIP}.
KHI also couples with RMI, which emerges at the multilayered target capsule interface due to ablation shockwaves and rarefaction waves induced during implosion. During the implosion acceleration phase, KHI also interacts with RTI, where a lighter material overlays the dense metal shell\cite{rikanati2000IOP}.
These instabilities disrupt capsule symmetry, impede ignition hotspot formation, reduce laser energy deposition, limit implosion velocity, and potentially rupture the fuel shell, ultimately leading to ICF ignition failure\cite{wang2017SCP,he2007TEPJ}.
Moreover, recent studies demonstrated that the kinetic effects, marked by significant discrete effects and nonequilibrium effects, have a substantial impact on the success of ignition in ICF\cite{rinderknecht2018kinetic,cai2021hybrid,lianqiang2020research}.

In combustion processes, KHI enhances fluid layer mixing and mass exchange\cite{papas2009POTC}, accelerating fuel-oxidizer reactions and improving combustion efficiency and stability\cite{som2010CAF}.
In climatology and environmental sciences, KHI influences atmospheric wind-cloud interactions, ocean surface wind fluctuations, and dynamic air-sea interface changes in climate models, thereby affecting climate change predictions and natural disaster assessments\cite{hasegawa2004NAT}.
In materials science, especially in nanomaterial fabrication and thin-film deposition\cite{liu2015POTN,bertevas2019POF}, KHI is crucial for understanding material interfaces\cite{liu2015IJOH}, surface tension \cite{zhang2001AIWR}, and mass transfer\cite{asthana2010IJOE}.
KHI is also significant in liquid-gas interfaces \cite{lee2015EJOM,sofonea2001PASM}, droplet collisions\cite{liu2015POTN}, supersonic flows\cite{wan2015PRL}, and advanced fluid control and microfluidics\cite{mieussens2000JOCP}.

Traditionally, KHI has been studied using three primary approaches: theoretical analysis, experimental research, and numerical simulation.
Theoretical analysis often depends on simplified assumptions, such as linearization and ideal fluid approximations, which cannot fully describe complex nonlinear phenomena. Furthermore, theoretical methods struggle to handle complex boundary conditions and multi-scale effects observed in real-world scenarios\cite{kim2022MDPI}.
Experimental research, however, faces challenges including equipment limitations, constraints in visualization techniques, and difficulties in controlling initial and boundary conditions, leading to discrepancies between experimental, theoretical, and numerical results\cite{leon1999APS}.
Compared to the other two approaches, numerical simulations offer significant advantages. They efficiently handle complex boundary conditions and multi-physics coupling, offer greater flexibility in adjusting initial conditions and precision, and enable real-time data extraction and analysis \cite{olson2011AIP,chaurasia2011CUP}.

Advances in computational fluid dynamics have solidified numerical simulations as an essential tool for studying KHI.
Over the past two decades, researchers have utilized diverse numerical methods to investigate KHI, offering valuable insights for engineering applications in complex flow conditions.
Early studies primarily examined the fundamental development of KHI using direct numerical simulation (DNS), large-eddy simulation (LES), and molecular dynamics (MD).
In 2004, Lund \textit{et al.} used DNS and LES to show that initial vortices at the KHI shear interface interact and evolve into complex turbulent structures over time\cite{lund2004IEEE}.
Building on this, Ganesh \textit{et al.} (2010) applied MD to observe KHI for the first time in a two-dimensional, strongly coupled Yukawa liquid. They found that the linear growth rate of KHI increases with the strength of the coupling\cite{ganesh2010PRL}.

Later studies investigated how external forces and environmental factors influence KHI.
In 2011, Shadloo \textit{et al.} employed the smoothed particle hydrodynamics  method and demonstrated that the Richardson number ($Ri$) primarily governs KHI development. They observed that when $Ri < 1$, lower values of $Ri$ enhance the probability and accelerate KHI growth \cite{shadloo2011SMSA}.
Building on this, Zellinger \textit{et al.} (2012) used magnetohydrodynamic simulations to demonstrate that gravity and density variations play a crucial role in KHI evolution. They found that increased density enhances KHI, while gravity influences vortex formation and instability propagation \cite{zellinger2012POP}.

Recent research has broadened the scope of KHI studies, incorporating diverse environmental and physical contexts.
In 2014, Rahmani \textit{et al.} employed DNS to show that the Reynolds number ($Re$) significantly influences KHI development and mixing. Their findings indicate that at low $Re$, mixing is primarily driven by the pairing of two KH vortices, whereas at high $Re$, this pairing is suppressed\cite{rahmani2014JOFM}.
In 2018, Liu \textit{et al.} conducted magnetohydrodynamic simulations to examine how magnetic fields influence KHI in parallel shear flows at high $Re$. They analyzed the stabilizing effect of magnetic fields aligned parallel to the main flow and their role in suppressing instability \cite{liu2018POF}.
Most recently, in 2024, Fritts \textit{et al.} used DNS to investigate KHI evolution at different $Re$, emphasizing the effects of tube and knot dynamics on turbulence intensity and energy dissipation in KHI vortices \cite{fritts2024JOTAS}.
Collectively, these studies demonstrate the broad applications of numerical simulations in KHI research, from fundamental vortex dynamics to environmental and material effects.

Physically, KHI is a prototypical nonlinear, nonequilibrium, and multiscale flow.
In its later stages, complex interactions among various nonequilibrium forces and kinetic modes lead to the emergence of diverse small-scale structures, cross-scale interfaces, significant discrete effects, as well as both hydrodynamic nonequilibrium effects (HNEs) and thermodynamic
nonequilibrium effects (TNEs).
Macroscopic fluid models, relying on the quasi-continuum assumption and near-equilibrium approximations, fail to capture discrete effects and TNE behaviors in such multiscale systems.
Microscopic methods, such as molecular dynamics, effectively capture TNEs but are limited by temporal and spatial scales due to their high computational cost.
Mesoscopic models bridge the gap between microscopic and macroscopic scales, balancing physical accuracy and computational efficiency.
These methods include the lattice Boltzmann method (LBM) \cite{succi2018lattice,guo2013lattice,Shan-JFM-2006,GLS-model-PRE2007,fei2023coupled}, discrete velocity method \cite{yang2016JOCP,Shu-POF-2018}, gas-kinetic unified algorithm\cite{lzh-2015}, unified gas-kinetic scheme\cite{xu2010JOCP}, discrete unified gas kinetic scheme\cite{guo2015PRE}, and the discrete Boltzmann modeling and complex physical field analysis method (DBM)\cite{Xu2022BSTP,xu2024FOP,Gan2022}, developed by our group.

{DBM evolved from the physical modeling branch of LBM. Before 2012, the two methods had no differences in physical functionality.
In 2012, Xu \textit{et al.} introduced the use of non-conserved kinetic moments of the distribution function to characterize a system's nonequilibrium state, effects, and behaviors.
This marked the inception of DBM research and its distinction from standard LBM in terms of physical functionality\cite{xu2012FOP}.
In 2024, Xu \textit{et al.} further expanded the DBM framework, refining its coarse-grained modeling strategies and complex physical field analysis. They elucidated its connections to traditional fluid modeling and other kinetic methods, providing insights for the continuous progress of DBM \cite{xu2024FOP}.

Currently, we are following the traditional research route where the physics model construction and the numerical scheme for solving model equations are investigated separately. As both a kinetic model construction method and a tool for analyzing complex physical fields, DBM focuses on two fundamental aspects:
(i) Model construction: This defines the governing equations, which include the evolution equation governing system behavior and the discrete velocity constraint equations.
(ii) Analysis framework: By leveraging higher-order non-conserved moments of $(f - f^{eq})$, DBM efficiently extracts discrete and nonequilibrium states, effects, and behaviors. Additionally, constructing the phase space via the independent components of $\mathbf{M}_{m,n}(f - f^{eq})$ provides an intuitive representation of these nonequilibrium characteristics.

In particular, DBM addresses two critical questions: ``how to include sufficient discrete/TNE states and effects in the modeling equations'', and ``how to extract and analyze the underlying discrete/TNE states and effects from the massive simulation data''. The former is point beyond the NS modeling. The latter is the point beyond current other kinetic simulation methods. The refinement of DBM strategies has further clarified its physical applicability.

In recent years, DBM has been extensively applied to a wide range of complex fluid systems, including: liquid-vapor phase-separating systems \cite{Gan2015,Gan2022},
hydrodynamic instabilities\cite{Lai2016PRE,JieChen,chen2024surface,xu2025influence},
reactive flows (e.g., detonation and combustion) \cite{zhang2016kinetic,ji2022JOCP},
Shock reflection and shock wave/boundary layer interaction \cite{Qiu-2022-POF,Qiu2020AP,Qiu2021APS},
micro-nanoscale flows\cite{zhang2019discrete,zhang2023lagrangian},
plasma kinetics\cite{liu2023POT,Song2024POF}, etc.
As an analytical tool for complex physical fields, DBM can be used to characterize the evolution stages of complex flows and analyze the physical mechanisms of flow systems.
For instance, the temporal evolution of non-equilibrium intensity \( D \) has been employed to distinguish different stages of phase separation, including the spinodal decomposition stage, the transition to domain growth, and the domain growth stage \cite{Gan2015}. Similarly, the evolution of the non-equilibrium component \( \Delta_{3,1y}^* \) has been used to characterize various stages of RTI development and serve as a physical marker for interface tracking in complex flow simulations \cite{Lai2016PRE}. The evolution of viscous stress has been utilized to classify different stages of bubble coalescence \cite{Sun2024POF}, etc.

In recent years, significant progress has also been made in studying KHI and its nonequilibrium effects using DBM \cite{gan2011PRE,lin2016CAF,gan2019FOP,lin2019CITP,lin2021multiple,li2024POF}.

\emph{\textbf{Early foundations}}:
In 2011, Gan \textit{et al.} employed the lattice Boltzmann kinetic method, a precursor to DBM, to investigate the competing effects of the velocity transition layer ($L_u$) and the density transition layer ($L_\rho$) on KHI growth\cite{gan2011PRE}. They introduced the dimensionless parameter $R = L_\rho / L_u$ to quantify these interactions, which was later validated through theoretical analysis\cite{wang2010POP}.
Building on this, Lin \textit{et al.} (2016) developed a compressible two-component DBM capable of capturing detailed interface deformations during KHI evolution, significantly improving multiscale flow modeling \cite{lin2016CAF}.

\textbf{\emph{Investigations into nonequilibrium effects}}:
In 2019, Gan \textit{et al.} applied DBM to examine the influence of viscosity, heat conduction, and the Prandtl number on KHI evolution\cite{gan2019FOP}. They introduced a novel approach combining nonequilibrium behavior tracking with morphological analysis. This method enabled precise characterization of characteristic structures, as well as the width and growth rate of the KH mixing layer.
Key findings included:
(i) Strong spatiotemporal correlations between the mixing layer width, nonequilibrium intensity, and interface boundary length.
(ii) Viscous stress suppresses KHI growth, prolongs the linear stage, reduces maximum disturbance energy, enhances both global and local thermodynamic nonequilibrium, and broadens the nonequilibrium region.
(iii) Heat conduction inhibits small-wavelength growth, favoring the formation of large-scale structures. Under high heat conduction, an initial suppression phase is followed by enhancement, with suppression dominating early stages and enhancement prevailing later.
That same year, Lin \textit{et al.} investigated the role of relaxation time in nonequilibrium behaviors during KHI evolution. Their results indicated that longer relaxation times increased global nonequilibrium intensity, mixing entropy, and mixing free energy. For smaller Atwood numbers, nonequilibrium peaked earlier, with rapid changes in mixing entropy and free energy\cite{lin2019CITP}.

\textbf{\emph{Advances in multicomponent and multi-relaxation-time modeling}}:
Further extending DBM applications, Lin \textit{et al.} (2021) employed a multi-relaxation-time, multi-component DBM to analyze nonequilibrium behaviors in KHI. Their findings revealed that heat conduction and initial temperature gradients had minimal impact on KHI morphology and evolution within the examined parameter range, a notable contrast to single-fluid models\cite{lin2021multiple}.
In 2022, Li \textit{et al.} examined KHI evolution under varying conditions, showing that global density gradients, nonequilibrium intensity, and related regions initially increased before declining. Additionally, tangential velocity accelerated KHI evolution and enhanced nonequilibrium intensity \cite{li2022FOP}.

\textbf{\emph{Recent insights}}:
Most recently, in 2024, Li \textit{et al.} used a two-component DBM to study the effects of initial amplitude and $Re$ on KHI in the presence of a gravitational field. Their results provided a comprehensive understanding of how density gradients, mixing degree, mixing width, viscous stress, and heat flux are influenced by these factors, offering new perspectives on KHI under complex initial and boundary conditions \cite{li2024POF}.

Despite significant progress in applying DBM to KHI studies, these models face severe challenges under extreme conditions. Such conditions include increased shear rates, narrowing of velocity, material, and mechanical interfaces, the emergence of additional nonequilibrium driving forces, the presence of multiple fluid instabilities, and enhanced nonlinear coupling effects. Under these circumstances, KHI exhibits persistent and pronounced TNEs and HNEs, posing substantial challenges to model validity.
Extreme flow environments frequently give rise to such conditions. For instance, in ICF, laser or heavy-ion beams induce steep density and velocity gradients at the shell interface, leading to rapid KHI development. In this regime, strong nonequilibrium effects dominate interface evolution and mixing behavior\cite{he2007TEPJ, soler2022FIA}.
In astrophysics, KHI rapidly develops during supernova explosions and stellar wind interactions with interstellar media. These processes produce complex vortex and turbulence structures due to sharp fluid interfaces and strong velocity shear\cite{burrows2000NAT, drake2010PT, hasegawa2004NAT}. Similarly, in the wake of high-speed aircraft, supersonic jets experience rapid KHI growth driven by strong shear layers, where nonequilibrium effects critically impact wake structures and flow evolution\cite{martens1994AIAA, shi2023POP}.

Therefore, in DBM physical modeling, it is essential to incorporate second-order or higher-order deviations of $f$ from $f^{eq}$. This requires the discrete equilibrium distribution function to satisfy additional higher-order kinetic moment constraints. Such advancements in mesoscopic modeling enable the derivation of higher-order cross-scale constitutive relations at the macroscopic level while enhancing the accuracy of the distribution function at the mesoscopic level. Consequently, developing and applying high-order DBM models is crucial for investigating discrete effects and TNEs in KHI under extreme conditions, as well as for uncovering the fundamental physical mechanisms governing complex flows.

Building on the high-order DBM framework proposed by Gan \textit{et al.} \cite{gan2018PRE}, this paper first constructs multiple discrete velocity sets (DVSs) to discretize phase space under constraints that account for second-order nonequilibrium effects. It then investigates TNE phenomena and their relative significance in the evolution of KHI.
The structure of this paper is as follows:
Section \ref{II} briefly introduces the DBM employed in this study. Section \ref{III} evaluates the performance and differences of various discrete velocity configurations in capturing TNEs and discrete distribution functions and provides recommendations for constructing DVSs. Section \ref{IV} examines the dominant roles of different-order TNEs in KHI development and compares the capabilities of higher-order and lower-order models in capturing TNEs. It further evaluates the effectiveness, extensibility, and limitations of a typical high-order model. Section \ref{V} presents the conclusions, discussions, and future outlooks.

\section{Discrete Boltzmann modeling and complex physical field analysis method}\label{II}

To simplify the complex collision term in the Boltzmann equation, the DBM adopts the Boltzmann-BGK equation:
\begin{equation}\label{e02}
	\partial_t f+\mathbf{v} \cdot \bm{\nabla} f=-\frac{1}{\tau}\left[f-f^{eq}\right],
\end{equation}
where $f$ and $f^{eq}$ denote the distribution function and the Maxwellian equilibrium distribution function, respectively:
\begin{equation}
	f^{eq}=\frac{\rho}{2 \pi R T (2 \pi n R T)^{1/2}} \exp \left[-\frac{(\mathbf{v}-\mathbf{u})^2}{2 R T}-\frac{\eta^2}{2 n R T}\right].
\end{equation}
Here, $\rho$, $\mathbf{v}$, $\mathbf{u}$, and $T$ represent the local density, particle velocity, flow velocity, and temperature, respectively. $R$ is the gas constant, while $n$ accounts for additional degrees of freedom from molecular rotation or vibration. The parameter $\eta$ describes these degrees of freedom.

For numerical simulations, the Boltzmann-BGK equation is discretized, yielding the discrete Boltzmann equation:
\begin{equation}\label{e04}
	\partial_t f_i+\mathbf{v}_i \cdot \bm{\nabla} f_i=-\frac{1}{\tau}\left[f_i-f_i^{eq}\right].
\end{equation}
In DBM, ``discrete'' refers to the discretization of phase space (velocity space). Traditional discretization methods are unsuitable because particle velocities span the entire range $(-\infty,+\infty)$ with arbitrary directions and magnitudes. To address this, DBM adopts a strategy that preserves the invariance of kinetic moment relations during phase-space discretization. Specifically, it ensures that kinetic moments computed via summation remain identical to their integral counterparts:
\begin{equation}\label{e05}
	\bm{\Phi}' = \sum_i f_i \bm{\Psi}'\left(\mathbf{v}_i, \eta_i\right) = \iint f \bm{\Psi}'(\mathbf{v}, \eta) d \mathbf{v} d \eta.
\end{equation}
Here, $\bm{\Psi}'\left(\mathbf{v}_i, \eta_i\right) = \left[1, \mathbf{v}_i, \frac{1}{2}\left(v_i^2+\eta_i^2\right), \mathbf{v}_i \mathbf{v}_i, \dots\right]^T$ represents the set of kinetic moments to be conserved. The number of terms in $\bm{\Psi}'$ increases as higher-order TNE effects are incorporated into the physical modeling.

From the Chapman-Enskog (CE) multi-scale analysis, the physical constraints imposed on $f$ can ultimately be transferred to $f^{eq}$. Consequently, Eq.~(\ref{e05}) simplifies to:
\begin{equation}\label{e06}
	\bm{\Phi} = \sum_i f_i^{eq} \bm{\Psi} \left(\mathbf{v}_i, \eta_i\right) = \iint f^{eq} \bm{\Psi}(\mathbf{v}, \eta) d \mathbf{v} d \eta.
\end{equation}
The accuracy of DBM modeling depends on the number of conserved kinetic moments during discretization. CE analysis indicates that when $f^{eq}$ satisfies the five moment constraints $\mathbf{M}_0$, $\mathbf{M}_1$, $\mathbf{M}_{2,0}$, $\mathbf{M}_2$, and $\mathbf{M}_{3,1}$, the following generalized fluid dynamics equations can be derived from the discrete Boltzmann equation:
\begin{equation}\label{e07}
	\partial_t \rho+\bm{\nabla} \cdot(\rho \mathbf{u})=0,
\end{equation}
\begin{equation}\label{e08}
	\partial_t(\rho \mathbf{u})+\bm{\nabla} \cdot\left(\rho \mathbf{u u}+P \mathbf{I}+\bm{\Delta}_2^*\right)=0,
\end{equation}
\begin{equation}\label{e09}
	\partial_t(\rho e) + \bm{\nabla} \cdot \left[(\rho e + P) \mathbf{u} + \bm{\Delta}_2^* \cdot \mathbf{u} + \bm{\Delta}_{3,1}^*\right] = 0.
\end{equation}
Here, $P=\rho R T$ is the pressure, $e=c_v T+u^2 / 2$ is the energy density, and $c_v=(n+2) R / 2$ represents the specific heat at constant volume. The terms $\bm{\Delta}_2^*$ and $\bm{\Delta}_{3,1}^*$ denote the non-organized momentum flux (NOMF) and non-organized energy flux (NOEF), respectively, which are associated with viscosity and heat conduction.

However, Eqs.~(\ref{e08}) and (\ref{e09}) are not closed, as the explicit forms of $\bm{\Delta}_2^*$ and $\bm{\Delta}_{3,1}^*$ remain unknown. To close the system and derive the first-order constitutive relations $\bm{\Delta}_2^{*(1)}$ and $\bm{\Delta}_{3,1}^{*(1)}$, additional constraints on $f_i^{eq}$, specifically $\mathbf{M}_3$ and $\mathbf{M}_{4,2}$, are required. Further imposing constraints on $\mathbf{M}_4$ and $\mathbf{M}_{5,3}$ enables the derivation of second-order constitutive relations $\bm{\Delta}_2^{*(2)}$ and $\bm{\Delta}_{3,1}^{*(2)}$, facilitating the recovery of Burnett-level fluid dynamics equations through the CE expansion.
In practice, deriving these intricate generalized hydrodynamic equations in DBM modeling is sufficient but not essential, as DBM does not directly solve such complex nonlinear equations. Instead, their derivation primarily serves to establish higher-order constitutive relations, providing a cross-scale constitutive framework for enhancing traditional fluid modeling.

From a kinetic macroscopic modeling perspective, the physical
functions of DBM correspond to the extended hydrodynamic
equations (EHEs).
These equations include not only the evolution of conserved moments, such as mass, momentum, and energy, but also that of closely related nonconserved moments. Examples include the evolution of viscous stress $\bm{\Delta}_2$ and heat flux $\bm{\Delta}_{3,1}$, along with their respective fluxes, $\bm{\Delta}_{3}$ and  $\bm{\Delta}_{4,2}$.
While low-order conserved moment equations form the foundation of hydrodynamics, high-order nonconserved moments become increasingly important as discretization levels rise and nonequilibrium effects intensify.

As nonequilibrium intensity increases, the derivation and solution of the EHEs become significantly more challenging due to their pronounced nonlinearity and the presence of huge amount of nonlinear terms. In multiphase systems, where different media possess distinct velocities and mean flow velocities, the EHEs' forms are not unique. In contrast, the complexity of DBM modeling increases only marginally under such conditions. This is achieved by incorporating the necessary elements into $\bm{\Psi}'\left(\mathbf{v}_i, \eta_i\right)$ \cite{Gan2022}.
Specifically, when the Knudsen number ($Kn$) increases by an order of magnitude, DBM requires the introduction of only two additional non-conserved moments. This advantage renders DBM particularly well-suited for modeling and simulating nonequilibrium complex fluids \cite{xu2024FOP}.

Next, to derive the discrete equilibrium distribution function (DEDF), the nine kinetic moment relations are expressed in matrix form:
\begin{equation}\label{e10}
	\bm{\Phi} = \mathbf{C} \cdot \mathbf{f}^{\mathrm{eq}},
\end{equation}
where $\mathbf{f}^{\mathrm{eq}} = \left(f_1^{\mathrm{eq}}, f_2^{\mathrm{eq}}, \cdots, f_{25}^{\mathrm{eq}}\right)^T$ represents the set of discrete equilibrium distribution functions, and $\bm{\Phi} = \left(\mathbf{M}_0, \mathbf{M}_1, \mathbf{M}_{2,0}, \cdots, \mathbf{M}_{5,3}\right)^T = \left(M_0, M_{1x}, M_{1y}, \cdots, M_{5,3yyy}\right)^T$ denotes the corresponding kinetic moments. The coefficient matrix $\mathbf{C} = \left(\mathbf{c}_1, \mathbf{c}_2, \cdots, \mathbf{c}_{25}\right)$ is a $25 \times 25$ transformation matrix linking the DEDF to the kinetic moments, where each column vector is defined as $\mathbf{c}_{i} = \left[1, v_{ix}, v_{iy}, \cdots, \frac{1}{2}\left(v_i^2 + \eta_i^2\right)v_{iy}v_{iy}v_{iy}\right]^T$.
The DEDF can then be determined as\cite{gan2013EPL}:
\begin{equation}\label{e11}
	\mathbf{f}^{\mathrm{eq}} = \mathbf{C}^{-1} \cdot \bm{\Phi},
\end{equation}
where $\mathbf{C}^{-1}$ is the inverse of $\mathbf{C}$. To ensure the invertibility of $\mathbf{C}$, a two-dimensional discrete velocity model with 25 discrete velocities (D2V25) is constructed for velocity space discretization.

In analyzing complex physical fields, the DBM serves as a coarse-grained descriptive framework rooted in statistical physics, enabling efficient extraction and characterization of nonequilibrium behaviors. Compared to LBM, DBM introduces a systematic approach for detecting, describing, and analyzing nonequilibrium states, effects, and behaviors using high-order nonconserved moments of $(f_i - f_i^{\mathrm{eq}})$:
\begin{equation}\label{e12}
	\begin{aligned}
		\bm{\Delta}_{m,n} &= \mathbf{M}_{m,n}\left(f - f^{\mathrm{eq}}\right) \\
		&= \sum_i \left(\frac{1}{2}\right)^{1 - \delta_{m,n}}
		\left(f_i - f_i^{\mathrm{eq}}\right)
		\underbrace{\mathbf{v}_i \mathbf{v}_i \cdots \mathbf{v}_i}_n
		&\left(\mathbf{v}_i^2 + \eta^2\right)^{\frac{m-n}{2}}.
	\end{aligned}
\end{equation}
and
\begin{equation}\label{e13}
	\begin{aligned}
		\bm{\Delta}_{m,n}^* &= \mathbf{M}_{m,n}^*\left(f - f^{\mathrm{eq}}\right) \\
		&= \sum_i \left(\frac{1}{2}\right)^{1 - \delta_{m,n}}
		\left(f_i - f_i^{\mathrm{eq}}\right)
		\underbrace{\mathbf{v}_i^* \mathbf{v}_i^* \cdots \mathbf{v}_i^*}_n
		& \left(\mathbf{v}_i^{*2} + \eta^2\right)^{\frac{m-n}{2}},
	\end{aligned}
\end{equation}
where $\bm{\Delta}_{m,n}^*$ represents the contraction of an $m$th-order tensor into an $n$th-order tensor, and $\mathbf{v}_i^* = \mathbf{v}_i - \mathbf{u}$ is the velocity relative to the macroscopic flow.
$\bm{\Delta}_{m,n}$  describes the combination of HNEs and TNEs. $\bm{\Delta}_{m,n}^*$ focuses on the thermal fluctuations of microscopic particles relative to $\mathbf{u}$, capturing purely TNEs.

To quantify the overall degree of nonequilibrium, a global TNE intensity is defined as:
\begin{equation}\label{e14}
	D_T^* = \sqrt{\sum_{m,n}\left(\bm{\Delta}_{m,n}^*\right)^2}.
\end{equation}
Its dimensionless counterpart is given by:
\begin{equation}\label{e15}
	\tilde{D}_T^* = \sqrt{\sum_{m,n}\left(\bm{\Delta}_{m,n}^* / T^{\frac{m}{2}}\right)^2}.
\end{equation}

Given the vast amount of data generated in numerical simulations, efficiently extracting and analyzing nonequilibrium information across multiple perspectives is a significant challenge. The degree of nonequilibrium depends on the chosen perspective, and different viewpoints may yield distinct conclusions, highlighting the complexity of nonequilibrium behaviors. To systematically address this, a nonequilibrium strength vector is introduced to describe the TNE intensity across different perspectives :
\begin{equation}\label{e16}
	S_{\mathrm{TNE}} = \left\{
	\begin{array}{c}
		\Delta_{2xx}^*, \Delta_{2xy}^*, \Delta_{2yy}^*, \Delta_{3,1x}^*, \Delta_{3,1y}^*, \cdots, \Delta_{5,3xyy}^*, \Delta_{5,3yyy}^*, \\
		\left|\bm{\Delta}_2\right|, \left|\bm{\Delta}_{3,1}\right|, \left|\bm{\Delta}_3\right|, \left|\bm{\Delta}_{4,2}\right|, \left|\bm{\Delta}_{4}\right|, \left|\bm{\Delta}_{5,3}\right|, D_T^*, \tilde{D}_T^*, \\
		\tau, |\bm{\nabla} u|, |\bm{\nabla} \rho|, |\bm{\nabla} T|, Kn, \cdots
	\end{array}
	\right\}.
\end{equation}
The first row lists individual components of $\bm{\Delta}_{m,n}^*$, the second row represents aggregate nonequilibrium intensities, and the third row includes traditional macroscopic gradients such as velocity, density, and temperature gradients.

It is important to note that the BGK model inherently assumes $\mathrm{Pr} = 1$ due to its single-relaxation-time formulation. To overcome this limitation, an external forcing term can be introduced into the BGK collision operator:
\begin{equation}
	I_i = \left[A R T + B\left(\mathbf{v}_i - \mathbf{u}\right)^2\right] f_i^{\mathrm{eq}},
\end{equation}
where $A = -2B$ and $B = \frac{1}{2\rho T^2} \bm{\nabla} \cdot \left[\frac{n+4}{2} \rho T q \bm{\nabla} T\right]$. This modification adjusts the thermal conductivity coefficient to $\kappa = c_p P(\tau + q)$, resulting in a tunable Prandtl number $\mathrm{Pr} = {\tau}/{(\tau + q)}$. This method is straightforward and avoids introducing additional kinetic moment constraints.

\section{Key considerations for constructing discrete velocity sets}\label{III}


\begin{figure}[htbp]
	\begin{center}
		\includegraphics[width=0.48\textwidth]{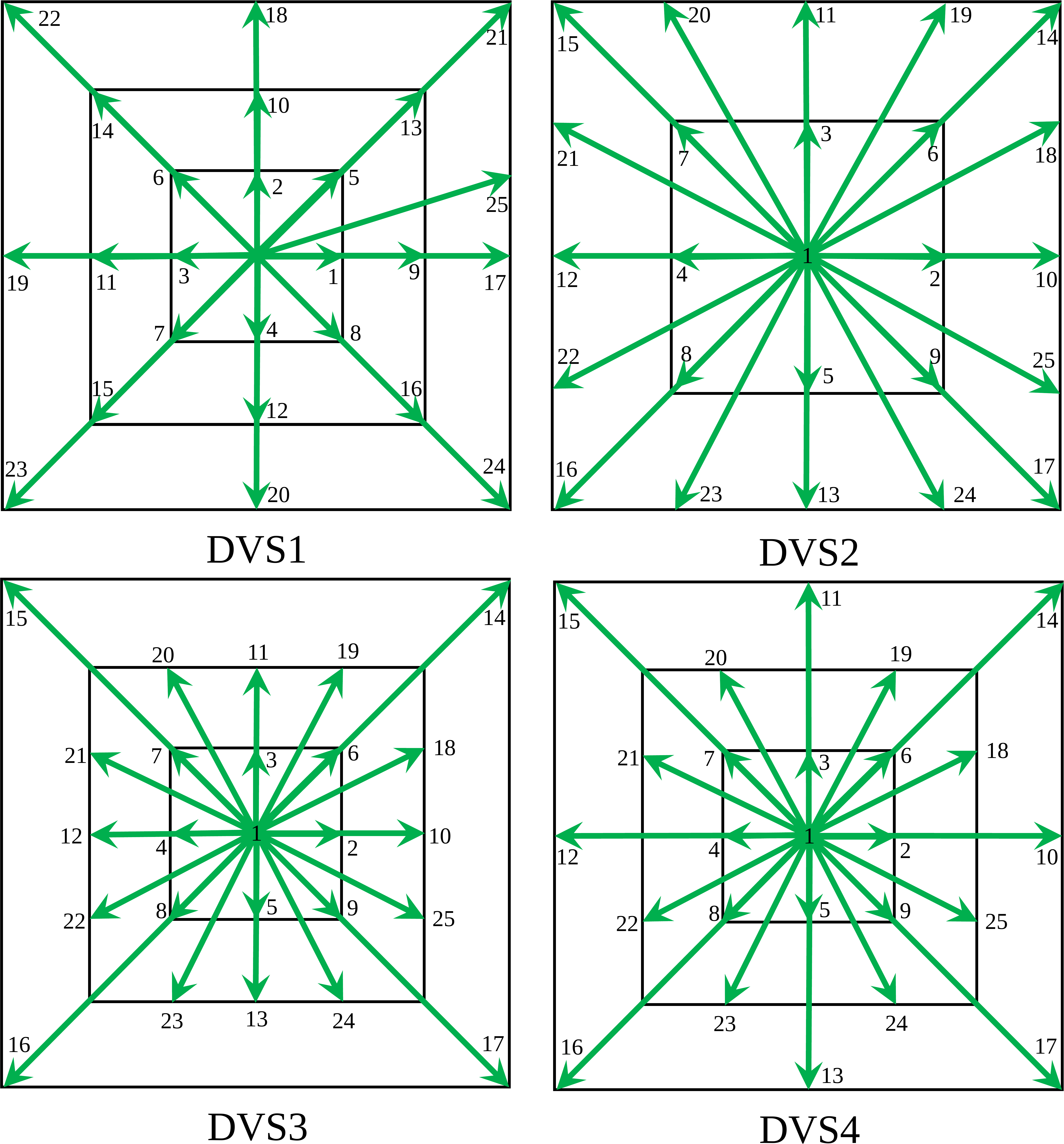}
	\end{center}
	\caption{Schematic representation of four two-dimensional discrete velocity sets, each comprising 25 discrete velocities.}
	\label{fig01}
\end{figure}
\begin{figure*}[htbp]
	\begin{center}
		\includegraphics[width=0.8\textwidth]{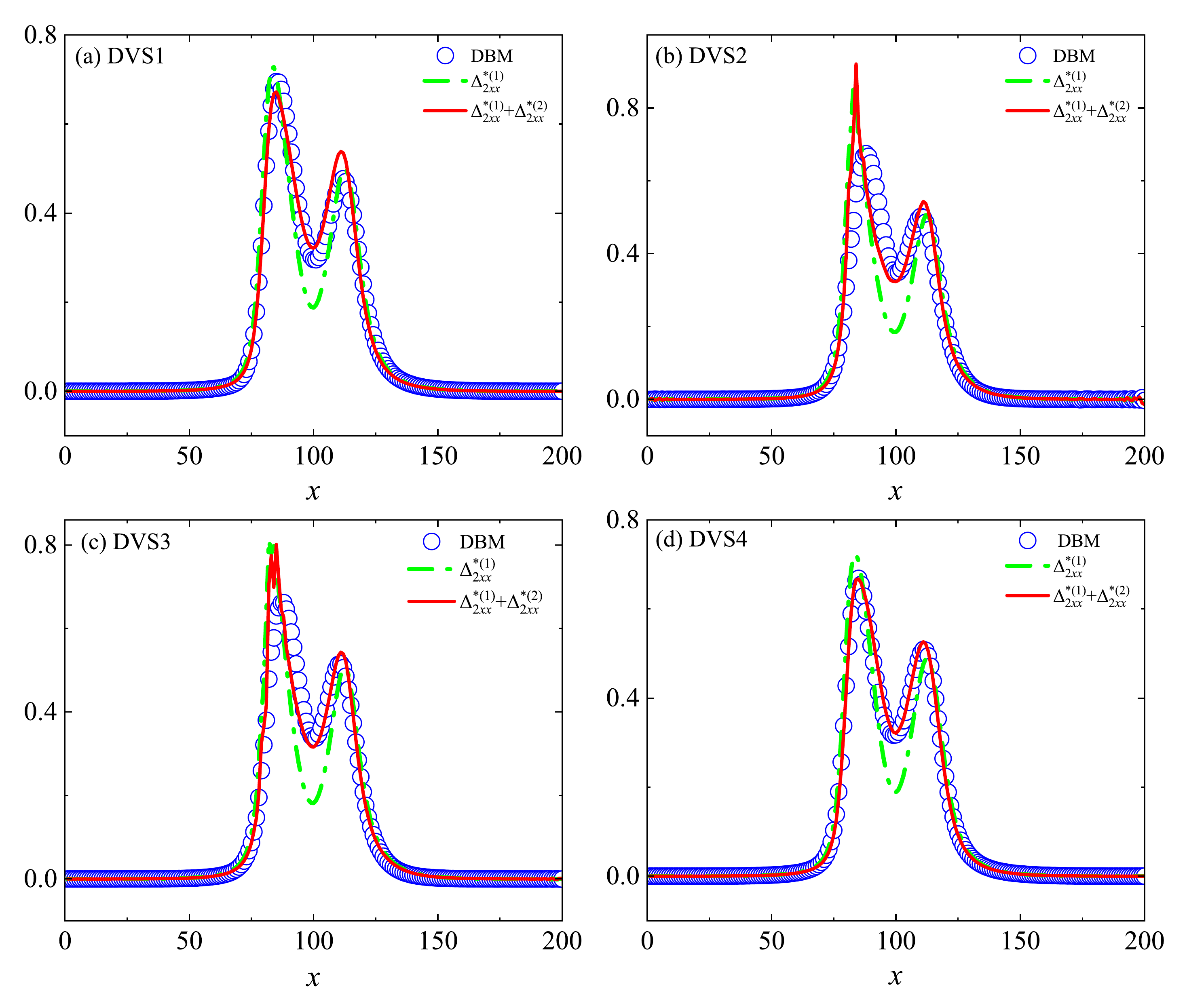}
	\end{center}
	\caption{Distributions of viscous stress along $y=0.3 L_y$ at $t=0.017$ for the four DVSs. (a) DVS1, (b) DVS2, (c) DVS3, (d) DVS4.}
	\label{fig02}
\end{figure*}

As a theoretical framework for model construction and complex physical field analysis, the DBM imposes only the essential physical constraints required to describe the considered TNEs.
It does not specify temporal, spatial, or phase-space discretization schemes.
However, numerically, the choice of DVS plays a crucial role in determining the ability to accurately capture physical effects in DBM simulation.
In 2018, Gan \textit{et al.} proposed a two-dimensional discrete velocity set with 26 discrete velocities (D2V26) for phase-space discretization \cite{gan2018PRE}. Building on this foundation, four new velocity sets (DVS1-DVS4) are developed to incorporate second-order nonequilibrium effects while optimizing computational efficiency and stability (Figure \ref{fig01}). The key characteristics of these sets are as follows:
(i) DVS1: Retains the first 25 discrete velocities from Ref. \onlinecite{gan2018PRE}.
(ii) DVS2: A more symmetric 25-velocity set designed for improved performance \cite{chen2025arXiv}.
(iii) DVS3: Extends DVS2 by introducing additional velocity magnitudes, resulting in a more diverse and evenly distributed velocity set.
(iv) DVS4: Further optimizes velocity distribution by increasing the diversity of magnitude combinations.
The detailed configurations of the four DVSs are given as follows:
\begin{equation}\label{e17}
	\text{DVS1: } (v_{ix}, v_{iy}) =
	\begin{cases}
		cyc:c(\pm 1, 0), &  1 \leq i \leq 4, \\
		c(\pm 1, \pm 1), & 5 \leq i \leq 8, \\
		cyc:c(\pm 2, 0), & 9 \leq i \leq 12, \\
		c(\pm 2, \pm 2), & 13 \leq i \leq 16, \\
		cyc:c(\pm 3, 0), & 17 \leq i \leq 20, \\
		c(\pm 3, \pm 3), & 21 \leq i \leq 24, \\
		c(2, 1), & i = 25.
	\end{cases}
\end{equation}
\begin{equation}\label{e18}
	\text{DVS2: } (v_{ix}, v_{iy}) =
	\begin{cases}
		c(0, 0), & i = 1, \\
		cyc:c(\pm 1, 0), & 2 \leq i \leq 5, \\
		c(\pm 1, \pm 1), & 6 \leq i \leq 9, \\
		cyc:c(\pm 2, 0), & 10 \leq i \leq 13, \\
		c(\pm 2, \pm 2), & 14 \leq i \leq 17, \\
		cyc:c(\pm 2, \pm 1), & 18 \leq i \leq 25.
	\end{cases}
\end{equation}
\begin{equation}\label{e19}
	\text{DVS3: } (v_{ix}, v_{iy}) =
	\begin{cases}
		c(0, 0), & i = 1, \\
		cyc:c(\pm 1, 0), & 2 \leq i \leq 5, \\
		c(\pm 1, \pm 1), & 6 \leq i \leq 9, \\
		cyc:c(\pm 2, 0), & 10 \leq i \leq 13, \\
		c(\pm 3, \pm 3), & 14 \leq i \leq 17, \\
		cyc:c(\pm 2, \pm 1), & 18 \leq i \leq 25.
	\end{cases}
\end{equation}
\begin{equation}\label{e20}
	\text{DVS4: } (v_{ix}, v_{iy}) =
	\begin{cases}
		c(0, 0), & i = 1, \\
		cyc:c(\pm 1, 0), & 2 \leq i \leq 5, \\
		c(\pm 1, \pm 1), & 6 \leq i \leq 9, \\
		cyc:c(\pm 3, 0), & 10 \leq i \leq 13, \\
		c(\pm 3, \pm 3), & 14 \leq i \leq 17, \\
		cyc:c(\pm 2, \pm 1), & 18 \leq i \leq 25.
	\end{cases}
\end{equation}
The parameters $\eta_i$ are defined as follows:
For DVS1: $\eta_i = i\eta_0$ $(1 \leq i \leq 4)$; $\eta_i = (i-1)\eta_0$ $(5 \leq i \leq 8)$; $\eta_i = 0$, otherwise.
For DVS2 and DVS3: $\eta_i = 10\eta_0$ $(i = 1)$; $\eta_i = \eta_0$ $(2 \leq i \leq 6)$; $\eta_i = 0$, otherwise.
For DVS4: $\eta_i = 4\eta_0$ $(1 \leq i \leq 2)$; $\eta_i = 3\eta_0$ $(i = 3)$; $\eta_i = 2\eta_0$ $(i = 4)$; $\eta_i = \eta_0$ $(i = 5$ or $14 \leq i \leq 17)$; $\eta_i = 0$, otherwise.
Here, ``cyc'' denotes cyclic permutation, and $c$ and $\eta_0$ are free parameters chosen to ensure the invertibility of matrix $\mathbf{C}^{-1}$, optimizing computational stability and efficiency.

The numerical stability of the four DVSs and their ability to describe TNEs are evaluated using two approaches:
(i) Comparing numerical solutions with analytical solutions of TNEs during a fluid collision process.
(ii) Analyzing the configuration of the discrete distribution function to identify factors affecting the stability and accuracy of the DVSs.
The initial conditions for the numerical simulation are defined as:
\begin{equation}\label{e21}
	\rho(x, y) = \frac{\rho_L + \rho_R}{2} - \frac{\rho_L - \rho_R}{2} \tanh \left( \frac{x - N_x \Delta x / 2}{L_\rho} \right),
\end{equation}
\begin{equation}\label{e22}
	u_x(x, y) = -u_0 \tanh \left( \frac{x - N_x \Delta x / 2}{L_u} \right),
\end{equation}
where \( L_\rho \) and \( L_u \) denote the widths of the density and velocity transition layers, respectively. The parameters \( \rho_L \) and \( \rho_R \) represent the fluid densities far from the interface on the left and right sides. The computational domain is a square with a side length of \( 0.3 \), discretized using a uniform \( 200 \times 200 \) grid.
The initial physical conditions are set as $4 \rho_L=\rho_R=4,4 P_L=P_R=4, u_y=0, L_\rho=L_u=20$, with parameters $\tau=2 \times 10^{-3}, u_0=1.5$, $\Delta t=10^{-5}$, $\Delta x=\Delta y=0.0015$.


Figure \ref{fig02} presents the distribution of viscous stress along $y=0.3L_y$ at $t=0.017$, computed using the four DVSs. The blue circles represent the DBM numerical solutions, while the green dashed line and red solid line correspond to the first- and second-order analytical solutions for viscous stress, respectively.
It is evident from Fig. \ref{fig02} that strong TNEs arise near the fluid interface, primarily driven by initial pressure and density gradients. Across all DVSs, the first-order analytical solution for viscous stress deviates from the numerical solution, underscoring the necessity of incorporating the second-order deviation $f^{(2)}$ in the distribution function $f$ from its equilibrium counterpart $f^{eq}$.
However, the numerical solutions obtained using DVS1, DVS2, and DVS3 exhibit noticeable discrepancies and numerical oscillations relative to their analytical counterparts, particularly near the local nonequilibrium peak. This finding highlights the critical role of phase-space discretization in determining the numerical stability and accuracy of the model.

In contrast, the simulation results obtained using DVS4 closely match the second-order analytical solution for viscous stress while demonstrating excellent numerical stability.
These findings underscore that a well-designed DVS is fundamental to the physical fidelity of DBM simulation, ensuring both stability and accuracy in modeling nonequilibrium effects.

\begin{figure*}[tbp]
	\begin{center}
		\includegraphics[width=1.0\textwidth]{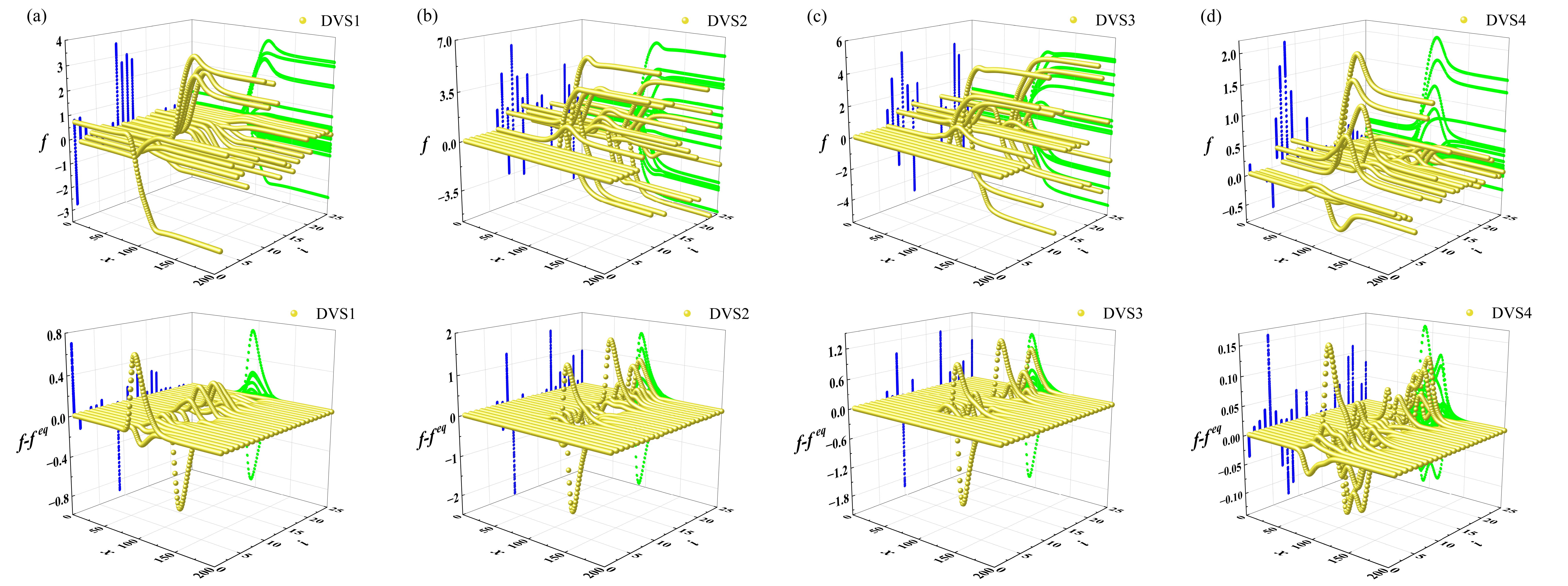}
	\end{center}
	\caption{Distribution of the distribution function $f$ and its deviation $\left(f-f^{e q}\right)$ along $y=0.3 L_y$ at $t=0.017$ for the four DVSs.}
	\label{fig03}
\end{figure*}

Numerical stability is a critical concern in computational fluid dynamics, particularly in nonequilibrium simulations. The primary factors influencing numerical stability can be broadly classified into (i) the accuracy of physical model, (ii) the numerical schemes for spatiotemporal derivative computations and (iii) the discretization of phase space.
Regarding the accuracy of physical models, all four DVSs belong to high-order models that surpass the NS level. High-order DBM integrates high-order constitutive relations to describe high-order nonequilibrium effects, enabling a more precise capture of small-scale structures and dynamical patterns, as well as their interactions. These models not only enhance physical functionality but also improve numerical stability by accurately representing the intrinsic driving mechanisms of system evolution.
Regarding numerical schemes, all four DVSs employ a fifth-order weighted essentially non-oscillatory finite difference scheme for spatial derivatives and a second-order implicit Runge-Kutta finite difference scheme for temporal derivatives. These high-order schemes are sufficiently robust to suppress numerical oscillations\cite{jiang1996JOCP,ascher1997ANM}.
Consequently, the observed oscillations primarily stem from the discretization of phase space, which plays a crucial role in preserving moment relations post-discretization.  This, in turn, influences the effectiveness of the various DVSs in implementing the functionality of the original physical model.

The preservation of kinetic moments can be further categorized into two aspects: theoretical preservation and numerical preservation. Theoretical preservation guarantees the existence of an invertible matrix for the DVS, ensuring a well-defined equilibrium distribution function. This condition is met by all the aforementioned DVSs.
Numerical preservation, however, is evaluated based on the smoothness of the distribution function $f$ throughout the simulation.
Figure \ref{fig03} presents distributions of  $f$ and  $\left(f-f^{e q}\right)$ along the horizontal line $y=0.3 L_y$ at $t=0.017$, as computed using the four DVSs.
Key observations are as follows:

(i) The distribution functions \( f \) computed using DVS1, DVS2, and DVS3 exhibit significant fluctuations, with maximum differences \( (f_{\max} - f_{\min}) \) of 0.7, 1.8, and 1.2, respectively. In contrast, DVS4 yields a significantly smaller maximum difference of 0.16, indicating superior continuity in \( f \). This improved continuity plays a crucial role in reducing numerical oscillations.

(ii) In DVS1, the number of contributing distribution functions \( f \) (i.e., those with \( f \neq 0 \)) is relatively small, suggesting an insufficient number of effective \( f \) values. This limitation hinders the enforcement of nonequilibrium constraints. Similar issues occur in DVS2 and DVS3, whereas DVS4 exhibits better coverage of effective \( f \) values, significantly enhancing its ability to capture TNEs.

(iii) The deviations \( (f - f^{eq}) \) obtained from DVS1, DVS2, and DVS3 exhibit pronounced oscillations, especially in DVS2. An ideal DVS should yield a smooth and continuous \( (f - f^{eq}) \) curve, ensuring the continuity of higher-order non-conservative moments \( \mathbf{M}_{m,n}(f - f^{eq}) \).

(iv) DVS4 exhibits superior performance by utilizing a greater number of \( f \) values, yielding smoother and more continuous \( f \) and \( (f - f^{eq}) \). The discrete velocity magnitudes in DVS4 are more uniformly distributed, with the following amplitudes:
For $i=1$, $\left|v_i\right|=0$; for $2 \leq i \leq 5$, $\left|v_i\right|=1$; for $6 \leq i \leq 9$, $\left|v_i\right|=\sqrt{2}$; for $10 \leq i \leq 13$, $\left|v_i\right|=3$; for $14 \leq i \leq 17$, $\left|v_i\right|=3\sqrt{2}$; and for $18 \leq i \leq 25$, $\left|v_i\right|=\sqrt{5}$.
This distribution better aligns with the actual particle velocity distribution in phase space, improving numerical stability and accuracy.

Based on this analysis, DVS4 is selected for subsequent numerical simulations. This study provides practical insights into optimizing phase space discretization and developing robust, high-order DBM simulation.
In future work, we will apply von Neumann stability analysis to theoretically investigate its impact on numerical stability \cite{Gan2008PA}.

\section{Multiscale TNEs and their relative importance in KHI}\label{IV}


In KHI, velocity shear induces a substantial velocity gradient in the fluid, which is closely associated with viscous stress. While various nonequilibrium effects contribute to KHI dynamics, viscous stress plays a dominant role, particularly during vortex formation and growth.
As a characteristic feature, viscous stress encapsulates the inherent complexity of the system. As a governing effect, its uneven distribution and localized dissipation profoundly influence the evolution of KHI. This section specifically investigates the roles of different orders of viscous stress in KHI, with similar methodologies applicable to the analysis of other thermodynamic nonequilibrium measures.

Physically, nonequilibrium behavior in a system is driven by factors such as relaxation time \( \tau \), density gradients \( \bm{\nabla} \rho \), temperature gradients \( \bm{\nabla} T \), and velocity gradients \( \bm{\nabla} \mathbf{u} \). 
To fully characterize the intensity of TNE, we introduce a TNE intensity vector with five components:
\begin{equation}
	\mathbf{S}_{\text{TNE}} = \left( \tau, \bm{\nabla} \mathbf{u}, \bm{\nabla} \rho, \bm{\nabla} T, \bm{\Delta}_{2}^* \right).
\end{equation}
By varying these factors, different orders of viscous stress can be generated during the evolution of KHI.

The following research methodology is employed to investigate the role of viscous stress and its dominance at various orders during the evolution of KHI:

\emph{\textbf{(i) Analysis of $\bm{\Delta}_2^{*(1)}$ and $\bm{\Delta}_2^{*(2)}$}}:
Each term in the expressions for \( \bm{\Delta}_2^{*(1)} \) and \( \bm{\Delta}_{2}^{*(2)} \) (see Appendix \ref{NCR}) is analyzed to reveal the underlying physical mechanisms in different scenarios. Based on this analysis, a criterion is established to determine the dominance of viscous stress at different orders during KHI evolution, using the relative intensity ratio:
$R_{\text{TNE}}={\bm{\Delta}_{2}^{*(2)}}/{\bm{\Delta}_{2}^{*(1)}}$.

\emph{\textbf{(ii) Comparison of high-order and low-order models}}:
When higher-order nonequilibrium effects become significant, the differences between high-order and low-order models are evaluated. The analysis focuses on macroscopic quantities, nonequilibrium effects, and kinetic moments, highlighting the limitations of low-order models in capturing complex nonequilibrium dynamics.

\textbf{\emph{(iii) Evaluation of high-order model capabilities}}:
The simulation capabilities of the high-order model are systematically assessed to demonstrate its effectiveness, extensibility, and limitations in describing higher-order nonequilibrium phenomena.

\textbf{\emph{(iv) Presentation of a nonequilibrium phase diagram}}:
A nonequilibrium phase diagram is constructed to visually illustrate the multiscale characteristics of KHI evolution. This diagram highlights the interplay, competition, and cooperation between nonequilibrium effects of different orders.

In the numerical simulation, all relevant physical quantities, including the discretized Boltzmann equation, particle velocities, and fluid mechanics variables, are nondimensionalized using appropriate reference variables. To achieve this, three fundamental reference variables are selected: the characteristic flow length \( L_0 \), the reference density \( \rho_0 \), and the reference temperature \( T_0 \). Derived variables, such as the reference flow velocity \( u_0 \) and pressure \( P_0 \), are defined as $u_0 = \sqrt{R T_0}$, $P_0 = \rho_0 R T_0$,
where \( R \) is the specific gas constant, equal to \( 287 \, \mathrm{J} / (\mathrm{kg} \cdot \mathrm{K}) \).
For the simulation, the fluid is assumed to be air under standard conditions. Accordingly, the reference density, temperature, and flow velocity are specified as
$\rho_0 = 1.165 \, \mathrm{kg} / \mathrm{m}^3$, $T_0 = 303 \, \mathrm{K}$, $u_0 \approx 294.892 \, \mathrm{m} / \mathrm{s}$.

The initial conditions for the simulation are defined as follows:
\begin{equation}\label{e23}
	\rho(x, y)=\frac{\rho_L+\rho_R}{2}-\frac{\rho_L-\rho_R}{2} \tanh \left(\frac{x}{D_\rho}\right),
\end{equation}
\begin{equation}\label{e24}
	v(x, y)=\frac{v_L+v_R}{2}-\frac{v_L-v_R}{2} \tanh \left(\frac{x}{D_v}\right),
\end{equation}
\begin{equation}\label{e25}
	P_L=P_R=P.
\end{equation}
Here, \( D_\rho = 18 \) and \( D_v = 2 \) represent the widths of the density and velocity transition layers, respectively. The left (\( \rho_L = 8.0 \)) and right (\( \rho_R = 2.0 \)) fluid densities, as well as the pressures on both sides (\( P_L = P_R = 3.7 \)), are specified. The velocities of the left and right fluids in the \( y \)-direction, far from the interface, are given by \( v_L = -v_R = v_0 \).
To initiate vortex formation in the KHI, a small velocity perturbation is introduced in the \( x \)-direction:
\begin{equation}\label{e26}
	u_x(x, y) = u_0 \sin (k y) \exp \left(-2 \pi \left| x - \frac{L_x}{2} \right|\right).
\end{equation}
The perturbation amplitude is set to \( u_0 = 50 \Delta y \), and the initial perturbation wavenumber is given by \( k = {2 \pi}/{L_y} \).

Boundary conditions for the KHI simulations are implemented as follows: periodic boundary conditions are applied in the \( y \)-direction, while outflow boundary conditions, commonly used in KHI studies, are enforced in the \( x \)-direction \cite{keppens1999POP, perucho2007PRE, li2009CITP}.
Boundary conditions play a critical role in fluid models, directly influencing nonequilibrium behaviors. Properly designed kinetic boundary conditions are essential for ensuring physical accuracy, numerical stability, and computational precision.

For the high-order model, the D2V25 DVS4 configuration described in Section \ref{III} is adopted. The low-order model employs the first 16 components from DVS1, corresponding to a Navier-Stokes (NS)-level DBM widely used in previous studies \cite{gan2013EPL, gan2019FOP}.
For analyzing the dominance of nonequilibrium effects at different orders, the computational domain is defined as a rectangle with a length of 1.2 and a height of 0.4, discretized into a uniform \( 600 \times 200 \) grid. For the comparison between high- and low-order models, the computational domain is adjusted to a square of size \( 0.4 \times 0.4 \), discretized into a \( 200 \times 200 \) grid. This adjustment is necessary to mitigate the numerical instability encountered in the low-order model.
The remaining parameters are set as \( c = 1.05 \), \( \eta_0 = 1.0 \),  \( \Delta t = 6 \times 10^{-5} \), and \( \gamma = 2.0 \). All parameters are kept identical for both high- and low-order models to ensure a fair comparison.

\subsection{First-order TNEs dominance case}\label{1stD}

\begin{figure}[htbp]
	\begin{center}
		\includegraphics[width=0.5\textwidth]{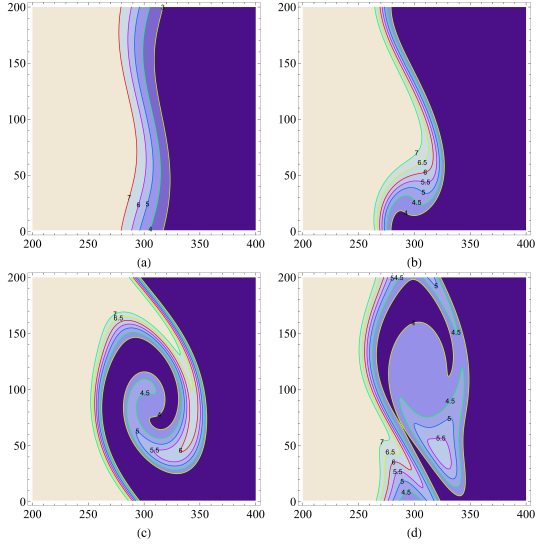}
	\end{center}
	\caption{Density evolution patterns of KHI at four moments for the first-order TNEs dominance case: (a) $t = 0.24$, (b) $t = 0.60$, (c) $t = 1.68$, (d) $t = 2.40$.}
	\label{fig04}
\end{figure}

\begin{figure*}[htbp]
	\begin{center}
		\includegraphics[width=0.78\textwidth]{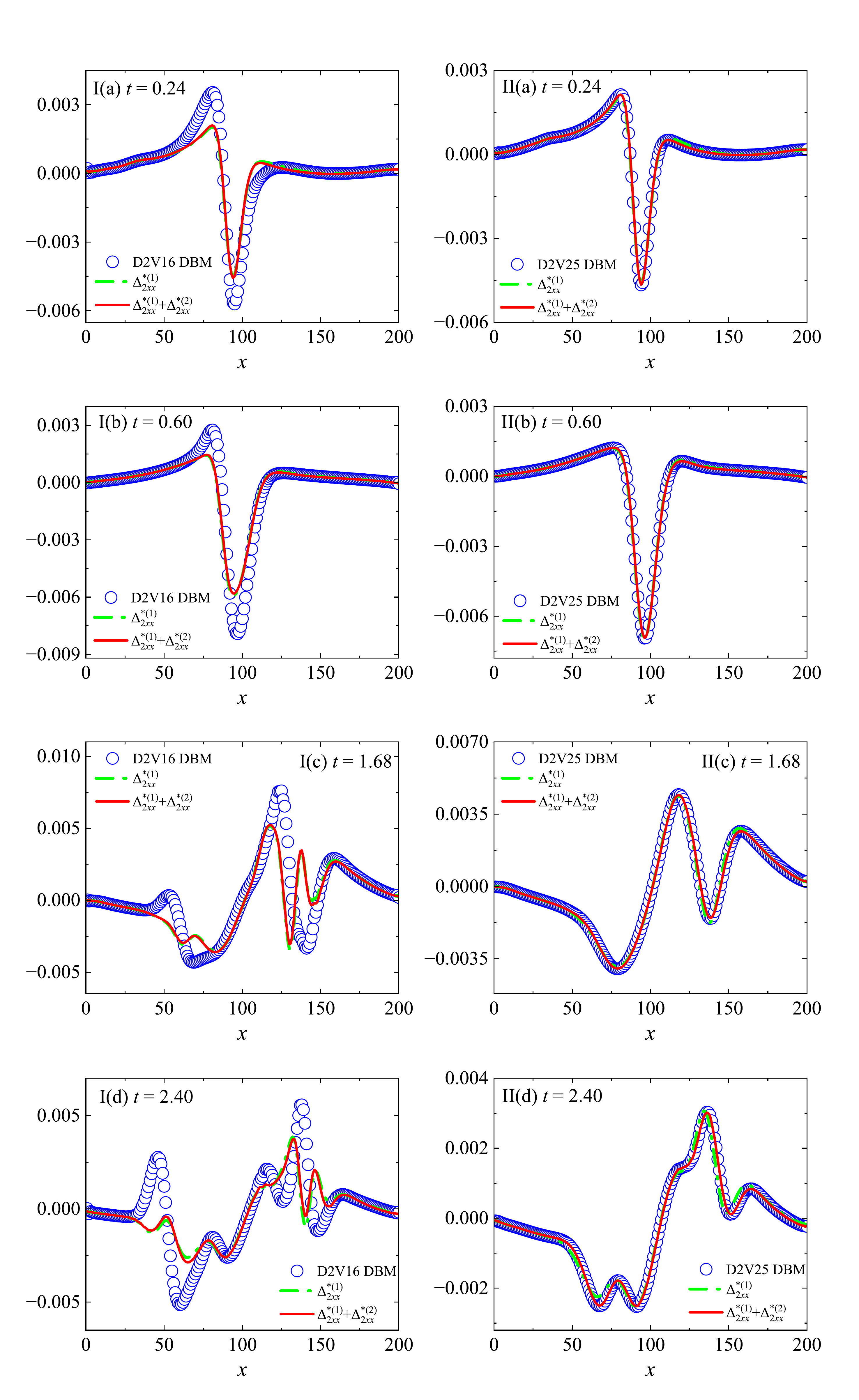}
	\end{center}
	\caption{Distributions of $\Delta_{2 x x}^*$ along $y = 0.7L_y$ at $t = 0.24$, $0.60$, $1.68$, and $2.40$, computed using D2V16 (left column) and D2V25 (right column) DVSs.}
	\label{fig05}
\end{figure*}

\begin{figure}[htbp]
	\begin{center}
		\includegraphics[width=0.5\textwidth]{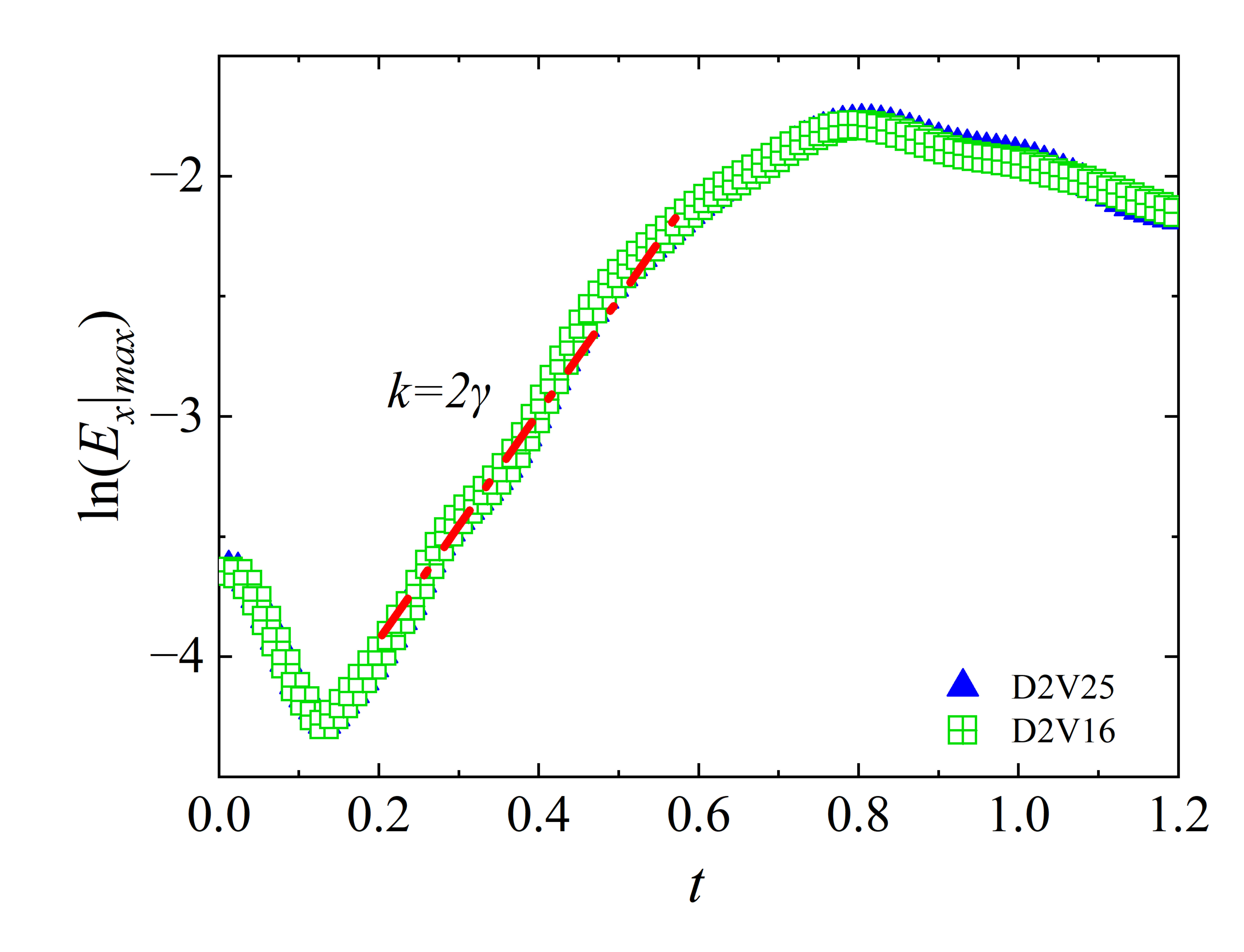}
	\end{center}
	\caption{Time evolution of peak disturbance kinetic energy for D2V25 and D2V16 DVSs in the first-order nonequilibrium dominant case.}
	\label{fig06}
\end{figure}

In this case, we set \(v_L = -v_R = 0.3\) and \(\tau = 3 \times 10^{-4}\). Figure \ref{fig04} illustrates the density fields at four time instances during the evolution of KHI.
At \( t = 0.24 \), the interface begins to oscillate due to shear forces induced by the initial perturbation and the velocity gradient. As the system progresses into the linear growth phase, small perturbations on the interface grow, and by \( t = 1.68 \), they develop into well-defined large-scale vortices. The strengthening of these vortices increases the velocity differences between fluid layers, amplifying shear stress and promoting mixing.
At \( t = 2.40 \), a mixing layer forms around the initial interface. The interface remains continuous and smooth, demonstrating the model's capability to accurately capture interface deformation.

Figure \ref{fig05} presents the distributions of viscous stress \( \Delta_{2xx}^* \) along the line \( y = 0.7L_y \) at the same time instances as in Fig. \ref{fig04}, computed using the D2V16 (left column) and D2V25 (right column) DVSs.
The right column of Fig. \ref{fig05} clearly shows that, throughout the evolution of KHI, viscous stress is predominantly governed by first-order nonequilibrium effects. This is because the initial shear velocity gradients primarily drive \( \Delta_{2xx}^{*(1)} \), while \( \Delta_{2xx}^{*(2)} \) remains negligible.
As a result, velocity gradients have a stronger influence than density and temperature gradients on the distribution of viscous stress during KHI evolution.

The linear growth rate of the KHI can be quantified by the slope of the perturbation kinetic energy peak over time \cite{amerstorfer2010POP, ganesh2010PRL, obergaulinger2010AA}.
As shown in Fig. \ref{fig06}, in the first-order nonequilibrium dominant regime, the linear growth rates from the first-order and second-order models are nearly identical. This suggests that, in terms of macroscopic quantities and slowly varying variables, the D2V16 and D2V25 DVSs achieve comparable numerical accuracy.

However, from a mesoscopic and rapidly varying perspective, as shown in the left column of Fig. \ref{fig05}, even when first-order nonequilibrium effects dominate viscous stress, a significant discrepancy remains between the viscous stress computed by the low-order model and the analytical solution.
This deviation initially appears in regions with strong nonequilibrium effects [Fig. \ref{fig05}I(a)-I(b)] and gradually spreads across the entire domain [Fig. \ref{fig05}I(c)-I(d)].
Over time, the discrepancy between the numerical and higher-order analytical solutions grows due to the accumulation of physical errors.

In contrast, the viscous stress computed using the D2V25 DVS remains highly consistent with the second-order analytical solution throughout the evolution of the KHI.
While the D2V16 DVS adequately captures the KHI growth rate, its representation of nonequilibrium effects exhibits significant deviations that are difficult to detect using traditional methods.
As a fundamental constitutive relation, errors in viscous stress accumulate over time and directly affect the accuracy of macroscopic quantities.
This discrepancy becomes particularly evident in Fig. \ref{fig06} for \( t > 0.8 \).

\subsection{Activation of second-order TNEs}\label{2ndD}

\begin{figure}[tbp]
	\begin{center}
		\includegraphics[width=0.52\textwidth]{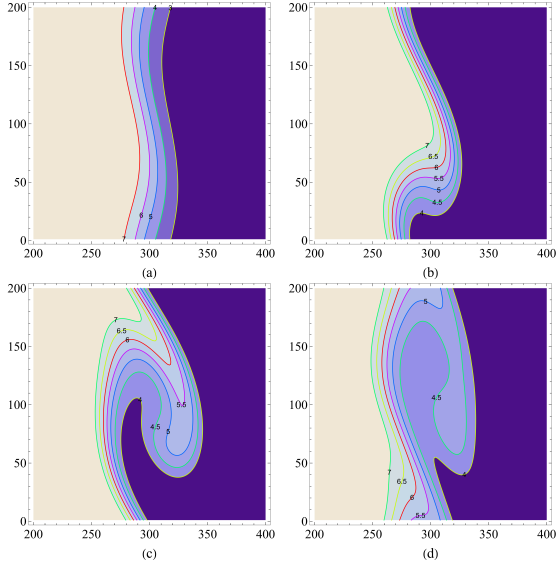}
	\end{center}
	\caption{Density evolution patterns of KHI at four moments with the activation of second-order TNEs: (a) $t = 0.24$, (b) $t = 0.72$, (c) $t = 1.68$, (d) $t = 2.40$.}
	\label{fig07}
\end{figure}

Next, with \( v_L = -v_R = 0.3 \) fixed, \( \tau \) is increased to \( 8 \times 10^{-4} \). Figure \ref{fig07} illustrates the temporal evolution of the density field at various stages of KHI.
Compared to Fig. \ref{fig04}, a larger relaxation time increases physical viscosity, thereby suppressing KHI development.
As a result, the KHI arms widen, the vortices shrink, and the system tends to form larger-scale structures \cite{kim2011JOF, awasthi2014IJO}.

\begin{figure*}[tbp]
	\begin{center}
		\includegraphics[width=0.78\textwidth]{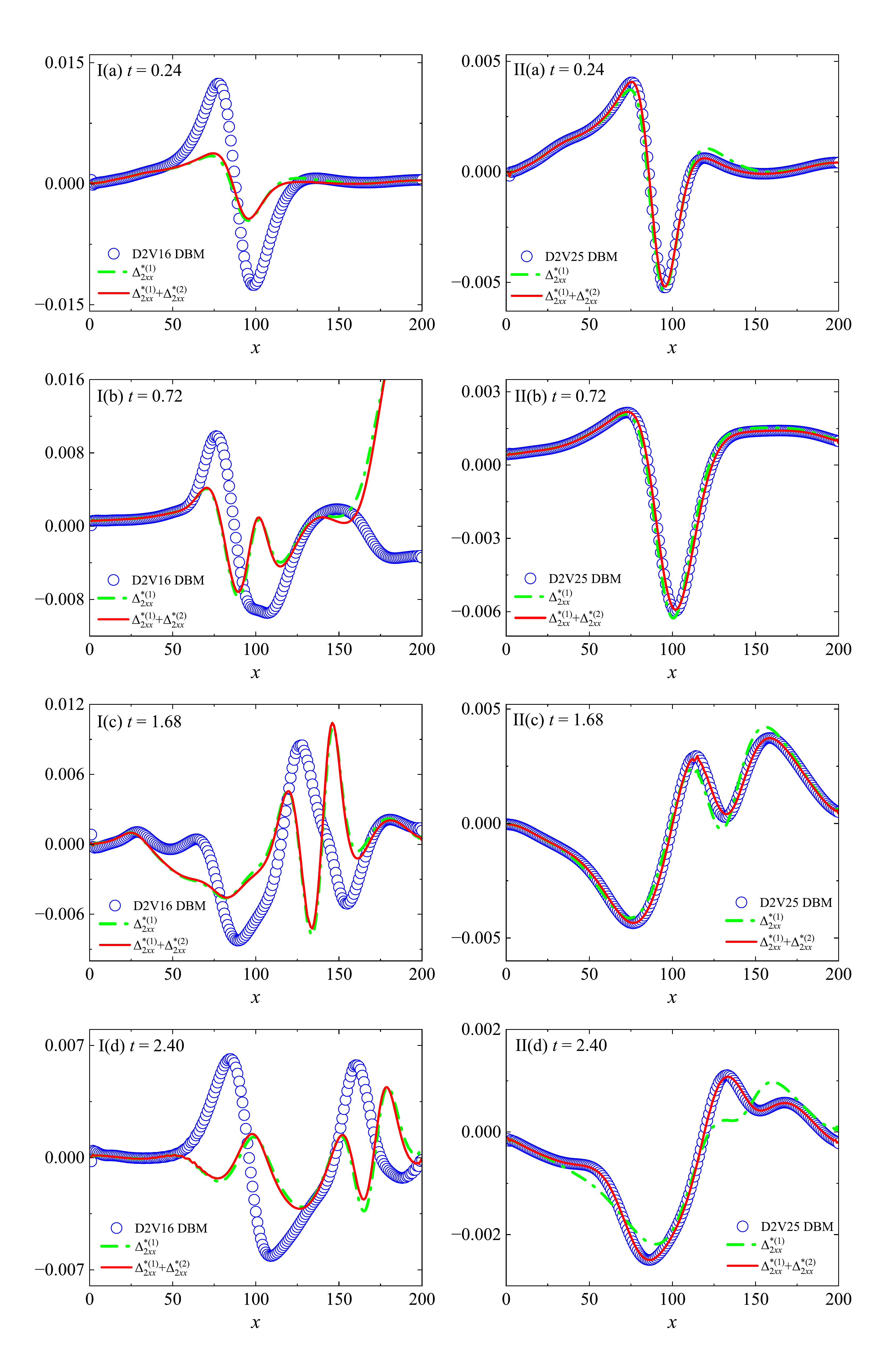}
	\end{center}
	\caption{Distributions of $\Delta_{2 x x}^*$ along $y = 0.7L_y$ at different times: $t = 0.24$, $0.72$, $1.68$, and $2.40$, computed using D2V16 (left column) and D2V25 (right column).}
	\label{fig08}
\end{figure*}

\begin{figure*}[tbp]
	\begin{center}
		\includegraphics[width=0.78\textwidth]{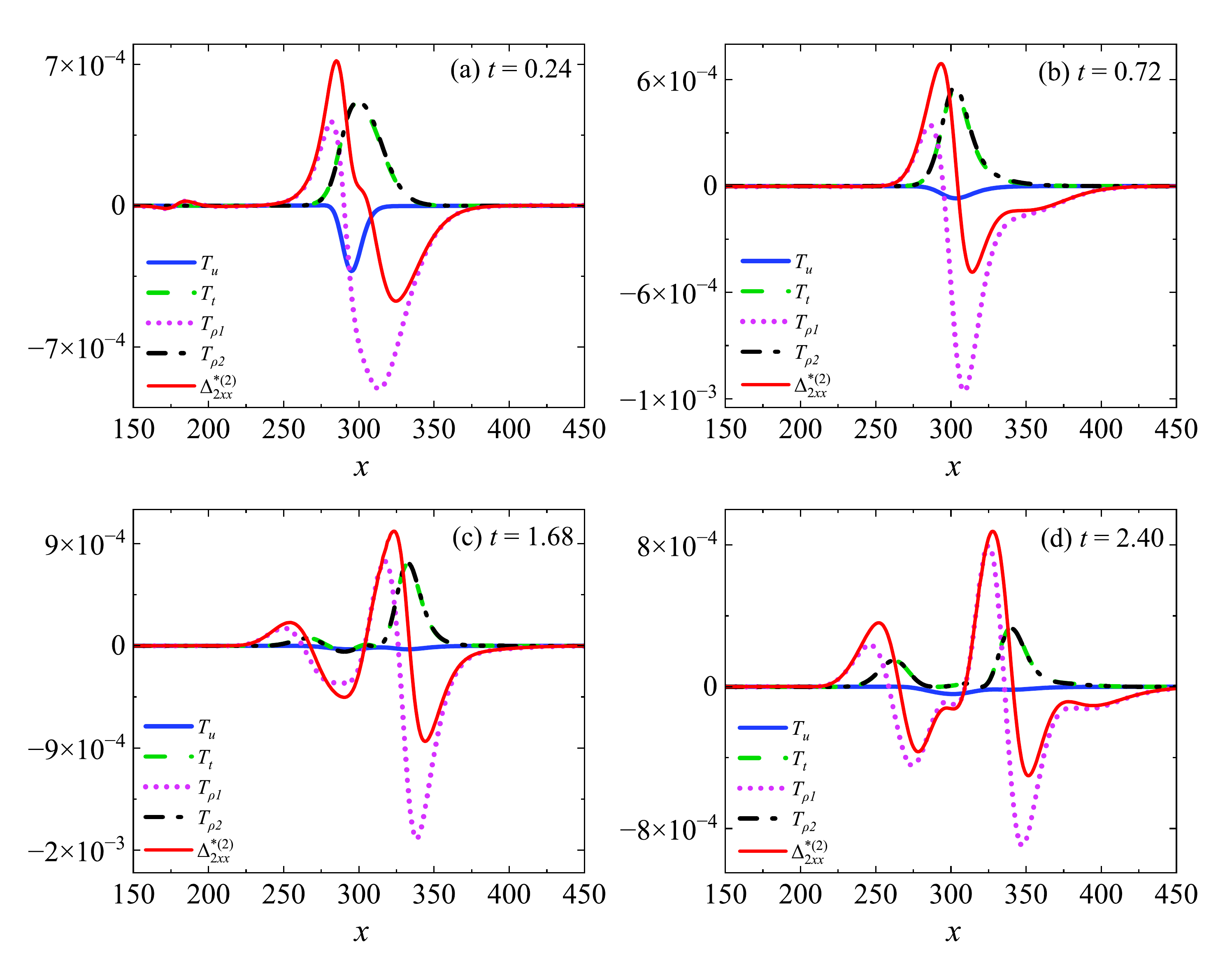}
	\end{center}
	\caption{Distributions of $\Delta_{2 x x}^{*(2)}$ and its four components along $y = 0.7L_y$ at different times: (a) $t = 0.24$, (b) $t = 0.72$, (c) $t = 1.68$, (d) $t = 2.40$.}
	\label{fig09}
\end{figure*}

\subsubsection{Comparison of viscous stress and its components}

Figure \ref{fig08} compares the viscous stress obtained from high-order (right column) and low-order (left column) models at various stages of KHI evolution with the analytical solution.
Compared to the weak nonequilibrium case, the deviation of D2V16 results from the analytical solution increases significantly.
In contrast, the D2V25 results closely align with the second-order nonequilibrium analytical solution, demonstrating its capability to capture stronger nonequilibrium effects.

The gap between the green dashed line and the red solid line in the right column of Fig. \ref{fig08} indicates that second-order nonequilibrium effects are significant, and even dominant, in both the early and late stages of the evolution.
At \( x = 325 \) and \( t = 2.40 \), the first-to-second-order TNE intensity ratio, \( R_{\text{TNE}} \), reaches 2.40, highlighting the local dominance of second-order nonequilibrium effects.
This underscores the multiscale nature of both the flow structures and TNEs.
Compared to the first-order TNE-dominant case, only the relaxation time \( \tau \) and the spatiotemporal ratio \( \tau/\Delta x \) are adjusted. This suggests that the dominance of second-order nonequilibrium effects is sensitive to \( \tau \) and/or \( \tau/\Delta x \).
This aspect will be further discussed in Sec. \ref{D2nd}.

Next, we analyze the physical mechanisms governing the onset of second-order nonequilibrium effects by decomposing the analytical expression of \( \Delta_{2xx}^{*(2)} \) into its fundamental components.
Based on the nature of the driving forces, second-order nonequilibrium effects can be decomposed into four terms\cite{Gan2025}.

(i) The velocity gradient term \( T_u \) is composed of the squares or products of velocity gradients, capturing the nonlinear effects of velocity variations:
\begin{equation}\label{e27}
	\begin{aligned}
		T_u= & 2 n_2^{-2} \tau^2 \rho R T\left[n_{-2} n_1\left(\partial_x u_x\right)^2+n_1 n_2\left(\partial_y u_x\right)^2\right. \\
		& \left.-4 n \partial_x u_x \partial_y u_y-n_2\left(\partial_x u_y\right)^2-n_{-2}\left(\partial_y u_y\right)^2\right].
	\end{aligned}
\end{equation}

(ii) The temperature gradient term \( T_t \) is related to the square of the temperature gradient, reflecting the intensity of temperature gradients within the system:
\begin{equation}\label{e28}
	T_t=2 n_2^{-2} \tau^2 \rho R^2\left[n_1 n_2\left(\partial_x T\right)^2-n_2\left(\partial_y T\right)^2\right].
\end{equation}

(iii) The second-order density gradient term \( T_{\rho 1} \) involves the second derivatives of density, reflecting the density curvature in the \( x \)- and \( y \)-directions:
\begin{equation}
	T_{\rho 1}=-2 n_2^{-2} \tau^2 R^2 T^2\left(n_1 n_2 \frac{\partial^2}{\partial x^2} \rho-n_2 \frac{\partial^2}{\partial y^2} \rho\right)
\end{equation}

(iv) The first-order density gradient term \( T_{\rho 2} \) is related to the square of the first-order density gradient, reflecting the role of density inhomogeneity in viscous stress:
\begin{equation}\label{e30}
	T_{\rho 2}=2 n_2^{-2} \tau^2 \frac{R^2 T^2}{\rho}\left[n_1 n_2\left(\partial_x \rho\right)^2-n_2\left(\partial_y \rho\right)^2\right].
\end{equation}
Here, \( n \) denotes the additional degrees of freedom, with \( n_a = n + a \).
Consequently, the total second-order nonequilibrium stress is expressed as:
\begin{equation}\label{D2xx2}
	\Delta_{2 x x}^{*(2)}=T_u+T_t+T_{\rho 1}+T_{\rho 2}.
\end{equation}

\begin{figure}[tbp]
	\begin{center}
		\includegraphics[width=0.45\textwidth]{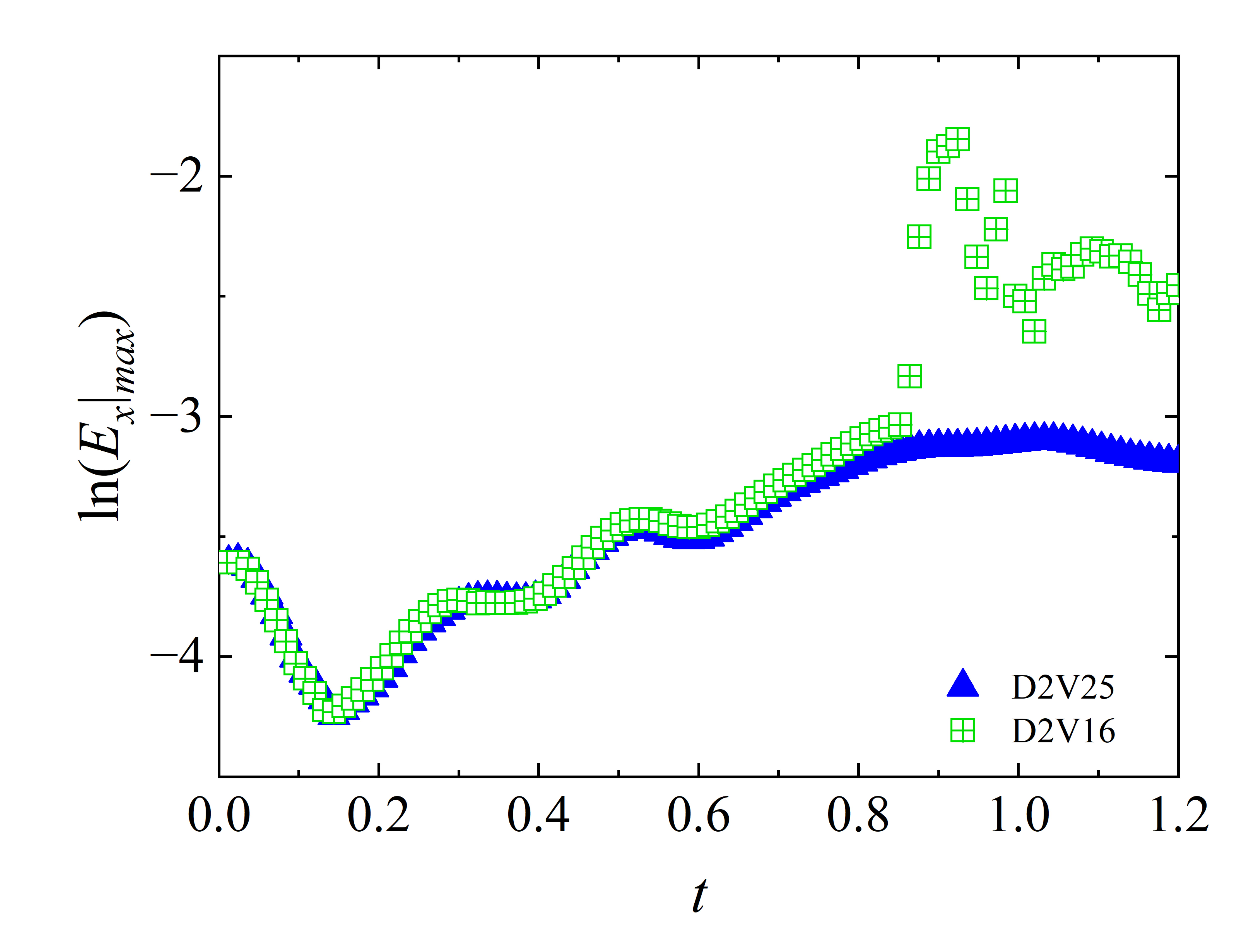}
	\end{center}
	\caption{Logarithmic variation of peak disturbance kinetic energy over time for D2V25 and D2V16 DVSs in the second-order nonequilibrium dominant case.}
	\label{fig10}
\end{figure}

Figure \ref{fig09} shows the distributions of \( \Delta_{2xx}^* \) and its four components along \( y = 0.7L_y \) at \( t = 0.24, 0.72, 1.68, \) and \( 2.40 \). The following key observations can be made:

(i) In the early stages of KHI evolution, the velocity gradient term (\( T_u \)) dominates. As the instability develops, its fluctuations gradually diminish and stabilize. This trend results from dissipative mechanisms, such as viscosity and heat conduction, which convert kinetic energy into internal energy, progressively reducing the velocity gradient.

(ii) The temperature gradient term, \( T_t \), and the first-order density gradient term, \( T_{\rho 2} \), remain comparable in magnitude throughout the evolution. This behavior arises from the nearly constant pressure, both initially and during evolution, which constrains the product of density and temperature gradients. Substituting the temperature \( T = \frac{P}{\rho} \) and density \( \rho = \frac{P}{T} \) into the analytical expressions for \( T_t \) and \( T_{\rho 2} \) yields consistent results.

(iii) The second-order density gradient term, \( T_{\rho 1} \), is consistently the largest, highlighting its dominant role in governing nonequilibrium effects.

(iv) Owing to the distinct driving mechanisms of each component in \( \Delta_{2xx}^{*(2)} \), the peaks of different nonequilibrium terms occur at varying spatial locations, exhibiting different fluctuation ranges.

(v) As shown in Fig. \ref{fig08}, second-order nonequilibrium effects emerge at \( t = 0.24 \), vanish at \( t = 0.72 \), and reappear after \( t = 1.68 \), progressively intensifying thereafter. This suggests that the dominance of different nonequilibrium driving forces varies across different stages of KHI evolution.

Figure \ref{fig09} further supports these findings. Throughout the evolution of KHI, \( T_{\rho 1} \) consistently dominates \( \Delta_{2xx}^{*(2)} \). At \( t = 0.24 \), \( T_u \) is the primary contributor to second-order nonequilibrium effects.
However, as \( T_u \) diminishes, the negative contribution of \( T_{\rho 1} \) at \( t = 0.72 \) offsets the positive contributions of other terms, temporarily eliminating second-order nonequilibrium effects.
As KHI evolves, the overall flow velocity decreases, reducing the influence of the velocity gradient term. Meanwhile, the effects of density and temperature gradients intensify.
Since first-order nonequilibrium effects are primarily driven by velocity gradients, they diminish relative to second-order effects, which become dominant in the later stages.

\begin{figure}[tbp]
	\begin{center}
		\includegraphics[width=0.5\textwidth]{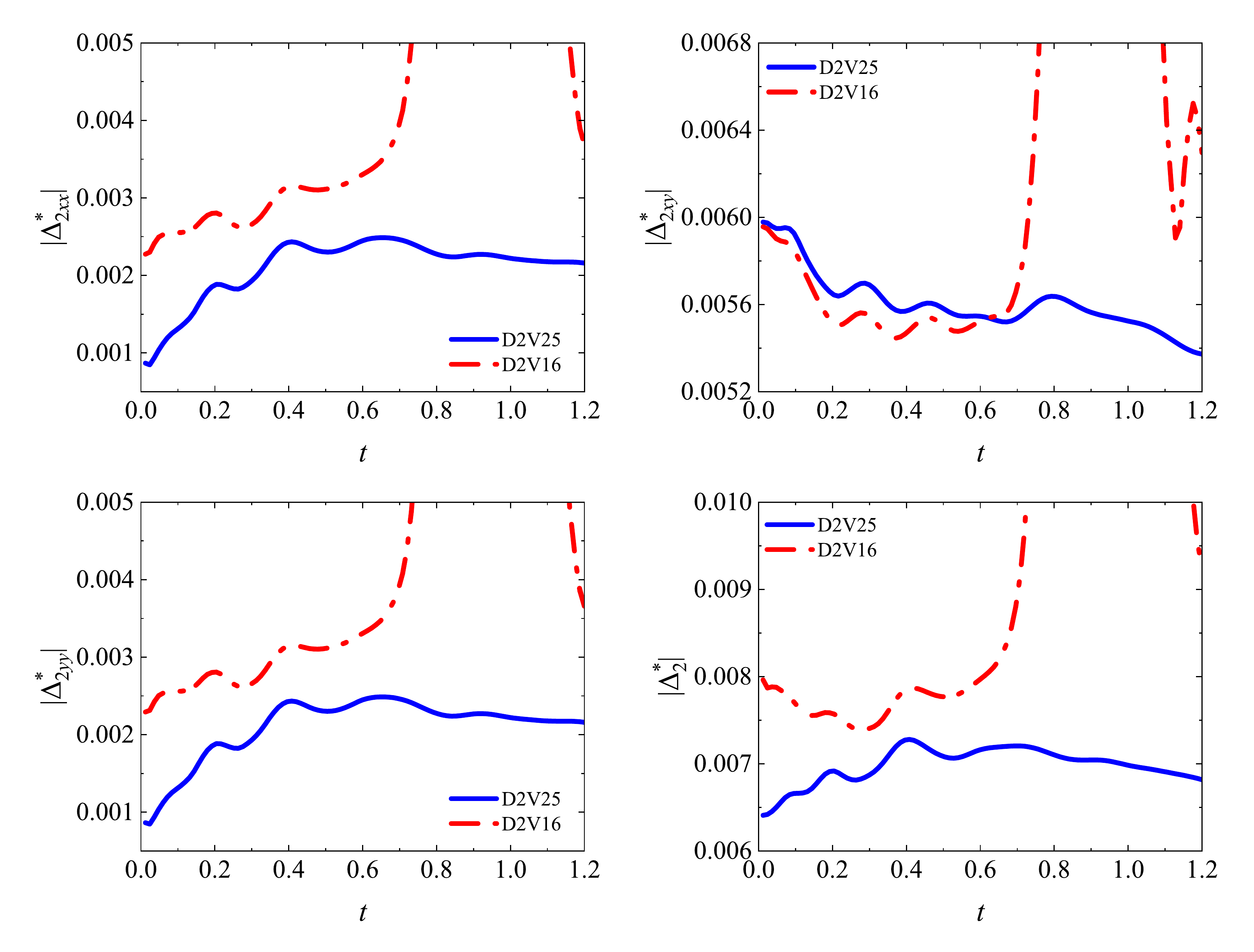}
	\end{center}
	\caption{Time evolution of global average viscous stress components $\left|\Delta_{2 x x}^*\right|$, $\left|\Delta_{2 x y}^*\right|$, $\left|\Delta_{2 y y}^*\right|$, and total $\left|\Delta_2^*\right|$ in the second-order nonequilibrium dominant case.}
	\label{fig11}
\end{figure}

Next, we compare the simulation results of the high- and low-order models from multiple perspectives.

\begin{figure*}[htbp]
	\begin{center}
		\includegraphics[width=0.95\textwidth]{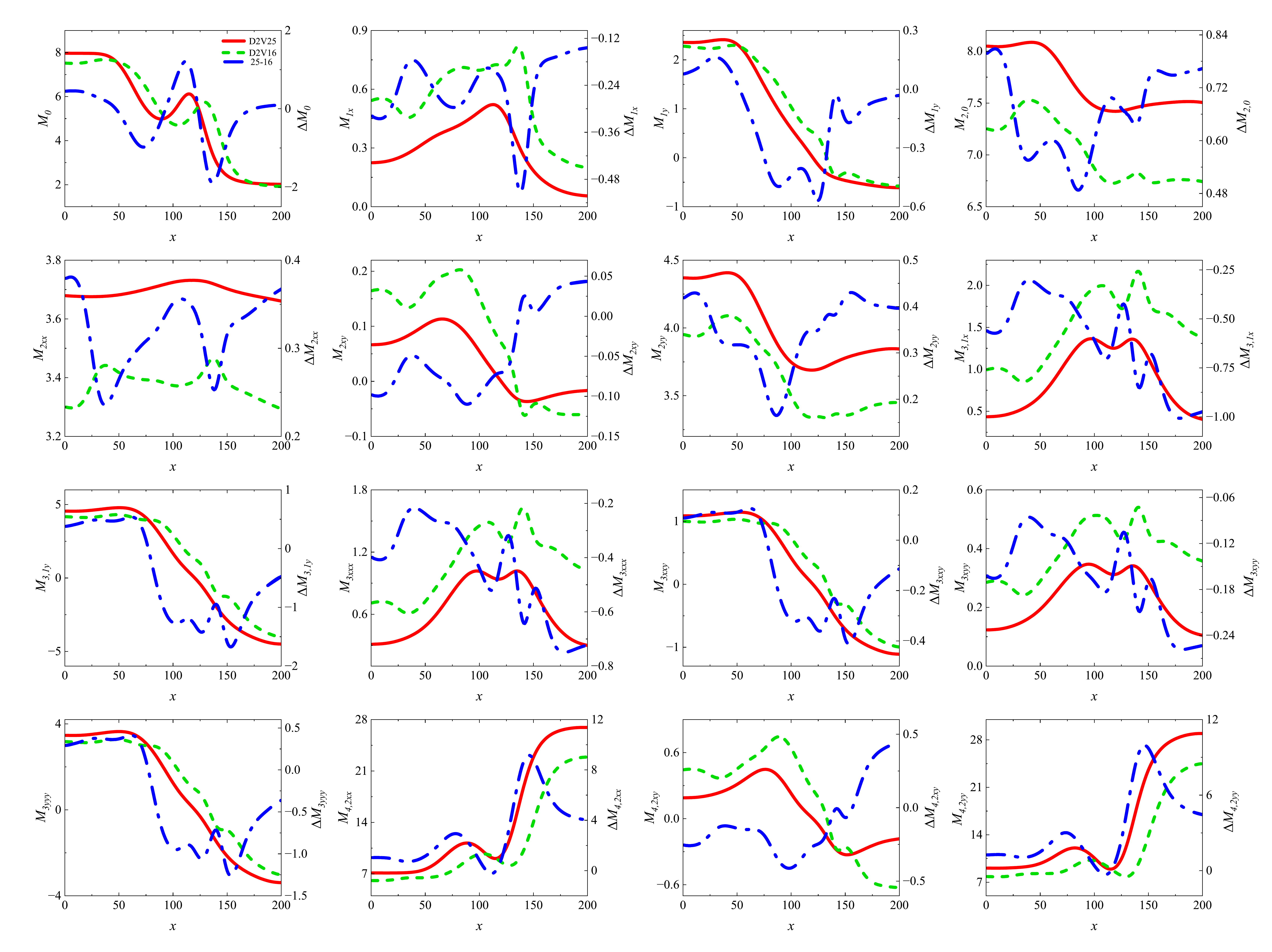}
	\end{center}
	\caption{Kinetic moments $\mathbf{M}_0$, $\mathbf{M}_1$, $\mathbf{M}_{2,0}$, $\mathbf{M}_2$, $\mathbf{M}_{3,1}$, $\mathbf{M}_3$, and $\mathbf{M}_{4,2}$, along with their differences between D2V25 and D2V16 DVSs, along $y = 0.7L_y$ at $t = 1.68$.}
	\label{fig12}
\end{figure*}

\begin{figure*}[tbp]
	\begin{center}
		\includegraphics[width=0.8\textwidth]{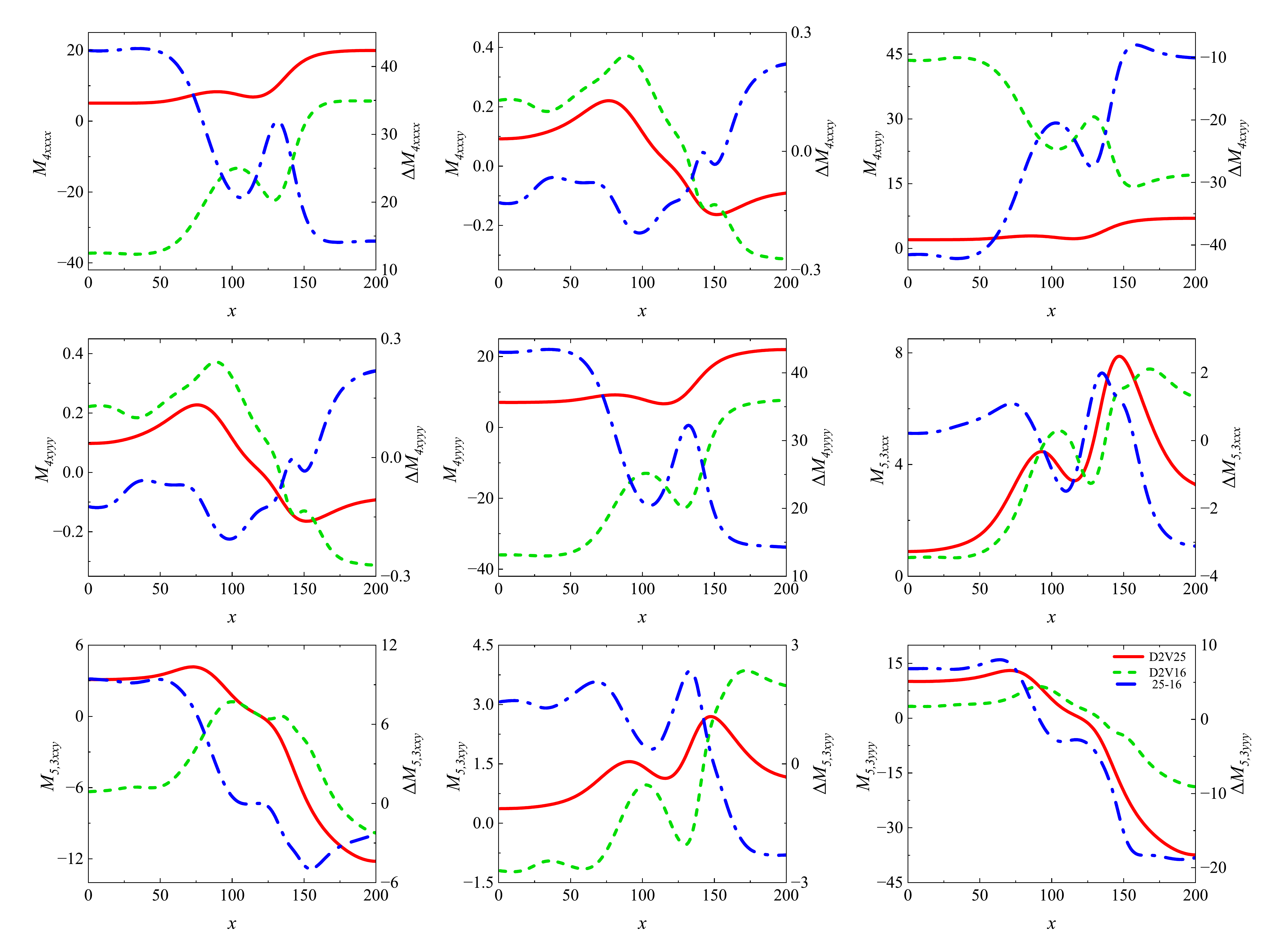}
	\end{center}
	\caption{Kinetic moments $\mathbf{M}_{4}$ and $\mathbf{M}_{5,3}$, along with their differences between D2V25 and D2V16 DVSs, along $y = 0.7L_y$ at $t = 1.68$.}
	\label{fig13}
\end{figure*}
\begin{figure}[tbp]
	\begin{center}
		\includegraphics[width=0.5\textwidth]{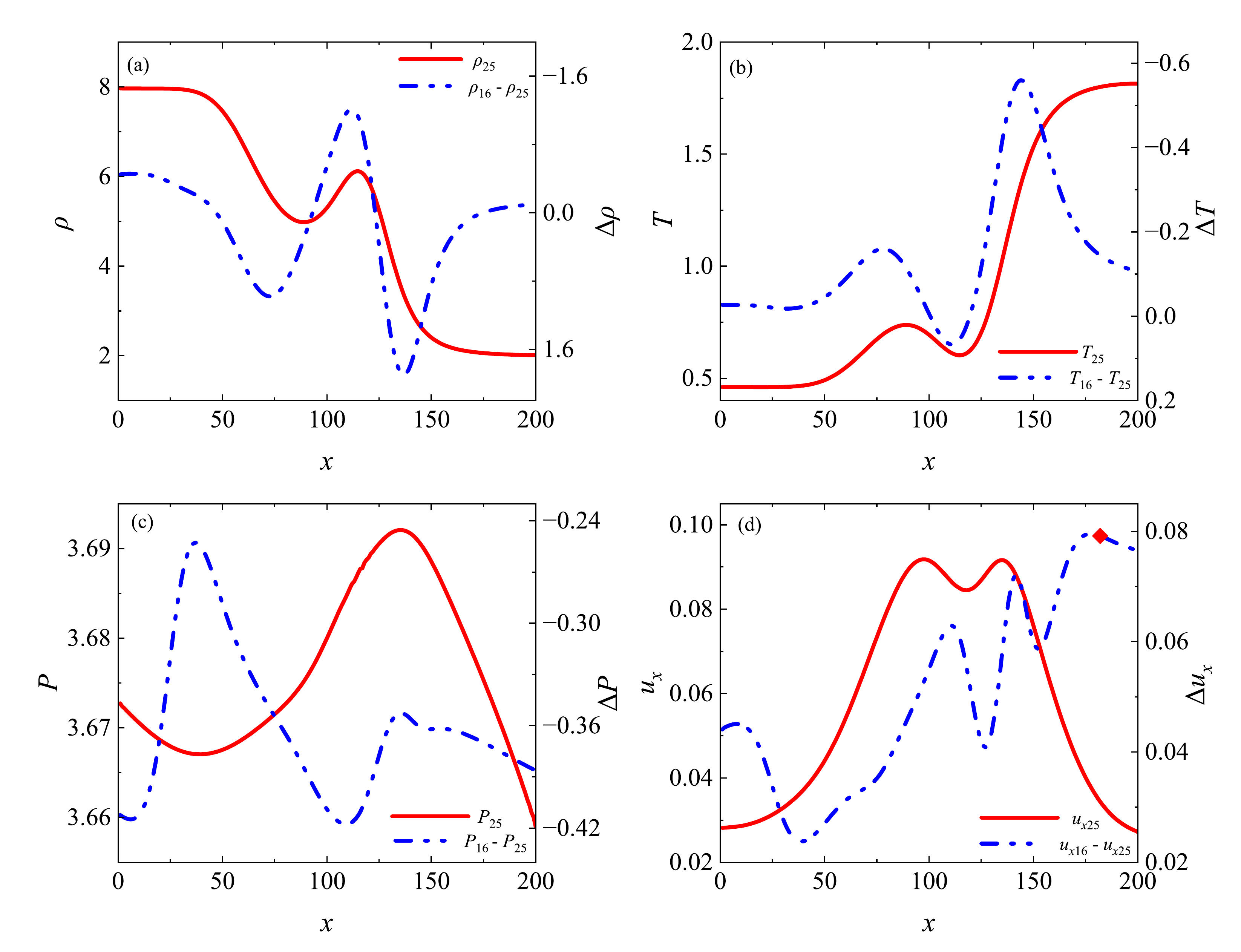}
	\end{center}
	\caption{Macroscopic quantities computed using the D2V25 along $y = 0.7L_y$ at $t = 1.68$, and the differences between higher-order and lower-order models: (a) density, (b) temperature, (c) pressure, and (d) velocity.}
	\label{fig14}
\end{figure}

\subsubsection{Comparison of KHI growth rate}

Figure \ref{fig10} shows the logarithmic variation of peak disturbance kinetic energy over time, obtained from the D2V25 and D2V16 DVSs. With the onset of second-order nonequilibrium effects, the evolution of KHI exhibits distinct stages. Figure \ref{fig11} illustrates the evolution of global average viscous stress components, which also follow a staged pattern.

From Fig. \ref{fig10} and Fig. \ref{fig11}, the following key observations are made:

(i) At \( t \approx 0.2 \), both \( \ln \left(\left.E_x\right|_{\max }\right) \) and \( \left|\Delta_{2 x y}^*\right| \) reach minima, followed by local maxima at \( t = 0.3 \). At \( t=0.4 \), \( \ln \left(\left.E_x\right|_{\max }\right) \) stabilizes, whereas \( \left|\Delta_{2 x y}^*\right| \) continues to decline. After \( t = 0.8 \), \( \left|\Delta_{2 x y}^*\right| \) decreases rapidly, while disturbance kinetic energy saturates, demonstrating a strong correlation between the two.

(ii) In the D2V16 DVS, \( \ln \left(\left.E_x\right|_{\max }\right) \) diverges for \( t>0.8 \), demonstrating the model's failure to capture viscous stress accurately. Throughout KHI evolution, the D2V16 DVS substantially overestimates TNEs, leading to simulation divergence.

(iii) The evolution of \( \left|\Delta_{2 x y}^*\right| \) exhibits an inverse correlation with other nonequilibrium components.

(iv) The evolution of \( \left|\Delta_{2 x x}^*\right| \) and \( \left|\Delta_{2 y y}^*\right| \) remains consistent, reflecting their symmetric analytical forms.

\subsubsection{Comparison of kinetic moments}

Kinetic moments are fundamental physical quantities that characterize the nonequilibrium state and dynamics of fluid systems. Increasing the number and order of preserved kinetic moments in physical modeling enhances the model's ability to capture the fluid's nonlinear dynamics. To directly assess the limitations of the low-order model and the advantages of the high-order model, we compare the kinetic moments computed by both approaches.

The D2V25 DVS preserves 25 independent kinetic moments, whereas the D2V16 DVS retains only the first 16. Figure \ref{fig12} shows the kinetic moments \( \mathbf{M}_0, \mathbf{M}_1, \mathbf{M}_{2,0}, \mathbf{M}_2, \mathbf{M}_{3,1}, \mathbf{M}_3 \), and \( \mathbf{M}_{4,2} \) along \( y = 0.7L_y \) at \( t = 1.68 \) for both D2V25 and D2V16 DVSs, along with their differences.
The relative error \( \varepsilon = {|M_{\alpha \beta 25} - M_{\alpha \beta 16}|}/{|M_{\alpha \beta 25}|} \) in the D2V16 DVS accumulates as KHI evolves. As shown in Fig. \ref{fig07}, at \( t = 1.68 \), large-scale vortices form along the fluid interface.
At this stage, the accumulated relative errors in the first 16 kinetic moments of D2V16 become significant. Specifically, in the distribution curves of \( M_{1y} \), \( M_{2xy} \), \( M_{3,1y} \), \( M_{3xxy} \), and \( M_{4,2xy} \), the maximum relative errors of D2V16 compared to D2V25 are as follows:
200\% at \( x = 130 \) for \( M_{1y} \),
110\% at \( x = 85 \) for \( M_{2xy} \),
 160\% at \( x = 110 \) for \( M_{3,1y} \),
100\% at \( x = 105 \) for \( M_{3xxy} \),
 240\% at \( x = 105 \) for \( M_{4,2xy} \).
Additionally, several other kinetic moments exhibit errors exceeding \( 50\% \), further highlighting the limitations of the low-order model.

The relative errors in the first 16 kinetic moments of the D2V16 DVS primarily arise from its failure to satisfy higher-order moment conditions, particularly \( \mathbf{M}_4 \) and \( \mathbf{M}_{5,3} \).
Satisfying the first 16 kinetic moments ensures the recovery of linear constitutive laws, while higher-order moment conditions govern nonlinear constitutive behavior.
As the relaxation time increases and shear velocity intensifies, TNEs exhibit strong nonlinearity. Consequently, the linear constitutive laws in the D2V16 DVS lead to significant macroscopic errors.
In contrast, the D2V25 DVS satisfies all the moment conditions necessary to accurately capture higher-order constitutive effects.

Figure \ref{fig13} presents the calculations of the kinetic moments \( \mathbf{M}_4 \) and \( \mathbf{M}_{5,3} \) along the line \( y = 0.7L_y \) at \( t = 1.68 \), obtained using the D2V25 and D2V16 DVSs, along with their differences. The description of higher-order kinetic moments by the low-order model is essentially a ``guess'', which proves highly inaccurate. As shown in Fig. \ref{fig13}, the ``guess'' made by the D2V16 DVS is highly inaccurate.
For \( \mathbf{M}_4 \), the relative error in the D2V16 DVS reaches:
300\% at \( x = 110 \) for \( M_{4xxxx} \) and \( M_{4yyyy} \),
340\% at \( x = 110 \) for \( M_{4xxyy} \).
Additionally, several other kinetic moments exhibit relative errors of approximately 100\%. This further highlights the limitations of the low-order model in capturing higher-order nonequilibrium effects.

\subsubsection{Comparison of macroscopic quantities}

Figure \ref{fig14} shows the distributions of macroscopic quantities computed by both models. Macroscopic quantities play a fundamental role in traditional fluid mechanics. During the evolution of the KHI, the relative error in macroscopic quantities computed by the low-order model accumulates over time.
As second-order nonequilibrium effects become significant, a complex interplay arises, with first- and second-order nonequilibrium effects alternating. This results in substantial errors in the macroscopic quantities computed by the D2V16 DVS. At \( t = 1.68 \), when large vortices fully develop, the accumulated errors in macroscopic quantities computed by D2V16 reach their peak. The relative errors in density, temperature, and pressure reach \( 43\% \), while the velocity error exceeds \( 200\% \) at \( x \approx 180 \), as shown in Fig. \ref{fig14}(d).

Two primary factors account for the significant errors in D2V16 DVS:

(1) Insufficient velocity nodes: With fewer velocity nodes, D2V16 oversimplifies microscopic velocity distributions in certain directions. This limitation hinders its ability to capture complex velocity variations during KHI evolution, particularly in shear layers and unstable vortex regions. In contrast, D2V25, with more velocity nodes, better captures microscopic velocity variations in all directions. This allows for a more accurate representation of complex vortices and velocity gradients in macroscopic quantities.

(2) Errors in higher-order kinetic moments: The primary limitation of D2V16 is its failure to preserve higher-order kinetic moments. Higher-order moments refine physical models by capturing nonequilibrium effects beyond the NS framework. They accurately capture fine-scale fluid structures and rapid dynamics, directly influencing the characterization of nonequilibrium effects. The long-term accumulation of these effects significantly impacts macroscopic behavior.

The computational costs of high-order and low-order models are compared. For the same computational task, the ratio of computational time between the two models is 0.61, which closely matches the ratio of their discrete velocity counts (0.64). This indicates that computational efficiency is approximately inversely proportional to the number of discrete velocities.
It is necessary to note that computational efficiency is not the primary criterion for model selection; rather, the corresponding physical functionality should guide the choice.

\begin{figure}[htbp]
	\begin{center}
		\includegraphics[width=0.52\textwidth]{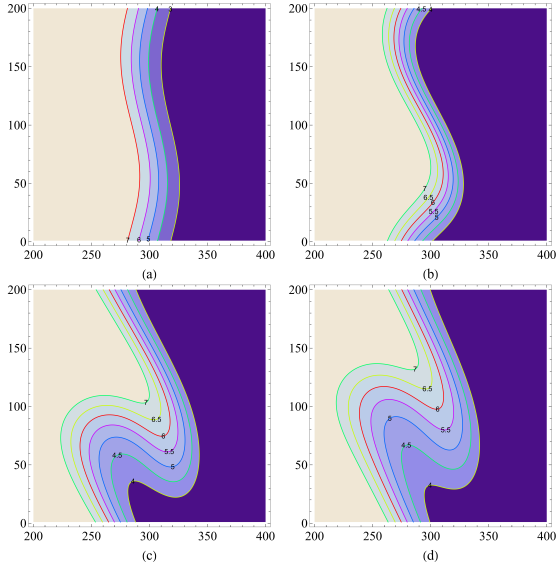}
	\end{center}
	\caption{Density evolution patterns of the KHI at four moments under the dominance of second-order TNEs: (a) $t = 0.24$, (b) $t = 0.72$, (c) $t = 2.70$, and (d) $t = 3.30$.}
	\label{fig15}
\end{figure}

\begin{figure*}[htbp]
	\begin{center}
		\includegraphics[width=0.78\textwidth]{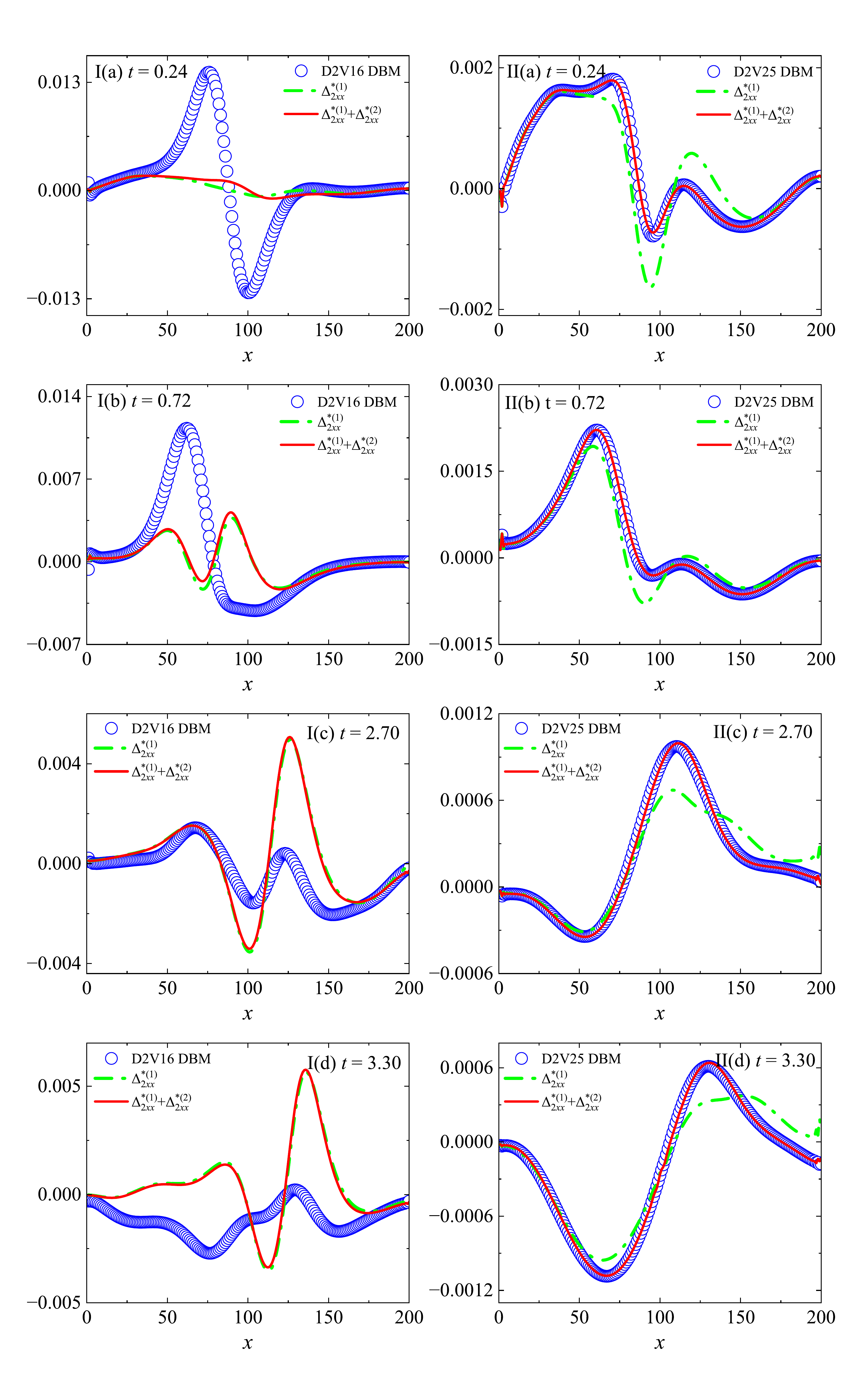}
	\end{center}
	\caption{Distributions of $\Delta_{2 x x}^*$ along $y = 0.7L_y$ at $t = 0.24$, $0.72$, $2.70$, and $3.30$, computed using D2V16 (left column) and D2V25 (right column) DVSs.}
	\label{fig16}
\end{figure*}

\begin{figure}[htbp]
	\begin{center}
		\includegraphics[width=0.5\textwidth]{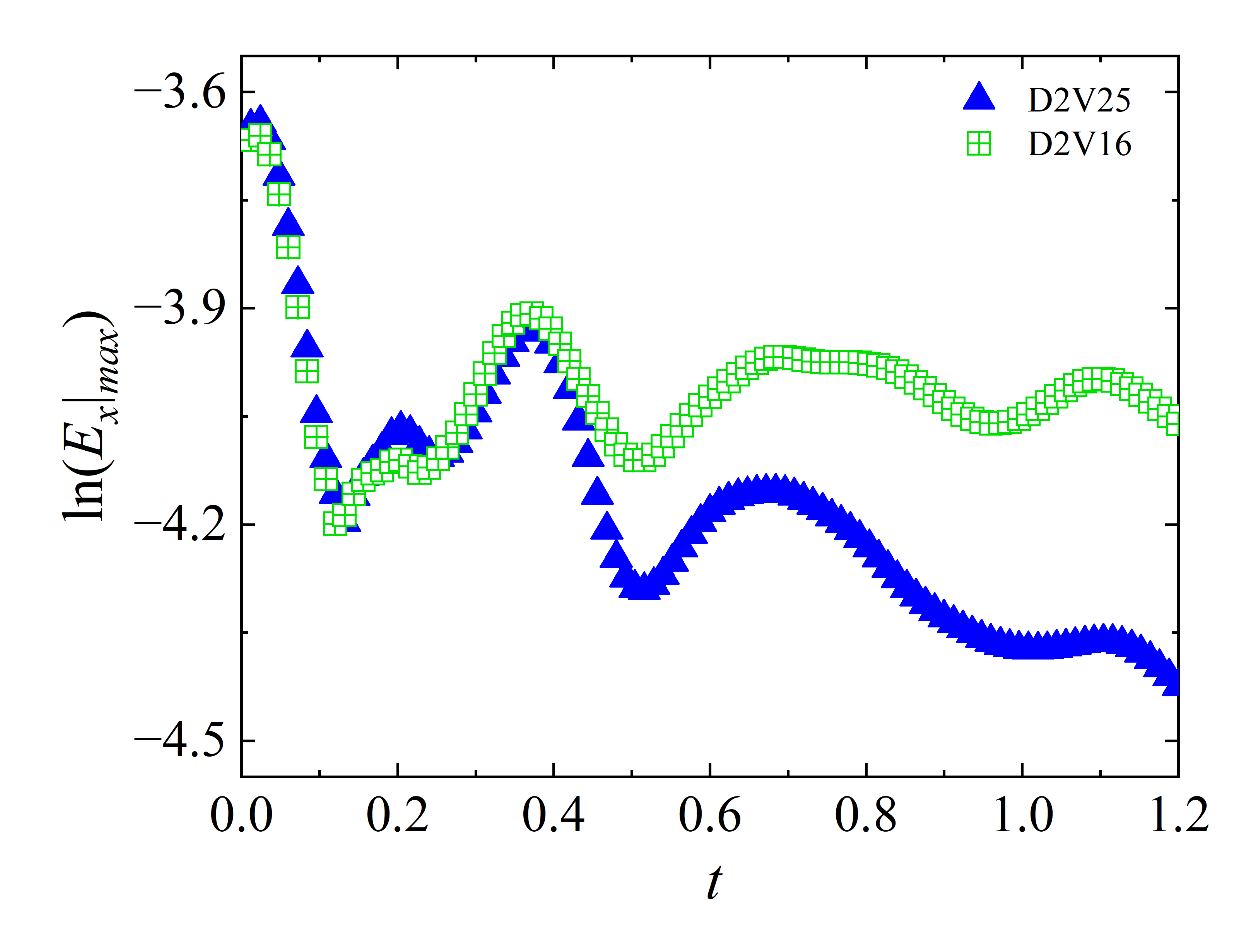}
	\end{center}
	\caption{Logarithmic variations of perturbation peak kinetic energy over time for D2V25 and D2V16 DVSs in the second-order nonequilibrium dominant case.}
	\label{fig17}
\end{figure}

\subsection{Dominance of second-order TNEs }\label{D2nd}

In this case, \( v_L = -v_R = 0.1 \) and \( \tau = 0.001 \) are set. The small initial shear velocity leads to rapid dissipation, which restricts the development of the KHI compared to the previous two cases.
As shown in Fig. \ref{fig15}, the density evolution indicates that KHI does not develop distinct vortices but instead forms a tentacle-like structure.
Figure \ref{fig16} shows the viscous stress distribution during KHI evolution, computed using both D2V25 and D2V16 DVSs. The D2V25 results show that second-order nonequilibrium effects dominate throughout the process. At \( t = 0.24 \), near \( x = 125 \), \( R_{\text{TNE}} \) reaches 0.98 and remains above 0.5 at other times.
The lower shear velocity weakens the velocity gradient, reducing first-order nonequilibrium effects. Meanwhile, larger gradients in other macroscopic quantities enhance second-order nonequilibrium effects.

Compared to the previous two cases (Figs. \ref{fig05} and \ref{fig08}), the discrepancy between the D2V16 simulation and the analytical solution in Fig. \ref{fig16} is more pronounced.
The constitutive relations in D2V16 omit density and temperature gradient contributions, limiting its ability to capture nonlinear second-order viscous stress and its feedback on first-order viscous stress.
In the early stages of KHI (\( t = 0.24 \) and \( t = 0.72 \)), both numerical and analytical solutions exhibit significant underestimation of viscous stress. In contrast, in the later stages (\( t = 2.70 \) and \( t = 3.30 \)), both solutions overestimate viscous stress.
Despite consistently large \( R_{\text{TNE}} \) values throughout the evolution, D2V25 accurately captures nonequilibrium effects in this complex dynamic process.

As shown in Fig. \ref{fig17}, with dominant second-order nonequilibrium effects, the evolution of the KHI follows a distinct staged progression. In this case, a relatively large relaxation time \( \tau \) accelerates dissipation, thereby restricting KHI development. Consequently, KHI fails to fully develop and saturates at approximately \( t \approx 0.4 \). Additionally, a significant discrepancy exists in the maximum disturbance kinetic energy between the two models.
In contrast, as shown in Fig. \ref{fig06}, with first-order nonequilibrium dominance, KHI undergoes continuous linear growth without distinct stages. This occurs because momentum transfer during this stage is primarily governed by the velocity gradient, which exerts a stable and uniform effect on perturbations. The fluid dynamics remain relatively simple, lacking complex nonlinear feedback mechanisms.

As second-order nonequilibrium effects emerge and gradually dominate, additional macroscopic gradients develop, enhancing the coupling and competition among different nonequilibrium driving forces. This increases the complexity of nonequilibrium characteristics, ultimately inducing irregularities and nonlinearities in momentum and energy transfer. Consequently, the disturbance growth rate exhibits acceleration, deceleration, or fluctuation at different stages, indicating distinct phase transitions.

In summary, with first-order nonequilibrium dominance, a single nonequilibrium driving force prevails. However, as second-order nonequilibrium effects dominate, multiple nonequilibrium driving forces interact, increasing system complexity and nonlinearity, and thereby influencing the staged development of KHI.

A key question remains: under what conditions do second-order nonequilibrium effects become significant or dominant? Reviewing the three scenarios, we observe that:

In Sec. \ref{1stD}, where \( {\tau}/{\Delta x} = 0.15 \), first-order nonequilibrium effects remain dominant throughout.

In Sec. \ref{2ndD}, where \( {\tau}/{\Delta x} = 0.40 \), first- and second-order nonequilibrium effects alternate.

In Sec. \ref{D2nd}, where \( {\tau}/{\Delta x} = 0.50 \), second-order nonequilibrium effects fully dominate.

To provide a comprehensive characterization, we analyze the relative nonequilibrium strength:
$R_{\text{TNE}} = {\bm{\Delta}_{2}^{*(2)}}/{\bm{\Delta}_{2}^{*(1)}}$.
The analytical expressions for \( \Delta_{2 x x}^{*(1)} \) and \( \Delta_{2 x x}^{*(2)} \) are provided in Eqs. \ref{B1} and \ref{D2xx2}.
Dividing each term in \( \Delta_{2 x x}^{*(2)} \) by \( \Delta_{2 x x}^{*(1)} \) yields:
\begin{equation}
	\frac{T_u}{\Delta_{2 x x}^{*(1)}} \sim \frac{n-2}{n+2} \tau \frac{\partial u_x}{\partial x},
\end{equation}
\begin{equation}
	\frac{T_t}{\Delta_{2 x x}^{*(1)}} \sim  \frac{\tau R}{T} \frac{\left(\partial_x T\right)^2}{\partial_x u_x},
\end{equation}
\begin{equation}
	\frac{T_{\rho 1}}{\Delta_{2 x x}^{*(1)}} \sim  \frac{\tau R T}{\rho} \frac{\partial^2 \rho}{\partial x^2} \frac{1}{\partial_x u_x},
\end{equation}
\begin{equation}
	\frac{T_{\rho 2}}{\Delta_{2 x x}^{*(1)}} \sim  \frac{\tau R T}{\rho^2} \frac{\left(\partial_x \rho\right)^2}{\partial_x u_x}.
\end{equation}
This leads to a simplified set of criteria for evaluating the relative strength of different orders of nonequilibrium effects:
\begin{equation}
	\mathbf{C}_{\text{TNE}}=\left\{ \frac{n-2}{n+2} \tau \frac{\partial u_x}{\partial x}, \frac{\tau R}{T} \frac{\left(\partial_x T\right)^2}{\partial_x u_x}, \frac{\tau R T}{\rho} \frac{\partial^2 \rho}{\partial x^2} \frac{1}{\partial_x u_x}, \frac{\tau R T}{\rho^2} \frac{\left(\partial_x \rho\right)^2}{\partial_x u_x}\right\}.
\end{equation}

In the first criterion, for our simulations, \( {(n-2)}/{(n+2)} \) and \( \Delta u_x \) are both of order unity, whereas \( \tau \) ranges from \( 10^{-5} \) to \( 10^{-2} \) or higher.
Therefore, the spatiotemporal ratio \( {\tau}/{\Delta x} \) quantifies the relative strength of temporal and spatial nonequilibrium effects and serves as a criterion for determining the dominant nonequilibrium order.
When the velocity difference is significant, this criterion can be generalized into a dimensionless vector analogous to the Knudsen number:
\begin{equation}
\bm{\Xi }_{\text{TNE}} = \frac{\tau \Delta \mathbf{u}}{\Delta \mathbf{r}}.
\end{equation}
This formulation explicitly accounts for velocity gradient variations, providing a comprehensive characterization of nonequilibrium dominance across different flow regimes.

Compared with the traditional \( Kn \), the criterion proposed in this study is based on \( R_{\text{TNE}} \). Although both quantify the degree of nonequilibrium within a system, they differ in terms of descriptive perspective, sensitivity, and information richness. \( Kn \) primarily relies on slow variables, such as macroscopic gradients, which makes it challenging to dynamically and precisely capture instantaneous changes and microscopic details within the fluid \cite{Gan2022}. In contrast, \( R_{\text{TNE}} \) reveals the relative significance of nonequilibrium effects at different orders through the ratios of nonequilibrium quantities, thereby offering a more accurate representation of the intrinsic characteristics of nonequilibrium states.

\subsection{Effectiveness, extensibility, and limitations of high-order DVS} \label{DDD}

The next step is to further increase \( {\tau}/{\Delta x} \) and evaluate the upper limit of the D2V25 DVS's ability to describe nonequilibrium effects when second-order nonequilibrium effects fully dominate.
The initial conditions are set as \( \rho_L = 6.0 \), \( \rho_R = 2.0 \), and \( P_L = P_R = 4.0 \). The velocity perturbation in the \( x \)-direction is prescribed by Eq. (\ref{e26}), with an amplitude of \( u_0 = 100 \Delta y \) and an initial wavenumber \( k = 2 \pi / L_y \).
All other parameters, boundary conditions, and finite difference schemes remain the same as those used in the previous cases.

\subsubsection{Absolute dominance of second-order TNEs}

\begin{figure}[htbp]
	\begin{center}
		\includegraphics[width=0.52\textwidth]{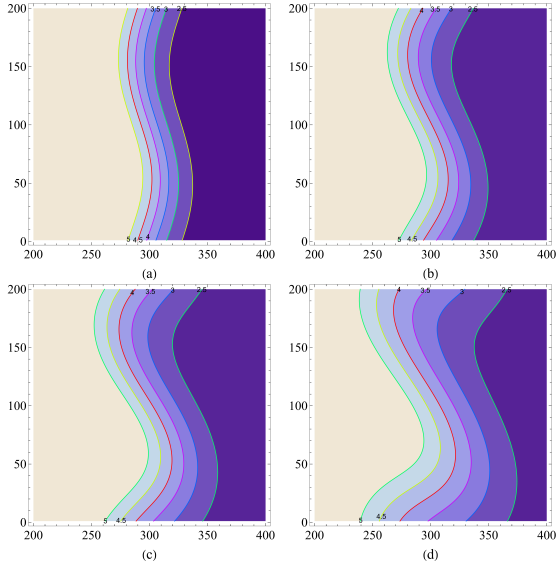}
	\end{center}
	\caption{Density evolution patterns of the KHI at four moments under the absolute dominance of second-order TNEs: (a) $t = 0.18$, (b) $t = 0.36$, (c) $t = 0.54$, (d) $t = 1.08$.}
	\label{fig18}
\end{figure}

\begin{figure*}[htbp]
	\begin{center}
		\includegraphics[width=0.78\textwidth]{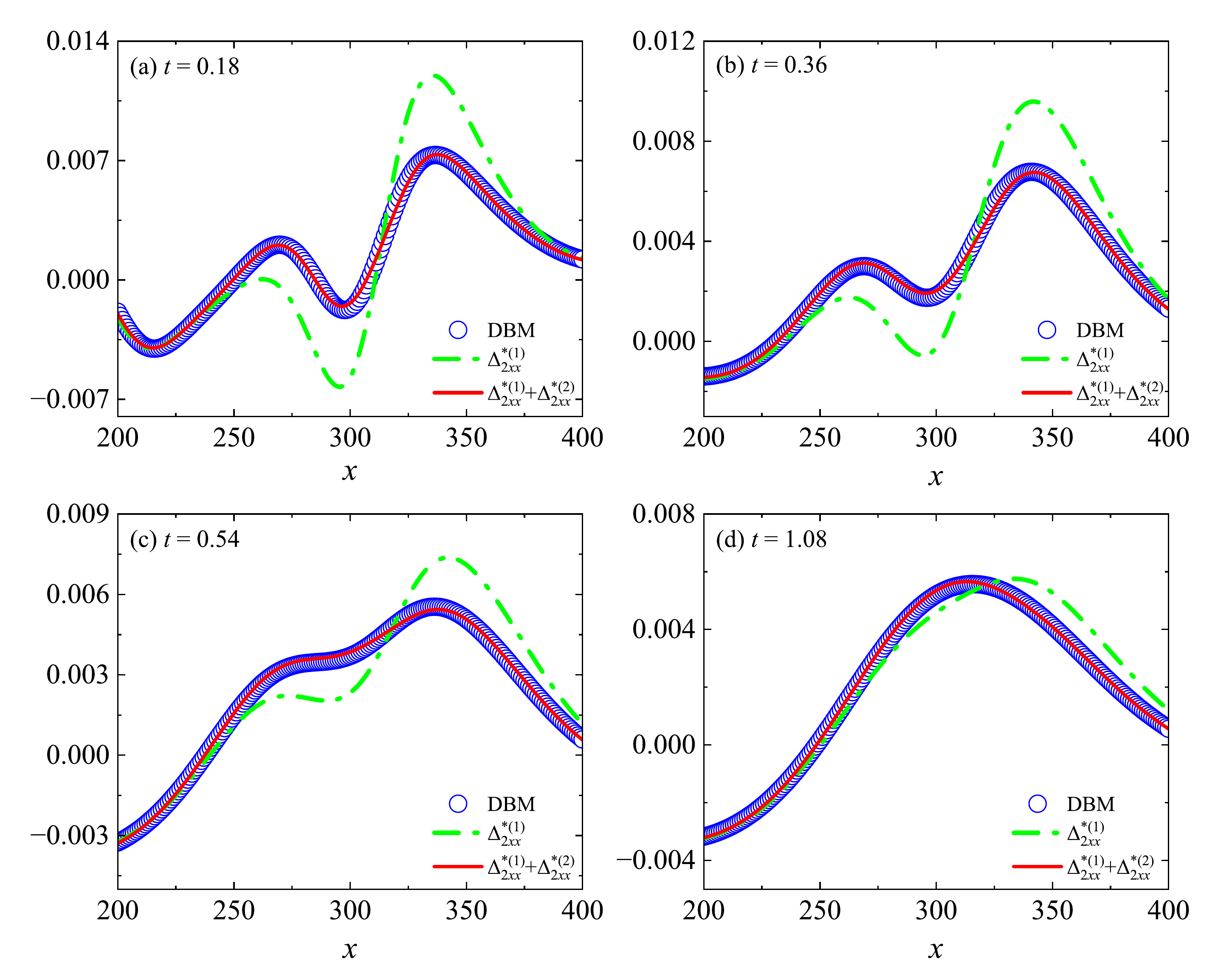}
	\end{center}
	\caption{Distributions of $\Delta_{2 x x}^*$ along $y = 0.5L_y$ at $t = 0.18$, $t = 0.36$, $t = 0.54$, and $t = 1.08$.}
	\label{fig19}
\end{figure*}

For \( \tau = 0.003 \) and \( {\tau}/{\Delta x} = 1.5 \), Fig. \ref{fig18} shows the density evolution of the KHI at \( t = 0.18 \), \( 0.36 \), \( 0.54 \), and \( 1.08 \). Initially, the interface remains nearly flat, with only minor perturbations.
Strong viscosity causes the shear velocity difference across the interface to dissipate rapidly, suppressing the shear-driven instability. Although perturbations develop, their growth is constrained by insufficient kinetic energy. Even at later stages, despite increased interface deformation, the characteristic KHI vortex structure does not emerge. Instead, the interface exhibits mild deformation and layering, with minimal mixing between adjacent fluids.

As shown in Fig. \ref{fig19}, strong second-order nonequilibrium effects, driven by viscous stress, dominate KHI development at this stage. At \( t = 0.18 \) and \( t = 0.36 \), \( R_{\text{TNE}} = 5 \), validating the previous criterion. In this regime, where second-order nonequilibrium effects dominate, the viscous stress predicted by D2V25 closely aligns with the analytical solution, demonstrating its strong multiscale capability.
Moreover, at this stage, the D2V16 simulation results have diverged, making direct comparison infeasible.

\subsubsection{Effectiveness, extensibility, and limitations of D2V25 DVS}


The D2V25 DVS in this study demonstrates excellent stability and accuracy in computing viscous stress during the evolution of the KHI.
Next, we evaluate its ability to capture other nonequilibrium effects.

\begin{figure*}[htbp]
	\begin{center}
		\includegraphics[width=0.75\textwidth]{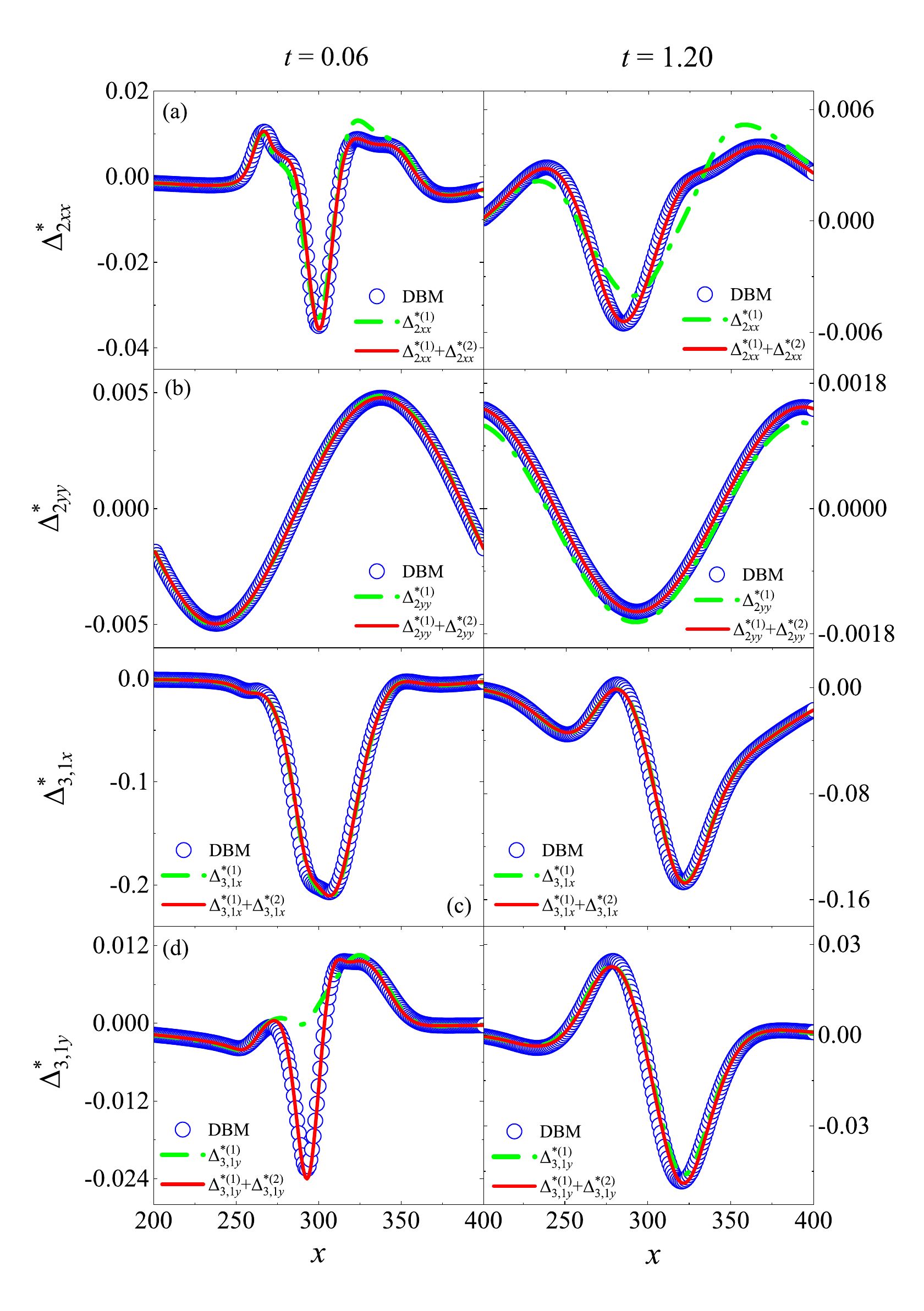}
	\end{center}
	\caption{Distributions of $\Delta_{2 x x}^*$, $\Delta_{3,1 x}^*$, and $\Delta_{3,1 y}^*$ along $y = 0.5L_y$, as well as $\Delta_{2 y y}^*$ along $x = 0.7L_x$, at $t = 0.06$ and $t = 1.20$, computed using the D2V25 DVS.}
	\label{fig20}
\end{figure*}

\subsubsubsection{Effectiveness}

Figure \ref{fig20} shows the distributions of \( \Delta_{2 x x}^* \), \( \Delta_{3,1 x}^* \), and \( \Delta_{3,1 y}^* \) along \( y=0.5L_y \), and \( \Delta_{2 y y}^* \) along \( x=0.7L_x \), at \( t=0.06 \) and \( t=1.20 \), computed using the D2V25 DVS.

(i) As discussed earlier, \( {\tau}/{\Delta x}=0.9 \) is relatively large, leading to the dominance of second-order nonequilibrium effects in KHI evolution, as shown in Fig. \ref{fig20}(a).

(ii) Figure \ref{fig20}(b) shows that D2V25 accurately predicts \( \Delta_{2 y y}^* \), with simulation results closely aligning with the second-order nonequilibrium analytical solution.

(iii) Figures \ref{fig20}(c) and \ref{fig20}(d) demonstrate that the D2V25 DVS accurately captures the system's heat flux. Throughout KHI evolution, \( \Delta_{3,1 x}^* \) consistently exhibits dominant first-order nonequilibrium effects. In contrast, \( \Delta_{3,1 y}^* \) transitions from second-order nonequilibrium dominance at \( t=0.06 \) to first-order dominance at \( t=1.20 \).

These findings highlight the effectiveness of the D2V25 DVS in accurately computing viscous stress and heat flux. Additionally, different nonequilibrium effects exhibit varying dominance in the same kinetic process, emphasizing their complexity.

\subsubsubsection{Extensibility}

\begin{figure*}[htbp]
	\begin{center}
		\includegraphics[width=0.75\textwidth]{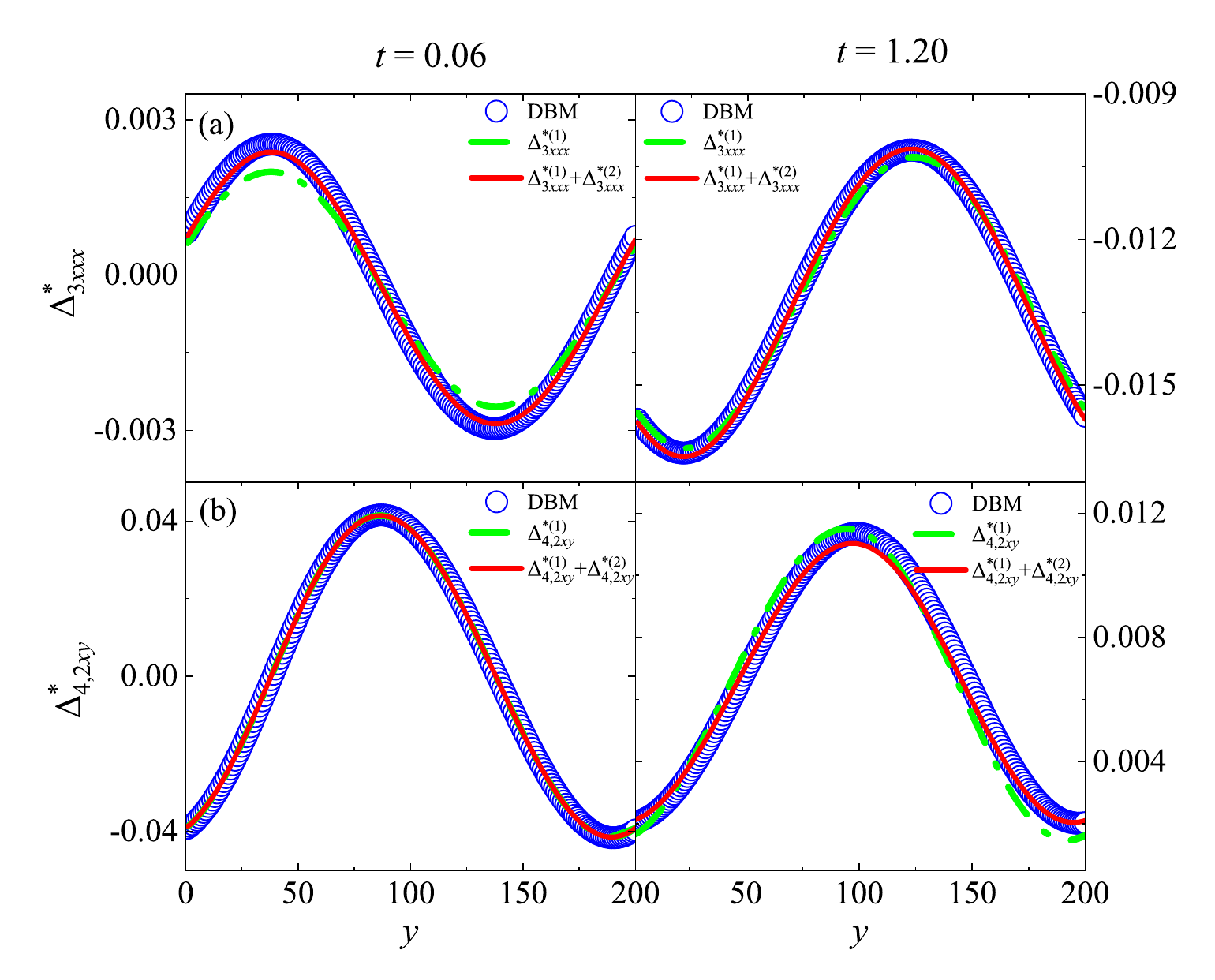}
	\end{center}
	\caption{Distributions of $\Delta_{3 x x x}^*$ and $\Delta_{4,2 x y}^*$ along $x = 0.7L_x$ at $t = 0.06$ and $t = 1.20$, computed using the D2V25 DVS.}
	\label{fig21}
\end{figure*}

Figure \ref{fig21} shows the distributions of \( \Delta_{3xxx}^* \) and \( \Delta_{4,2xy}^* \) along \( x = 0.7L_x \) at \( t = 0.06 \) and \( t = 1.2 \).
The nonequilibrium measures \( \bm{\Delta}_3^* \) and \( \bm{\Delta}_{4,2}^* \), along with their components, influence the accuracy of the constitutive relation evolution equations \cite{Gan2022}, necessitating higher-order kinetic moment relations for precise computation.
However, as shown in Fig. \ref{fig21}, the D2V25 DVS accurately captures higher-order nonequilibrium effects, including \( \Delta_{3xxx}^* \) and \( \Delta_{4,2xy}^* \), with numerical results closely aligning with the second-order analytical solutions.
This demonstrates the capability of the D2V25 DVS to capture higher-order nonequilibrium effects, suggesting that its practical simulation performance exceeds its theoretical design to some extent.

\subsubsubsection{Limitations}

\begin{figure*}[htbp]
	\begin{center}
		\includegraphics[width=0.75\textwidth]{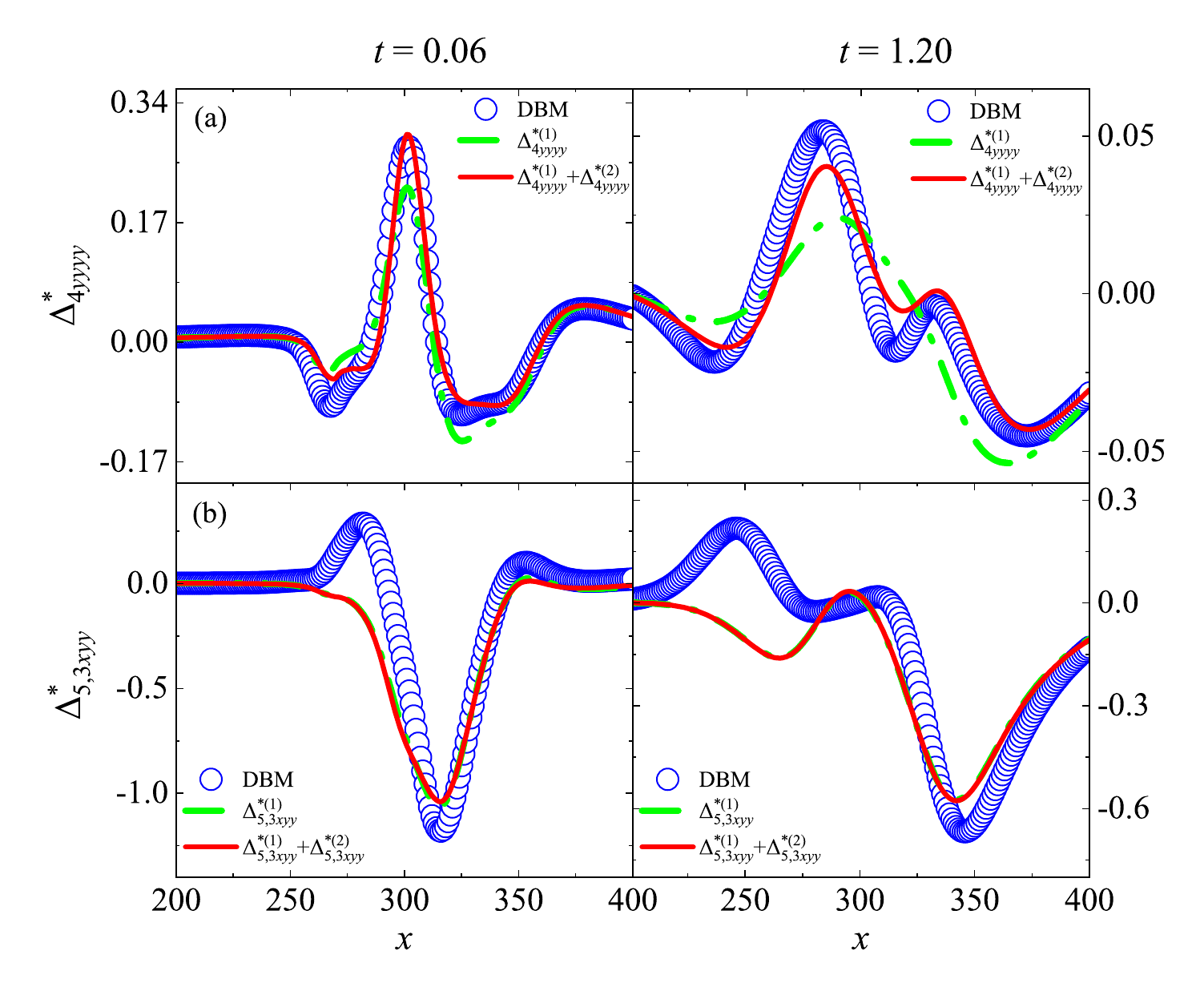}
	\end{center}
	\caption{Distributions of $\Delta_{4 y y y y}^*$ and $\Delta_{5,3 x y y}^*$ along $y = 0.3L_y$ at $t = 0.06$ and $t = 1.20$, computed using the D2V25 DVS.}
	\label{fig22}
\end{figure*}

As shown in Fig. \ref{fig22}, the simulation results for \( \Delta_{4yyyy}^* \) and \( \Delta_{5,3xyy}^* \) using the D2V25 DVS are unsatisfactory.
This indicates that while the D2V25 DVS demonstrates a certain level of extensibility, it remains inherently limited.
To address higher-order nonequilibrium effects, additional kinetic moment relations for \( f_i^{eq} \) are required, along with the adoption of higher-order models.

\subsection{Nonequilibrium phase diagram}

\begin{figure*}[htbp]
	\begin{center}
		\includegraphics[width=1.0\textwidth]{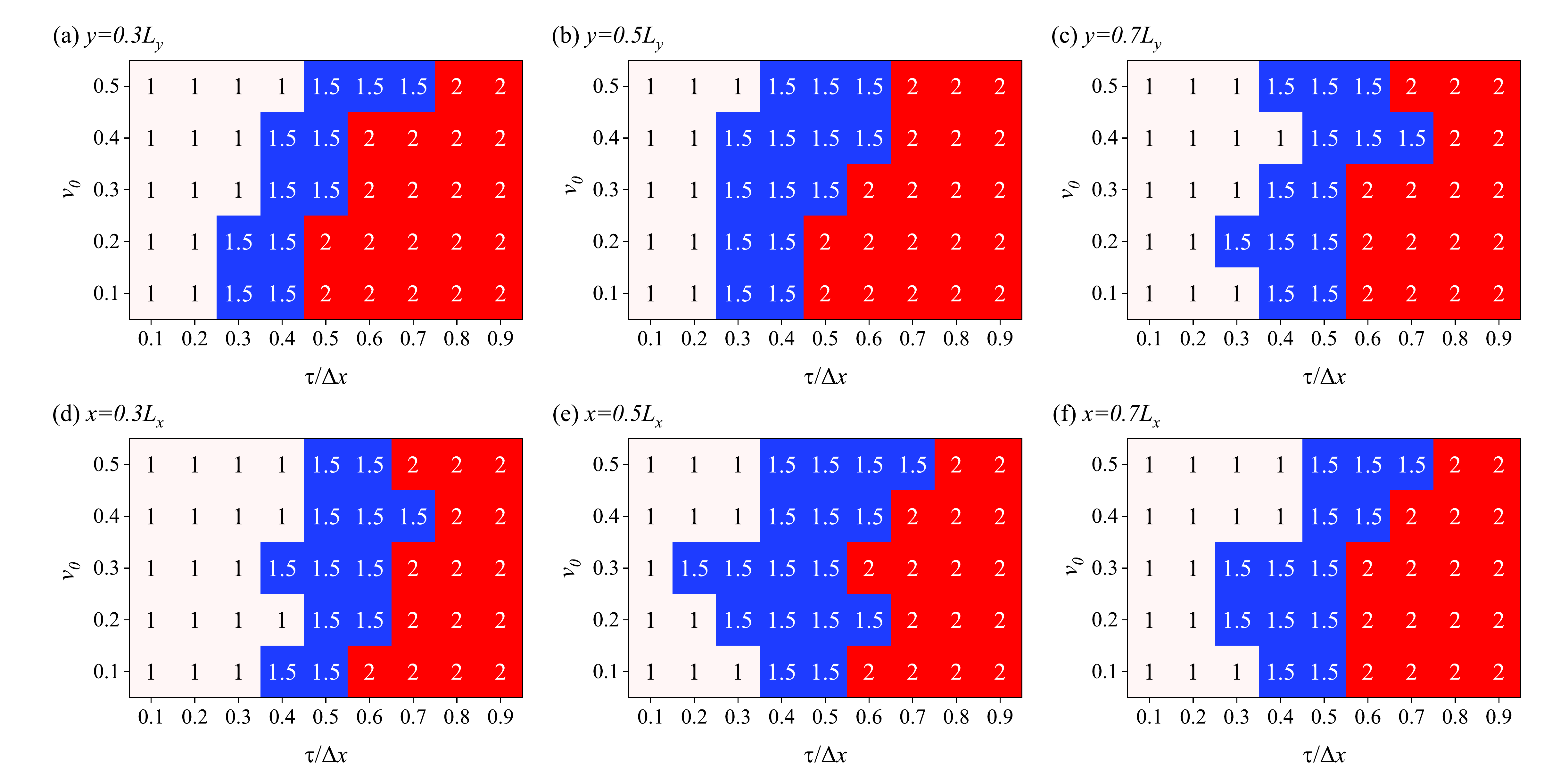}
	\end{center}
	\caption{Influence of shear velocity $v_0$ and ${\tau}/{\Delta x}$ on the dominance of nonequilibrium effects at different orders across various regions.}
	\label{fig23}
\end{figure*}

KHI is a multi-scale phenomenon where nonequilibrium effects of different orders dominate under varying conditions, including spatial location, time, and parameters, and may alternate in dominance. Identifying these patterns is crucial for selecting an appropriate model before simulation.
Next, we investigate how spatial location, shear velocity, and the spatiotemporal ratio influence the dominance of different orders of nonequilibrium effects in KHI, offering key insights for model selection. The initial conditions for the numerical simulations remain identical to those in Section \ref{DDD}.

Figure \ref{fig23} illustrates how spatial location, ${\tau}/{\Delta x}$, and shear velocity $v_0$ influence the dominance of different orders of nonequilibrium effects during KHI development. The notation ``1'' represents the dominance of first-order nonequilibrium effects, ``1.5'' indicates alternating first- and second-order effects, and ``2'' signifies the dominance of second-order nonequilibrium effects. It can be concluded that:

(i) Spatial location, shear velocity, and ${\tau}/{\Delta x}$ significantly influence the dominance of different orders of nonequilibrium effects.

(ii) Among these factors, ${\tau}/{\Delta x}$ plays the most critical role. Based on its value, the nonequilibrium phase diagram can be classified into three regions: for ${\tau}/{\Delta x}<R_1$, first-order nonequilibrium effects dominate; when $R_1<{\tau}/{\Delta x}<R_2$, first- and second-order effects alternate; and when ${\tau}/{\Delta x}>R_2$, second-order effects dominate. In the phase diagram, $R_1 \approx 0.3$ and $R_2 \approx 0.5$..

(iii) Shear velocity and spatial location influence only the specific values of $R_1$ and $R_2$. For instance, in Fig. \ref{fig23}(a), when $v_0<0.2$, $R_1 \approx 0.25$; when $0.2<v_0<0.4$, $R_1 \approx 0.35$; and when $v_0>0.4$, $R_1 \approx 0.45$. This is because shear velocity gradients predominantly drive first-order nonequilibrium effects, whereas ${\tau}/{\Delta x}$ primarily governs second-order effects.

(iv) Spatial position strongly affects the regions where first- and second-order nonequilibrium effects alternate in dominance, as seen in Figs. \ref{fig23}(b) and \ref{fig23}(e), which correspond to $y=0.5L_y$ and $x=0.5L_x$. These locations intersect the KHI vortex core, where multi-scale effects are most pronounced.

This analysis of the nonequilibrium phase diagram provides insights into the relative contributions of first- and second-order nonequilibrium effects in KHI and establishes a theoretical foundation for accurate model selection in multi-scale DBM simulations.

\section{Conclusions, discussions, and outlooks}\label{V}

The Kelvin-Helmholtz instability (KHI) is a battleground where competing nonequilibrium driving forces of different orders continuously reshape its evolution.
In this study, we employ a higher-order discrete Boltzmann model (DBM) as a powerful tool to examine the intricate dynamics of thermodynamic nonequilibrium effects (TNEs) in KHI. Our simulations reveal that, depending on the initial conditions, KHI evolves through three distinct scenarios:
(i) a regime where first-order TNEs dominate,
(ii) a transitional phase where first- and second-order TNEs alternate in dominance, and
(iii) a phase where second-order TNEs govern the instability's evolution.
To quantify and track these transitions, we introduce the relative nonequilibrium intensity, \( R_{\text{TNE}} \), establishing a robust criterion for identifying the dominant nonequilibrium order at different stages. The physical validity and numerical accuracy of this criterion are confirmed through simulations.

When second-order nonequilibrium effects come to the forefront, the battle between low- and high-order models becomes evident. Our comparative analysis reveals critical shortcomings in low-order models: they fail to accurately predict KHI growth rates, capture macroscopic evolution, preserve kinetic moment relationships, and faithfully reproduce thermodynamic nonequilibrium characteristics. In contrast, high-order models emerge as the superior contenders, offering sharper physical accuracy, greater cross-scale adaptability, and enhanced numerical stability in the presence of strong nonequilibrium effects.
Notably, although high-order models exhibit some degree of extensibility in capturing strong nonequilibrium effects, their applicability remains fundamentally constrained.
\emph{In physical modeling, the key lies in dynamically adjusting the preserved kinetic moment relations during phase-space discretization and tailoring discretization strategies to specific research objectives-only then can we ensure a precise and robust kinetic description.}
To further unravel the multiscale nature of KHI, we construct a nonequilibrium phase diagram-a roadmap that visually charts the interplay between different nonequilibrium orders. This diagram not only deepens our understanding of the instability's intricate dynamics but also serves as a valuable guide for selecting suitable kinetic models across a broad spectrum of nonequilibrium conditions in both scientific research and engineering applications.

Despite these advances, several frontiers remain unexplored.
(i) Our current model captures only second-order TNEs, whereas extreme conditions may require the inclusion of third- and higher-order effects.
(ii) Our analysis primarily focused on the role of viscous stress in shaping KHI. However, other nonequilibrium effects—driven by different physical mechanisms—may follow entirely different dominance pathways. For example, the evolution of heat flux follows distinct mechanisms: first-order heat flux is primarily driven by the temperature gradient, second-order heat flux involves the combined effects of temperature and velocity gradients, and third-order or higher-order heat flux is associated with more complex nonlinear effects.
(iii) Our study is limited to the two-dimensional case.
Expanding from two to three dimensions presents several challenges, including the design of a three-dimensional discrete velocity set, the theoretical derivation of nonequilibrium effects, and the analysis of three-dimensional physical fields.
This paper does not analyze the impact of external fields on KHI. For instance, in astrophysics and plasma physics, electromagnetic field effects are often significant.

To further advance the boundaries of kinetic modeling, future research will focus on developing a three-dimensional DBM model that incorporates third-order and higher-order TNEs, as well as multiphysics coupling. This model will aid in exploring the competition and coordination mechanisms among different orders and types of TNEs during the evolution of KHI. Additionally, by modifying the external force term, electromagnetic field effects will be introduced to further investigate their influence on the development of KHI. These efforts will not only enrich the theoretical foundation of kinetic modeling and provide precise simulations and analyses in broader physical contexts, but also offer new insights and techniques for addressing complex phenomena in plasma physics, astrophysics, and related fields.


\section*{Acknowledgements}
The authors sincerely thank Yanhong Wu, Zhaowen Zhuang and Jiahui Song for their valuable discussions and insights.
This work was supported by the National Natural Science Foundation of China (Grant Nos. U2242214, 52278119, 11875001, and 12172061), the Hebei Outstanding Youth Science Foundation (Grant No. A2023409003), the Central Guidance on Local Science and Technology Development Fund of Hebei Province (Grant No. 226Z7601G), the Fujian Provincial Units Special Funds for Education and Research (Grant No. 2022639), and the Foundation of the National Key Laboratory of Shock Wave and Detonation Physics (Grant No. JCKYS2023212003).

\appendix
\section{Kinetic moment relations}\label{KMR}

\begin{equation}
	\mathbf{M}_0=\sum_i f_i^{e q}=\rho,
\end{equation}
\begin{equation}
	\mathbf{M}_1=\sum_i f_i^{e q} \mathbf{v}_i=\rho \mathbf{u},
\end{equation}
\begin{equation}
	\mathbf{M}_{2,0}=\sum_i \frac{1}{2} f_i^{e q}\left(v_i^2+\eta_i^2\right)=\frac{1}{2} \rho\left[(n+2) R T+u^2\right],
\end{equation}
\begin{equation}
	\mathbf{M}_2=\sum_i f^{e q} \mathbf{v}_i \mathbf{v}_i=\rho(R T \mathbf{I}+\mathbf{u u}),
\end{equation}
\begin{equation}
	\mathbf{M}_{3,1}=\sum_i \frac{1}{2} f_i^{e q}\left(v_i^2+\eta_i^2\right) \mathbf{v}_i=\frac{1}{2} \rho \mathbf{u}\left[(n+4) R T+u^2\right],
\end{equation}
\begin{equation}
	\begin{aligned}
		\mathbf{M}_3 & =\sum_i f_i^{e q} \mathbf{v}_i \mathbf{v}_i \mathbf{v}_i \\
		& =\rho\left[\left(u_\alpha \delta_{\beta \gamma}+u_\beta \delta_{\alpha \gamma}+u_\gamma \delta_{\alpha \beta}\right) \mathbf{e}_\alpha \mathbf{e}_\beta \mathbf{e}_\gamma R T+\mathbf{u u u}\right],
	\end{aligned}
\end{equation}
\begin{equation}
	\begin{aligned}
		\mathbf{M}_{4,2} & =\sum_i f_i^{e q} \frac{v_i^2+\eta_i^2}{2} \mathbf{v}_i \mathbf{v}_i \\
		& =\rho\left[\left(\frac{n+4}{2} R T+\frac{u^2}{2}\right) R T \mathbf{I}+\left(\frac{n+6}{2} R T+\frac{u^2}{2}\right) \mathbf{u u}\right],
	\end{aligned}
\end{equation}
\begin{equation}
	\begin{aligned}
		\mathbf{M}_4= & \sum_i f_i^{e q} \mathbf{v}_i \mathbf{v}_i \mathbf{v}_i \mathbf{v}_i \\
		= & \rho\left[R^2 T^2\left(\delta_{\alpha \beta} \delta_{\gamma \lambda}+\delta_{\alpha \gamma} \delta_{\beta \lambda}+\delta_{\alpha \lambda} \delta_{\beta \gamma}\right) \mathbf{e}_\alpha \mathbf{e}_\beta \mathbf{e}_\gamma \mathbf{e}_\lambda\right. \\
		& +R T\left(u_\alpha u_\beta \delta_{\gamma \lambda}+u_\alpha u_\gamma \delta_{\beta \lambda}+u_\alpha u_\lambda \delta_{\beta \gamma}+u_\beta u_\gamma \delta_{\alpha \lambda}\right. \\
		& \left.\left.+u_\beta u_\lambda \delta_{\alpha \gamma}+u_\gamma u_\lambda \delta_{\alpha \beta}\right) \mathbf{e}_\alpha \mathbf{e}_\beta \mathbf{e}_\gamma \mathbf{e}_\lambda+\mathbf{u u u u}\right],
	\end{aligned}
\end{equation}
\begin{equation}
	\begin{aligned}
		\mathbf{M}_{5,3} & =\sum_i \frac{1}{2} f_i^{e q}\left(v_i^2+\eta_i^2\right) \mathbf{v}_i \mathbf{v}_i \mathbf{v}_i \\
		& =\rho\left[\left(\frac{n+8}{2} R T+\frac{u^2}{2}\right) \mathbf{u u u}+\left(\frac{n+6}{2} R T\right.\right. \\
		& \left.\left.+\frac{u^2}{2}\right)\left(u_\alpha \delta_{\beta \gamma}+u_\beta \delta_{\alpha \gamma}+u_\gamma \delta_{\alpha \beta}\right) \mathbf{e}_\alpha \mathbf{e}_\beta \mathbf{e}_\gamma R T\right].
	\end{aligned}
\end{equation}

\section{Nonlinear constitutive relations}\label{NCR}

\begin{equation}\label{B1}
	\bm{\Delta}^{*(1)}_{2} = -\mu \left[ \bm{\nabla} u + (\bm{\nabla} u)^{T} - \frac{2}{n+2} \mathbf{I} \bm{\nabla} \cdot \mathbf{u} \right],
\end{equation}
\begin{equation}
		\bm{\Delta}_{3,1}^{*(1)}=-\kappa \bm{\nabla} T,
\end{equation}
\begin{equation}
	\begin{aligned}
		\Delta_{2 x x}^{*(2)}= & 2 n_2^{-2} \tau^2\left\{\rho R T \left[n_{-2} n_1\left(\partial_x u_x\right)^2+n_1 n_2\left(\partial_y u_x\right)^2\right.\right. \\
		& -4 n \partial_x u_x \partial_y u_y-n_2\left(\partial_x u_y\right)^2 \\
		& \left.-n_{-2}\left(\partial_y u_y\right)^2\right]+\rho R^2\left[n_1 n_2\left(\partial_x T\right)^2-n_2\left(\partial_y T\right)^2\right] \\
		& -R^2 T^2\left[n_1 n_2 \frac{\partial^2}{\partial x^2} \rho-n_2 \frac{\partial^2}{\partial y^2} \rho\right] \\
		& \left.+\frac{R^2 T^2}{\rho}\left[n_1 n_2\left(\partial_x \rho\right)^2-n_2\left(\partial_y \rho\right)^2\right]\right\},
	\end{aligned}
\end{equation}\label{B2}
\begin{equation}
    \begin{aligned}
        \Delta_{2xy}^{*(2)} = & \ 2 \tau^2 \bigg[ n_2^{-1} \rho T \Big(n \partial_x u_x \partial_x u_y + n \partial_y u_x \partial_y u_y \\
        & -2 \partial_x u_y \partial_y u_y - 2 \partial_x u_x \partial_y u_x \Big) + \rho R^2 \partial_x T \partial_y T \\
        & - R^2 T^2 \frac{\partial^2}{\partial x \partial y} \rho + \frac{R^2 T^2}{\rho} \partial_x \rho \partial_y \rho \bigg]
    \end{aligned}
\end{equation}
\begin{equation}
	\begin{aligned}
		\Delta_{2 y y}^{*(2)}= & -2 n_2^{-2} \tau^2\left\{\rho R T \left[n_{-2}\left(\partial_x u_x\right)^2+n_2\left(\partial_y u_x\right)^2\right.\right. \\
		& +4 n \partial_x u_x \partial_y u_y-n_1 n_2\left(\partial_x u_y\right)^2 \\
		& \left.-n_{-2} n_1\left(\partial_y u_y\right)^2\right]+\rho R^2\left[n_2\left(\partial_x T\right)^2-n_1 n_2\left(\partial_y T\right)^2\right] \\
		& -R^2 T^2\left[n_2 \frac{\partial^2}{\partial x^2} \rho-n_1 n_2 \frac{\partial^2}{\partial y^2} \rho\right] \\
		& \left.+\frac{R^2 T^2}{\rho}\left[n_2\left(\partial_x \rho\right)^2-n_1 n_2\left(\partial_y \rho\right)^2\right]\right\},
	\end{aligned}
\end{equation}
\begin{equation}
    \begin{aligned}
        \Delta_{3,1 x}^{*(2)}= & n_2^{-1} \tau^2 \bigg\{ \rho R^2 T^2 \bigg[n_{-2} \frac{\partial^2}{\partial x^2} u_x + n_2 \frac{\partial^2}{\partial y^2} u_x - 4 \frac{\partial^2}{\partial x \partial y} u_y \bigg] \\
        & + \rho R^2 T \bigg[ (n_2^2 + 4 n) \partial_x u_x \partial_x T + n_2 n_6 \partial_y u_x \partial_y T \\
        & - 2 n_6 \partial_y u_y \partial_x T + 2 n_2 \partial_x u_y \partial_y T \bigg] \bigg\},
    \end{aligned}
\end{equation}
\begin{equation}
    \begin{aligned}
        \Delta_{3,1 y}^{*(2)}= & n_2^{-1} \tau^2 \bigg\{ \rho R^2 T^2 \bigg[n_2 \frac{\partial^2}{\partial x^2} u_y + n_{-2} \frac{\partial^2}{\partial y^2} u_y - 4 \frac{\partial^2}{\partial x \partial y} u_x \bigg] \\
        & + \rho R^2 T \bigg[ (n_2^2 + 4 n) \partial_y u_y \partial_y T + n_2 n_6 \partial_x u_y \partial_x T \\
        & - 2 n_6 \partial_x u_x \partial_y T + 2 n_2 \partial_y u_x \partial_x T \bigg] \bigg\},
    \end{aligned}
\end{equation}
where $n_a=n+a$.

\section*{Data Availability}
The data that support the findings of this study are available from the corresponding author upon reasonable request.

\section*{References}
\bibliography{KHI-DBM}

\begin{thebibliography}{99}%
\makeatletter
\providecommand \@ifxundefined [1]{%
 \@ifx{#1\undefined}
}%
\providecommand \@ifnum [1]{%
 \ifnum #1\expandafter \@firstoftwo
 \else \expandafter \@secondoftwo
 \fi
}%
\providecommand \@ifx [1]{%
 \ifx #1\expandafter \@firstoftwo
 \else \expandafter \@secondoftwo
 \fi
}%
\providecommand \natexlab [1]{#1}%
\providecommand \enquote  [1]{``#1''}%
\providecommand \bibnamefont  [1]{#1}%
\providecommand \bibfnamefont [1]{#1}%
\providecommand \citenamefont [1]{#1}%
\providecommand \href@noop [0]{\@secondoftwo}%
\providecommand \href [0]{\begingroup \@sanitize@url \@href}%
\providecommand \@href[1]{\@@startlink{#1}\@@href}%
\providecommand \@@href[1]{\endgroup#1\@@endlink}%
\providecommand \@sanitize@url [0]{\catcode `\\12\catcode `\$12\catcode
  `\&12\catcode `\#12\catcode `\^12\catcode `\_12\catcode `\%12\relax}%
\providecommand \@@startlink[1]{}%
\providecommand \@@endlink[0]{}%
\providecommand \url  [0]{\begingroup\@sanitize@url \@url }%
\providecommand \@url [1]{\endgroup\@href {#1}{\urlprefix }}%
\providecommand \urlprefix  [0]{URL }%
\providecommand \Eprint [0]{\href }%
\providecommand \doibase [0]{http://dx.doi.org/}%
\providecommand \selectlanguage [0]{\@gobble}%
\providecommand \bibinfo  [0]{\@secondoftwo}%
\providecommand \bibfield  [0]{\@secondoftwo}%
\providecommand \translation [1]{[#1]}%
\providecommand \BibitemOpen [0]{}%
\providecommand \bibitemStop [0]{}%
\providecommand \bibitemNoStop [0]{.\EOS\space}%
\providecommand \EOS [0]{\spacefactor3000\relax}%
\providecommand \BibitemShut  [1]{\csname bibitem#1\endcsname}%
\let\auto@bib@innerbib\@empty
\bibitem [{\citenamefont {Chandrasekhar}(2013)}]{chandrasekhar2013CC}%
  \BibitemOpen
  \bibfield  {author} {\bibinfo {author} {\bibfnamefont {S.}~\bibnamefont
  {Chandrasekhar}},\ }\href@noop {} {\emph {\bibinfo {title} {Hydrodynamic and
  hydromagnetic stability}}}\ (\bibinfo  {publisher} {Courier Corporation},\
  \bibinfo {year} {2013})\BibitemShut {NoStop}%
\bibitem [{\citenamefont {Drazin}\ and\ \citenamefont
  {Reid}(2004)}]{drazin2004CUP}%
  \BibitemOpen
  \bibfield  {author} {\bibinfo {author} {\bibfnamefont {P.~G.}\ \bibnamefont
  {Drazin}}\ and\ \bibinfo {author} {\bibfnamefont {W.~H.}\ \bibnamefont
  {Reid}},\ }\href@noop {} {\emph {\bibinfo {title} {Hydrodynamic stability}}}\
  (\bibinfo  {publisher} {Cambridge university press},\ \bibinfo {year}
  {2004})\BibitemShut {NoStop}%
\bibitem [{\citenamefont {Gan}\ \emph {et~al.}(2011)\citenamefont {Gan},
  \citenamefont {Xu}, \citenamefont {Zhang},\ and\ \citenamefont
  {Li}}]{gan2011PRE}%
  \BibitemOpen
  \bibfield  {author} {\bibinfo {author} {\bibfnamefont {Y.}~\bibnamefont
  {Gan}}, \bibinfo {author} {\bibfnamefont {A.}~\bibnamefont {Xu}}, \bibinfo
  {author} {\bibfnamefont {G.}~\bibnamefont {Zhang}}, \ and\ \bibinfo {author}
  {\bibfnamefont {Y.}~\bibnamefont {Li}},\ }\bibfield  {title} {\enquote
  {\bibinfo {title} {Lattice {B}oltzmann study on {K}elvin-{H}elmholtz
  instability: Roles of velocity and density gradients},}\ }\href@noop {}
  {\bibfield  {journal} {\bibinfo  {journal} {Phys. Rev. E: Stat. Nonlinear
  Soft Matter Phys.}\ }\textbf {\bibinfo {volume} {83}},\ \bibinfo {pages}
  {056704} (\bibinfo {year} {2011})}\BibitemShut {NoStop}%
\bibitem [{\citenamefont {Li}\ \emph {et~al.}(2018)\citenamefont {Li},
  \citenamefont {Geng}, \citenamefont {Liu}, \citenamefont {Wang},\ and\
  \citenamefont {Zang}}]{li2018IP}%
  \BibitemOpen
  \bibfield  {author} {\bibinfo {author} {\bibfnamefont {Y.}~\bibnamefont
  {Li}}, \bibinfo {author} {\bibfnamefont {X.}~\bibnamefont {Geng}}, \bibinfo
  {author} {\bibfnamefont {Z.}~\bibnamefont {Liu}}, \bibinfo {author}
  {\bibfnamefont {H.}~\bibnamefont {Wang}}, \ and\ \bibinfo {author}
  {\bibfnamefont {D.}~\bibnamefont {Zang}},\ }\bibfield  {title} {\enquote
  {\bibinfo {title} {Simulating {K}elvin--{H}elmholtz instability using
  dissipative particle dynamics},}\ }\href@noop {} {\bibfield  {journal}
  {\bibinfo  {journal} {Fluid Dyn. Res.}\ }\textbf {\bibinfo {volume} {50}},\
  \bibinfo {pages} {045512} (\bibinfo {year} {2018})}\BibitemShut {NoStop}%
\bibitem [{\citenamefont {Wang}\ \emph {et~al.}(2017)\citenamefont {Wang},
  \citenamefont {Ye}, \citenamefont {He}, \citenamefont {Wu}, \citenamefont
  {Fan}, \citenamefont {Xue}, \citenamefont {Guo}, \citenamefont {Miao},
  \citenamefont {Yuan}, \citenamefont {Dong} \emph {et~al.}}]{wang2017SCP}%
  \BibitemOpen
  \bibfield  {author} {\bibinfo {author} {\bibfnamefont {L.}~\bibnamefont
  {Wang}}, \bibinfo {author} {\bibfnamefont {W.}~\bibnamefont {Ye}}, \bibinfo
  {author} {\bibfnamefont {X.}~\bibnamefont {He}}, \bibinfo {author}
  {\bibfnamefont {J.}~\bibnamefont {Wu}}, \bibinfo {author} {\bibfnamefont
  {Z.}~\bibnamefont {Fan}}, \bibinfo {author} {\bibfnamefont {C.}~\bibnamefont
  {Xue}}, \bibinfo {author} {\bibfnamefont {H.}~\bibnamefont {Guo}}, \bibinfo
  {author} {\bibfnamefont {W.}~\bibnamefont {Miao}}, \bibinfo {author}
  {\bibfnamefont {Y.}~\bibnamefont {Yuan}}, \bibinfo {author} {\bibfnamefont
  {J.}~\bibnamefont {Dong}},  \emph {et~al.},\ }\bibfield  {title} {\enquote
  {\bibinfo {title} {Theoretical and simulation research of hydrodynamic
  instabilities in inertial-confinement fusion implosions},}\ }\href@noop {}
  {\bibfield  {journal} {\bibinfo  {journal} {Sci. China Phys., Mech. Astron.}\
  }\textbf {\bibinfo {volume} {60}},\ \bibinfo {pages} {1--35} (\bibinfo {year}
  {2017})}\BibitemShut {NoStop}%
\bibitem [{\citenamefont {Akula}\ \emph {et~al.}(2017)\citenamefont {Akula},
  \citenamefont {Suchandra}, \citenamefont {Mikhaeil},\ and\ \citenamefont
  {Ranjan}}]{akula2017JOFM}%
  \BibitemOpen
  \bibfield  {author} {\bibinfo {author} {\bibfnamefont {B.}~\bibnamefont
  {Akula}}, \bibinfo {author} {\bibfnamefont {P.}~\bibnamefont {Suchandra}},
  \bibinfo {author} {\bibfnamefont {M.}~\bibnamefont {Mikhaeil}}, \ and\
  \bibinfo {author} {\bibfnamefont {D.}~\bibnamefont {Ranjan}},\ }\bibfield
  {title} {\enquote {\bibinfo {title} {Dynamics of unstably stratified free
  shear flows: an experimental investigation of coupled {K}elvin--{H}elmholtz
  and {R}ayleigh--{T}aylor instability},}\ }\href@noop {} {\bibfield  {journal}
  {\bibinfo  {journal} {J. Fluid Mech.}\ }\textbf {\bibinfo {volume} {816}},\
  \bibinfo {pages} {619--660} (\bibinfo {year} {2017})}\BibitemShut {NoStop}%
\bibitem [{\citenamefont {Zhou}(2017{\natexlab{a}})}]{zhou2017rayleigh1}%
  \BibitemOpen
  \bibfield  {author} {\bibinfo {author} {\bibfnamefont {Y.}~\bibnamefont
  {Zhou}},\ }\bibfield  {title} {\enquote {\bibinfo {title}
  {{R}ayleigh-{T}aylor and {R}ichtmyer-{M}eshkov instability induced flow,
  turbulence, and mixing. i},}\ }\href@noop {} {\bibfield  {journal} {\bibinfo
  {journal} {Phys. Rep.}\ }\textbf {\bibinfo {volume} {720-722}},\ \bibinfo
  {pages} {1--136} (\bibinfo {year} {2017}{\natexlab{a}})}\BibitemShut
  {NoStop}%
\bibitem [{\citenamefont {Zhou}(2017{\natexlab{b}})}]{zhou2017rayleigh2}%
  \BibitemOpen
  \bibfield  {author} {\bibinfo {author} {\bibfnamefont {Y.}~\bibnamefont
  {Zhou}},\ }\bibfield  {title} {\enquote {\bibinfo {title}
  {{R}ayleigh-{T}aylor and {R}ichtmyer-{M}eshkov instability induced flow,
  turbulence, and mixing. ii},}\ }\href@noop {} {\bibfield  {journal} {\bibinfo
   {journal} {Phys. Rep.}\ }\textbf {\bibinfo {volume} {723-725}},\ \bibinfo
  {pages} {1--160} (\bibinfo {year} {2017}{\natexlab{b}})}\BibitemShut
  {NoStop}%
\bibitem [{\citenamefont {Zhou}\ \emph {et~al.}(2019)\citenamefont {Zhou},
  \citenamefont {Clark}, \citenamefont {Clark}, \citenamefont
  {Gail~Glendinning}, \citenamefont {Aaron~Skinner}, \citenamefont
  {Huntington}, \citenamefont {Hurricane}, \citenamefont {Dimits},\ and\
  \citenamefont {Remington}}]{zhou2019turbulent}%
  \BibitemOpen
  \bibfield  {author} {\bibinfo {author} {\bibfnamefont {Y.}~\bibnamefont
  {Zhou}}, \bibinfo {author} {\bibfnamefont {T.~T.}\ \bibnamefont {Clark}},
  \bibinfo {author} {\bibfnamefont {D.~S.}\ \bibnamefont {Clark}}, \bibinfo
  {author} {\bibfnamefont {S.}~\bibnamefont {Gail~Glendinning}}, \bibinfo
  {author} {\bibfnamefont {M.}~\bibnamefont {Aaron~Skinner}}, \bibinfo {author}
  {\bibfnamefont {C.~M.}\ \bibnamefont {Huntington}}, \bibinfo {author}
  {\bibfnamefont {O.~A.}\ \bibnamefont {Hurricane}}, \bibinfo {author}
  {\bibfnamefont {A.~M.}\ \bibnamefont {Dimits}}, \ and\ \bibinfo {author}
  {\bibfnamefont {B.~A.}\ \bibnamefont {Remington}},\ }\bibfield  {title}
  {\enquote {\bibinfo {title} {Turbulent mixing and transition criteria of
  flows induced by hydrodynamic instabilities},}\ }\href@noop {} {\bibfield
  {journal} {\bibinfo  {journal} {Phys. Plasmas}\ }\textbf {\bibinfo {volume}
  {26}} (\bibinfo {year} {2019})}\BibitemShut {NoStop}%
\bibitem [{\citenamefont {Smyth}\ and\ \citenamefont
  {Moum}(2012)}]{smyth2012ocean}%
  \BibitemOpen
  \bibfield  {author} {\bibinfo {author} {\bibfnamefont {W.~D.}\ \bibnamefont
  {Smyth}}\ and\ \bibinfo {author} {\bibfnamefont {J.~N.}\ \bibnamefont
  {Moum}},\ }\bibfield  {title} {\enquote {\bibinfo {title} {Ocean mixing by
  {K}elvin-{H}elmholtz instability},}\ }\href@noop {} {\bibfield  {journal}
  {\bibinfo  {journal} {Adv. Oceanogr. Limnol.}\ }\textbf {\bibinfo {volume}
  {25}},\ \bibinfo {pages} {140--149} (\bibinfo {year} {2012})}\BibitemShut
  {NoStop}%
\bibitem [{\citenamefont {Martens}, \citenamefont {Kinzie},\ and\ \citenamefont
  {McLaughlin}(1994)}]{martens1994AIAA}%
  \BibitemOpen
  \bibfield  {author} {\bibinfo {author} {\bibfnamefont {S.}~\bibnamefont
  {Martens}}, \bibinfo {author} {\bibfnamefont {K.~W.}\ \bibnamefont {Kinzie}},
  \ and\ \bibinfo {author} {\bibfnamefont {D.~K.}\ \bibnamefont {McLaughlin}},\
  }\bibfield  {title} {\enquote {\bibinfo {title} {Measurements of
  {K}elvin-{H}elmholtz instabilities in a supersonic shear layer},}\
  }\href@noop {} {\bibfield  {journal} {\bibinfo  {journal} {AIAA J. Air
  Transp.}\ }\textbf {\bibinfo {volume} {32}},\ \bibinfo {pages} {1633--1639}
  (\bibinfo {year} {1994})}\BibitemShut {NoStop}%
\bibitem [{\citenamefont {Jiang}(2021)}]{jiang2021impact}%
  \BibitemOpen
  \bibfield  {author} {\bibinfo {author} {\bibfnamefont {Q.}~\bibnamefont
  {Jiang}},\ }\bibfield  {title} {\enquote {\bibinfo {title} {Impact of
  elevated {K}elvin--{H}elmholtz billows on the atmospheric boundary layer},}\
  }\href@noop {} {\bibfield  {journal} {\bibinfo  {journal} {J. Atmos. Sci.}\
  }\textbf {\bibinfo {volume} {78}},\ \bibinfo {pages} {3983--3999} (\bibinfo
  {year} {2021})}\BibitemShut {NoStop}%
\bibitem [{\citenamefont {Howson}, \citenamefont {De~Moortel},\ and\
  \citenamefont {Antolin}(2017)}]{howson2017AA}%
  \BibitemOpen
  \bibfield  {author} {\bibinfo {author} {\bibfnamefont {T.}~\bibnamefont
  {Howson}}, \bibinfo {author} {\bibfnamefont {I.}~\bibnamefont {De~Moortel}},
  \ and\ \bibinfo {author} {\bibfnamefont {P.}~\bibnamefont {Antolin}},\
  }\bibfield  {title} {\enquote {\bibinfo {title} {The effects of resistivity
  and viscosity on the {K}elvin-{H}elmholtz instability in oscillating coronal
  loops},}\ }\href@noop {} {\bibfield  {journal} {\bibinfo  {journal} {Astron.
  Astrophys.}\ }\textbf {\bibinfo {volume} {602}},\ \bibinfo {pages} {A74}
  (\bibinfo {year} {2017})}\BibitemShut {NoStop}%
\bibitem [{\citenamefont {Hasegawa}\ \emph {et~al.}(2004)\citenamefont
  {Hasegawa}, \citenamefont {Fujimoto}, \citenamefont {Phan}, \citenamefont
  {Reme}, \citenamefont {Balogh}, \citenamefont {Dunlop}, \citenamefont
  {Hashimoto},\ and\ \citenamefont {TanDokoro}}]{hasegawa2004NAT}%
  \BibitemOpen
  \bibfield  {author} {\bibinfo {author} {\bibfnamefont {H.}~\bibnamefont
  {Hasegawa}}, \bibinfo {author} {\bibfnamefont {M.}~\bibnamefont {Fujimoto}},
  \bibinfo {author} {\bibfnamefont {T.~D.}\ \bibnamefont {Phan}}, \bibinfo
  {author} {\bibfnamefont {H.}~\bibnamefont {Reme}}, \bibinfo {author}
  {\bibfnamefont {A.}~\bibnamefont {Balogh}}, \bibinfo {author} {\bibfnamefont
  {M.}~\bibnamefont {Dunlop}}, \bibinfo {author} {\bibfnamefont
  {C.}~\bibnamefont {Hashimoto}}, \ and\ \bibinfo {author} {\bibfnamefont
  {R.}~\bibnamefont {TanDokoro}},\ }\bibfield  {title} {\enquote {\bibinfo
  {title} {Transport of solar wind into {E}arth's magnetosphere through
  rolled-up {K}elvin--{H}elmholtz vortices},}\ }\href@noop {} {\bibfield
  {journal} {\bibinfo  {journal} {Nature}\ }\textbf {\bibinfo {volume} {430}},\
  \bibinfo {pages} {755--758} (\bibinfo {year} {2004})}\BibitemShut {NoStop}%
\bibitem [{\citenamefont {Nomoto}, \citenamefont {Iwamoto},\ and\ \citenamefont
  {Kishimoto}(1997)}]{nomoto1997type}%
  \BibitemOpen
  \bibfield  {author} {\bibinfo {author} {\bibfnamefont {K.}~\bibnamefont
  {Nomoto}}, \bibinfo {author} {\bibfnamefont {K.}~\bibnamefont {Iwamoto}}, \
  and\ \bibinfo {author} {\bibfnamefont {N.}~\bibnamefont {Kishimoto}},\
  }\bibfield  {title} {\enquote {\bibinfo {title} {Type ia supernovae: their
  origin and possible applications in cosmology},}\ }\href@noop {} {\bibfield
  {journal} {\bibinfo  {journal} {Science}\ }\textbf {\bibinfo {volume}
  {276}},\ \bibinfo {pages} {1378--1382} (\bibinfo {year} {1997})}\BibitemShut
  {NoStop}%
\bibitem [{\citenamefont {Gamezo}\ \emph {et~al.}(2003)\citenamefont {Gamezo},
  \citenamefont {Khokhlov}, \citenamefont {Oran}, \citenamefont
  {Chtchelkanova},\ and\ \citenamefont {Rosenberg}}]{gamezo2003SCI}%
  \BibitemOpen
  \bibfield  {author} {\bibinfo {author} {\bibfnamefont {V.~N.}\ \bibnamefont
  {Gamezo}}, \bibinfo {author} {\bibfnamefont {A.~M.}\ \bibnamefont
  {Khokhlov}}, \bibinfo {author} {\bibfnamefont {E.~S.}\ \bibnamefont {Oran}},
  \bibinfo {author} {\bibfnamefont {A.~Y.}\ \bibnamefont {Chtchelkanova}}, \
  and\ \bibinfo {author} {\bibfnamefont {R.~O.}\ \bibnamefont {Rosenberg}},\
  }\bibfield  {title} {\enquote {\bibinfo {title} {Thermonuclear supernovae:
  Simulations of the deflagration stage and their implications},}\ }\href@noop
  {} {\bibfield  {journal} {\bibinfo  {journal} {Science}\ }\textbf {\bibinfo
  {volume} {299}},\ \bibinfo {pages} {77--81} (\bibinfo {year}
  {2003})}\BibitemShut {NoStop}%
\bibitem [{\citenamefont {Perucho}\ \emph {et~al.}(2007)\citenamefont
  {Perucho}, \citenamefont {Hanasz}, \citenamefont {Mart{\'\i}},\ and\
  \citenamefont {Miralles}}]{perucho2007PRE}%
  \BibitemOpen
  \bibfield  {author} {\bibinfo {author} {\bibfnamefont {M.}~\bibnamefont
  {Perucho}}, \bibinfo {author} {\bibfnamefont {M.}~\bibnamefont {Hanasz}},
  \bibinfo {author} {\bibfnamefont {J.-M.}\ \bibnamefont {Mart{\'\i}}}, \ and\
  \bibinfo {author} {\bibfnamefont {J.-A.}\ \bibnamefont {Miralles}},\
  }\bibfield  {title} {\enquote {\bibinfo {title} {Resonant
  {K}elvin-{H}elmholtz modes in sheared relativistic flows},}\ }\href@noop {}
  {\bibfield  {journal} {\bibinfo  {journal} {Phys. Rev. E: Stat. Nonlinear
  Soft Matter Phys.}\ }\textbf {\bibinfo {volume} {75}},\ \bibinfo {pages}
  {056312} (\bibinfo {year} {2007})}\BibitemShut {NoStop}%
\bibitem [{\citenamefont {Yang}\ \emph
  {et~al.}(2018{\natexlab{a}})\citenamefont {Yang}, \citenamefont {Xu},
  \citenamefont {Lim}, \citenamefont {Kim}, \citenamefont {Cho}, \citenamefont
  {Kim}, \citenamefont {Chae}, \citenamefont {Cho},\ and\ \citenamefont
  {Ji}}]{yang2018TAJ}%
  \BibitemOpen
  \bibfield  {author} {\bibinfo {author} {\bibfnamefont {H.}~\bibnamefont
  {Yang}}, \bibinfo {author} {\bibfnamefont {Z.}~\bibnamefont {Xu}}, \bibinfo
  {author} {\bibfnamefont {E.~K.}\ \bibnamefont {Lim}}, \bibinfo {author}
  {\bibfnamefont {S.}~\bibnamefont {Kim}}, \bibinfo {author} {\bibfnamefont
  {K.~S.}\ \bibnamefont {Cho}}, \bibinfo {author} {\bibfnamefont {Y.~H.}\
  \bibnamefont {Kim}}, \bibinfo {author} {\bibfnamefont {J.}~\bibnamefont
  {Chae}}, \bibinfo {author} {\bibfnamefont {K.}~\bibnamefont {Cho}}, \ and\
  \bibinfo {author} {\bibfnamefont {K.}~\bibnamefont {Ji}},\ }\bibfield
  {title} {\enquote {\bibinfo {title} {Observation of the {K}elvin--{H}elmholtz
  instability in a solar prominence},}\ }\href@noop {} {\bibfield  {journal}
  {\bibinfo  {journal} {Astrophys. J.}\ }\textbf {\bibinfo {volume} {857}},\
  \bibinfo {pages} {115} (\bibinfo {year} {2018}{\natexlab{a}})}\BibitemShut
  {NoStop}%
\bibitem [{\citenamefont {Sadler}\ \emph {et~al.}(2022)\citenamefont {Sadler},
  \citenamefont {Green}, \citenamefont {Li}, \citenamefont {Zhou},
  \citenamefont {Flippo},\ and\ \citenamefont {Li}}]{sadler2022AIP}%
  \BibitemOpen
  \bibfield  {author} {\bibinfo {author} {\bibfnamefont {J.~D.}\ \bibnamefont
  {Sadler}}, \bibinfo {author} {\bibfnamefont {S.}~\bibnamefont {Green}},
  \bibinfo {author} {\bibfnamefont {S.}~\bibnamefont {Li}}, \bibinfo {author}
  {\bibfnamefont {Y.}~\bibnamefont {Zhou}}, \bibinfo {author} {\bibfnamefont
  {K.~A.}\ \bibnamefont {Flippo}}, \ and\ \bibinfo {author} {\bibfnamefont
  {H.}~\bibnamefont {Li}},\ }\bibfield  {title} {\enquote {\bibinfo {title}
  {Faster ablative {K}elvin--{H}elmholtz instability growth in a magnetic
  field},}\ }\href@noop {} {\bibfield  {journal} {\bibinfo  {journal} {Phys.
  Plasma}\ }\textbf {\bibinfo {volume} {29}} (\bibinfo {year}
  {2022})}\BibitemShut {NoStop}%
\bibitem [{\citenamefont {Rikanati}\ \emph {et~al.}(2000)\citenamefont
  {Rikanati}, \citenamefont {Oron}, \citenamefont {Alon},\ and\ \citenamefont
  {Shvarts}}]{rikanati2000IOP}%
  \BibitemOpen
  \bibfield  {author} {\bibinfo {author} {\bibfnamefont {A.}~\bibnamefont
  {Rikanati}}, \bibinfo {author} {\bibfnamefont {D.}~\bibnamefont {Oron}},
  \bibinfo {author} {\bibfnamefont {U.}~\bibnamefont {Alon}}, \ and\ \bibinfo
  {author} {\bibfnamefont {D.}~\bibnamefont {Shvarts}},\ }\bibfield  {title}
  {\enquote {\bibinfo {title} {Statistical mechanics merger model for
  hydrodynamic instabilities},}\ }\href@noop {} {\bibfield  {journal} {\bibinfo
   {journal} {Astrophys. J. Suppl. Ser.}\ }\textbf {\bibinfo {volume} {127}},\
  \bibinfo {pages} {451} (\bibinfo {year} {2000})}\BibitemShut {NoStop}%
\bibitem [{\citenamefont {He}\ and\ \citenamefont {Zhang}(2007)}]{he2007TEPJ}%
  \BibitemOpen
  \bibfield  {author} {\bibinfo {author} {\bibfnamefont {X.}~\bibnamefont
  {He}}\ and\ \bibinfo {author} {\bibfnamefont {W.}~\bibnamefont {Zhang}},\
  }\bibfield  {title} {\enquote {\bibinfo {title} {Inertial fusion research in
  {C}hina},}\ }\href@noop {} {\bibfield  {journal} {\bibinfo  {journal} {Eur.
  Phys. J. D}\ }\textbf {\bibinfo {volume} {44}},\ \bibinfo {pages} {227--231}
  (\bibinfo {year} {2007})}\BibitemShut {NoStop}%
\bibitem [{\citenamefont {Rinderknecht}\ \emph {et~al.}(2018)\citenamefont
  {Rinderknecht}, \citenamefont {Amendt}, \citenamefont {Wilks},\ and\
  \citenamefont {Collins}}]{rinderknecht2018kinetic}%
  \BibitemOpen
  \bibfield  {author} {\bibinfo {author} {\bibfnamefont {H.~G.}\ \bibnamefont
  {Rinderknecht}}, \bibinfo {author} {\bibfnamefont {P.}~\bibnamefont
  {Amendt}}, \bibinfo {author} {\bibfnamefont {S.}~\bibnamefont {Wilks}}, \
  and\ \bibinfo {author} {\bibfnamefont {G.}~\bibnamefont {Collins}},\
  }\bibfield  {title} {\enquote {\bibinfo {title} {Kinetic physics in {ICF}:
  present understanding and future directions},}\ }\href@noop {} {\bibfield
  {journal} {\bibinfo  {journal} {Plasma Phys. Control. Fusion}\ }\textbf
  {\bibinfo {volume} {60}},\ \bibinfo {pages} {064001} (\bibinfo {year}
  {2018})}\BibitemShut {NoStop}%
\bibitem [{\citenamefont {Cai}\ \emph {et~al.}(2021)\citenamefont {Cai},
  \citenamefont {Yan}, \citenamefont {Yao},\ and\ \citenamefont
  {Zhu}}]{cai2021hybrid}%
  \BibitemOpen
  \bibfield  {author} {\bibinfo {author} {\bibfnamefont {H.}~\bibnamefont
  {Cai}}, \bibinfo {author} {\bibfnamefont {X.}~\bibnamefont {Yan}}, \bibinfo
  {author} {\bibfnamefont {P.}~\bibnamefont {Yao}}, \ and\ \bibinfo {author}
  {\bibfnamefont {S.}~\bibnamefont {Zhu}},\ }\bibfield  {title} {\enquote
  {\bibinfo {title} {Hybrid fluid--particle modeling of shock-driven
  hydrodynamic instabilities in a plasma},}\ }\href@noop {} {\bibfield
  {journal} {\bibinfo  {journal} {Matter Radiat. Extremes}\ }\textbf {\bibinfo
  {volume} {6}} (\bibinfo {year} {2021})}\BibitemShut {NoStop}%
\bibitem [{\citenamefont {Shan}\ \emph {et~al.}(2021)\citenamefont {Shan},
  \citenamefont {Wu}, \citenamefont {Yuan}, \citenamefont {Wang}, \citenamefont
  {Cai}, \citenamefont {Tian}, \citenamefont {Zhang}, \citenamefont {Zhang},
  \citenamefont {Deng}, \citenamefont {Zhang} \emph
  {et~al.}}]{lianqiang2020research}%
  \BibitemOpen
  \bibfield  {author} {\bibinfo {author} {\bibfnamefont {L.}~\bibnamefont
  {Shan}}, \bibinfo {author} {\bibfnamefont {F.}~\bibnamefont {Wu}}, \bibinfo
  {author} {\bibfnamefont {Z.}~\bibnamefont {Yuan}}, \bibinfo {author}
  {\bibfnamefont {W.}~\bibnamefont {Wang}}, \bibinfo {author} {\bibfnamefont
  {H.}~\bibnamefont {Cai}}, \bibinfo {author} {\bibfnamefont {C.}~\bibnamefont
  {Tian}}, \bibinfo {author} {\bibfnamefont {F.}~\bibnamefont {Zhang}},
  \bibinfo {author} {\bibfnamefont {T.}~\bibnamefont {Zhang}}, \bibinfo
  {author} {\bibfnamefont {Z.}~\bibnamefont {Deng}}, \bibinfo {author}
  {\bibfnamefont {W.}~\bibnamefont {Zhang}},  \emph {et~al.},\ }\bibfield
  {title} {\enquote {\bibinfo {title} {Research progress of kinetic effects in
  laser inertial confinement fusion},}\ }\href@noop {} {\bibfield  {journal}
  {\bibinfo  {journal} {High Power Laser and Particle Beams}\ }\textbf
  {\bibinfo {volume} {33}},\ \bibinfo {pages} {012004} (\bibinfo {year}
  {2021})}\BibitemShut {NoStop}%
\bibitem [{\citenamefont {Papas}\ and\ \citenamefont
  {Rais}(2009)}]{papas2009POTC}%
  \BibitemOpen
  \bibfield  {author} {\bibinfo {author} {\bibfnamefont {P.}~\bibnamefont
  {Papas}}\ and\ \bibinfo {author} {\bibfnamefont {R.}~\bibnamefont {Rais}},\
  }\bibfield  {title} {\enquote {\bibinfo {title} {Thermo-diffusive
  instabilities in axisymmetric, non-premixed jet flames},}\ }\href@noop {}
  {\bibfield  {journal} {\bibinfo  {journal} {Proc. Combust. Inst.}\ }\textbf
  {\bibinfo {volume} {32}},\ \bibinfo {pages} {1181--1189} (\bibinfo {year}
  {2009})}\BibitemShut {NoStop}%
\bibitem [{\citenamefont {Som}\ and\ \citenamefont
  {Aggarwal}(2010)}]{som2010CAF}%
  \BibitemOpen
  \bibfield  {author} {\bibinfo {author} {\bibfnamefont {S.}~\bibnamefont
  {Som}}\ and\ \bibinfo {author} {\bibfnamefont {S.~K.}\ \bibnamefont
  {Aggarwal}},\ }\bibfield  {title} {\enquote {\bibinfo {title} {Effects of
  primary breakup modeling on spray and combustion characteristics of
  compression ignition engines},}\ }\href@noop {} {\bibfield  {journal}
  {\bibinfo  {journal} {Combust. Flame}\ }\textbf {\bibinfo {volume} {157}},\
  \bibinfo {pages} {1179--1193} (\bibinfo {year} {2010})}\BibitemShut {NoStop}%
\bibitem [{\citenamefont {Liu}, \citenamefont {Tan},\ and\ \citenamefont
  {Xu}(2015)}]{liu2015POTN}%
  \BibitemOpen
  \bibfield  {author} {\bibinfo {author} {\bibfnamefont {Y.}~\bibnamefont
  {Liu}}, \bibinfo {author} {\bibfnamefont {P.}~\bibnamefont {Tan}}, \ and\
  \bibinfo {author} {\bibfnamefont {L.}~\bibnamefont {Xu}},\ }\bibfield
  {title} {\enquote {\bibinfo {title} {{K}elvin--{H}elmholtz instability in an
  ultrathin air film causes drop splashing on smooth surfaces},}\ }\href@noop
  {} {\bibfield  {journal} {\bibinfo  {journal} {Proc. Natl. Acad. Sci., India,
  Sect. A Phys. Sci.}\ }\textbf {\bibinfo {volume} {112}},\ \bibinfo {pages}
  {3280--3284} (\bibinfo {year} {2015})}\BibitemShut {NoStop}%
\bibitem [{\citenamefont {Bertevas}\ \emph {et~al.}(2019)\citenamefont
  {Bertevas}, \citenamefont {Tran~Duc}, \citenamefont {Le~Cao}, \citenamefont
  {Khoo},\ and\ \citenamefont {Phan~Thien}}]{bertevas2019POF}%
  \BibitemOpen
  \bibfield  {author} {\bibinfo {author} {\bibfnamefont {E.}~\bibnamefont
  {Bertevas}}, \bibinfo {author} {\bibfnamefont {T.}~\bibnamefont {Tran~Duc}},
  \bibinfo {author} {\bibfnamefont {K.}~\bibnamefont {Le~Cao}}, \bibinfo
  {author} {\bibfnamefont {B.~C.}\ \bibnamefont {Khoo}}, \ and\ \bibinfo
  {author} {\bibfnamefont {N.}~\bibnamefont {Phan~Thien}},\ }\bibfield  {title}
  {\enquote {\bibinfo {title} {A smoothed particle hydrodynamics ({SPH})
  formulation of a two-phase mixture model and its application to turbulent
  sediment transport},}\ }\href@noop {} {\bibfield  {journal} {\bibinfo
  {journal} {Phys. Fluids}\ }\textbf {\bibinfo {volume} {31}} (\bibinfo {year}
  {2019})}\BibitemShut {NoStop}%
\bibitem [{\citenamefont {Liu}\ \emph {et~al.}(2015)\citenamefont {Liu},
  \citenamefont {Wang}, \citenamefont {Zang},\ and\ \citenamefont
  {Zhao}}]{liu2015IJOH}%
  \BibitemOpen
  \bibfield  {author} {\bibinfo {author} {\bibfnamefont {G.}~\bibnamefont
  {Liu}}, \bibinfo {author} {\bibfnamefont {Y.}~\bibnamefont {Wang}}, \bibinfo
  {author} {\bibfnamefont {G.}~\bibnamefont {Zang}}, \ and\ \bibinfo {author}
  {\bibfnamefont {H.}~\bibnamefont {Zhao}},\ }\bibfield  {title} {\enquote
  {\bibinfo {title} {Viscous {K}elvin--{H}elmholtz instability analysis of
  liquid--vapor two-phase stratified flow for condensation in horizontal
  tubes},}\ }\href@noop {} {\bibfield  {journal} {\bibinfo  {journal} {Int. J.
  Heat Mass Transfer}\ }\textbf {\bibinfo {volume} {84}},\ \bibinfo {pages}
  {592--599} (\bibinfo {year} {2015})}\BibitemShut {NoStop}%
\bibitem [{\citenamefont {Zhang}\ \emph {et~al.}(2001)\citenamefont {Zhang},
  \citenamefont {He}, \citenamefont {Doolen},\ and\ \citenamefont
  {Chen}}]{zhang2001AIWR}%
  \BibitemOpen
  \bibfield  {author} {\bibinfo {author} {\bibfnamefont {R.}~\bibnamefont
  {Zhang}}, \bibinfo {author} {\bibfnamefont {X.}~\bibnamefont {He}}, \bibinfo
  {author} {\bibfnamefont {G.}~\bibnamefont {Doolen}}, \ and\ \bibinfo {author}
  {\bibfnamefont {S.}~\bibnamefont {Chen}},\ }\bibfield  {title} {\enquote
  {\bibinfo {title} {Surface tension effects on two-dimensional two-phase
  {K}elvin--{H}elmholtz instabilities},}\ }\href@noop {} {\bibfield  {journal}
  {\bibinfo  {journal} {Adv. Water Resour.}\ }\textbf {\bibinfo {volume}
  {24}},\ \bibinfo {pages} {461--478} (\bibinfo {year} {2001})}\BibitemShut
  {NoStop}%
\bibitem [{\citenamefont {Asthana}\ and\ \citenamefont
  {Agrawal}(2010)}]{asthana2010IJOE}%
  \BibitemOpen
  \bibfield  {author} {\bibinfo {author} {\bibfnamefont {R.}~\bibnamefont
  {Asthana}}\ and\ \bibinfo {author} {\bibfnamefont {G.}~\bibnamefont
  {Agrawal}},\ }\bibfield  {title} {\enquote {\bibinfo {title} {Viscous
  potential flow analysis of electrohydrodynamic {K}elvin--{H}elmholtz
  instability with heat and mass transfer},}\ }\href@noop {} {\bibfield
  {journal} {\bibinfo  {journal} {Int. J. Eng. Sci.}\ }\textbf {\bibinfo
  {volume} {48}},\ \bibinfo {pages} {1925--1936} (\bibinfo {year}
  {2010})}\BibitemShut {NoStop}%
\bibitem [{\citenamefont {Lee}\ and\ \citenamefont {Kim}(2015)}]{lee2015EJOM}%
  \BibitemOpen
  \bibfield  {author} {\bibinfo {author} {\bibfnamefont {H.~G.}\ \bibnamefont
  {Lee}}\ and\ \bibinfo {author} {\bibfnamefont {J.}~\bibnamefont {Kim}},\
  }\bibfield  {title} {\enquote {\bibinfo {title} {Two-dimensional
  {K}elvin--{H}elmholtz instabilities of multi-component fluids},}\ }\href@noop
  {} {\bibfield  {journal} {\bibinfo  {journal} {Eur. J. Mech. B. Fluids}\
  }\textbf {\bibinfo {volume} {49}},\ \bibinfo {pages} {77--88} (\bibinfo
  {year} {2015})}\BibitemShut {NoStop}%
\bibitem [{\citenamefont {Sofonea}\ and\ \citenamefont
  {Sekerka}(2001)}]{sofonea2001PASM}%
  \BibitemOpen
  \bibfield  {author} {\bibinfo {author} {\bibfnamefont {V.}~\bibnamefont
  {Sofonea}}\ and\ \bibinfo {author} {\bibfnamefont {R.~F.}\ \bibnamefont
  {Sekerka}},\ }\bibfield  {title} {\enquote {\bibinfo {title} {{BGK} models
  for diffusion in isothermal binary fluid systems},}\ }\href@noop {}
  {\bibfield  {journal} {\bibinfo  {journal} {Physica A}\ }\textbf {\bibinfo
  {volume} {299}},\ \bibinfo {pages} {494--520} (\bibinfo {year}
  {2001})}\BibitemShut {NoStop}%
\bibitem [{\citenamefont {Wan}\ \emph {et~al.}(2015)\citenamefont {Wan},
  \citenamefont {Malamud}, \citenamefont {Shimony}, \citenamefont {Di~Stefano},
  \citenamefont {Trantham}, \citenamefont {Klein}, \citenamefont {Shvarts},
  \citenamefont {Kuranz},\ and\ \citenamefont {Drake}}]{wan2015PRL}%
  \BibitemOpen
  \bibfield  {author} {\bibinfo {author} {\bibfnamefont {W.}~\bibnamefont
  {Wan}}, \bibinfo {author} {\bibfnamefont {G.}~\bibnamefont {Malamud}},
  \bibinfo {author} {\bibfnamefont {A.}~\bibnamefont {Shimony}}, \bibinfo
  {author} {\bibfnamefont {C.}~\bibnamefont {Di~Stefano}}, \bibinfo {author}
  {\bibfnamefont {M.}~\bibnamefont {Trantham}}, \bibinfo {author}
  {\bibfnamefont {S.}~\bibnamefont {Klein}}, \bibinfo {author} {\bibfnamefont
  {D.}~\bibnamefont {Shvarts}}, \bibinfo {author} {\bibfnamefont
  {C.}~\bibnamefont {Kuranz}}, \ and\ \bibinfo {author} {\bibfnamefont
  {R.}~\bibnamefont {Drake}},\ }\bibfield  {title} {\enquote {\bibinfo {title}
  {Observation of single-mode, {K}elvin-{H}elmholtz instability in a supersonic
  flow},}\ }\href@noop {} {\bibfield  {journal} {\bibinfo  {journal} {Phys.
  Rev. Lett.}\ }\textbf {\bibinfo {volume} {115}},\ \bibinfo {pages} {145001}
  (\bibinfo {year} {2015})}\BibitemShut {NoStop}%
\bibitem [{\citenamefont {Mieussens}(2000)}]{mieussens2000JOCP}%
  \BibitemOpen
  \bibfield  {author} {\bibinfo {author} {\bibfnamefont {L.}~\bibnamefont
  {Mieussens}},\ }\bibfield  {title} {\enquote {\bibinfo {title}
  {Discrete-velocity models and numerical schemes for the {B}oltzmann-{BGK}
  equation in plane and axisymmetric geometries},}\ }\href@noop {} {\bibfield
  {journal} {\bibinfo  {journal} {J. Comput. Phys.}\ }\textbf {\bibinfo
  {volume} {162}},\ \bibinfo {pages} {429--466} (\bibinfo {year}
  {2000})}\BibitemShut {NoStop}%
\bibitem [{\citenamefont {Kim}\ and\ \citenamefont {Bang}(2022)}]{kim2022MDPI}%
  \BibitemOpen
  \bibfield  {author} {\bibinfo {author} {\bibfnamefont {M.~S.}\ \bibnamefont
  {Kim}}\ and\ \bibinfo {author} {\bibfnamefont {K.~H.}\ \bibnamefont {Bang}},\
  }\bibfield  {title} {\enquote {\bibinfo {title} {Numerical simulation of
  {K}elvin--{H}elmholtz instability and boundary layer stripping for an
  interpretation of melt jet breakup mechanisms},}\ }\href@noop {} {\bibfield
  {journal} {\bibinfo  {journal} {Energ. Wasser Prax.}\ }\textbf {\bibinfo
  {volume} {15}},\ \bibinfo {pages} {7517} (\bibinfo {year}
  {2022})}\BibitemShut {NoStop}%
\bibitem [{\citenamefont {Leon}\ and\ \citenamefont
  {Manna}(1999)}]{leon1999APS}%
  \BibitemOpen
  \bibfield  {author} {\bibinfo {author} {\bibfnamefont {J.}~\bibnamefont
  {Leon}}\ and\ \bibinfo {author} {\bibfnamefont {M.}~\bibnamefont {Manna}},\
  }\bibfield  {title} {\enquote {\bibinfo {title} {Discrete instability in
  nonlinear lattices},}\ }\href@noop {} {\bibfield  {journal} {\bibinfo
  {journal} {Phys. Rev. Lett.}\ }\textbf {\bibinfo {volume} {83}},\ \bibinfo
  {pages} {2324} (\bibinfo {year} {1999})}\BibitemShut {NoStop}%
\bibitem [{\citenamefont {Olson}\ \emph {et~al.}(2011)\citenamefont {Olson},
  \citenamefont {Larsson}, \citenamefont {Lele},\ and\ \citenamefont
  {Cook}}]{olson2011AIP}%
  \BibitemOpen
  \bibfield  {author} {\bibinfo {author} {\bibfnamefont {B.~J.}\ \bibnamefont
  {Olson}}, \bibinfo {author} {\bibfnamefont {J.}~\bibnamefont {Larsson}},
  \bibinfo {author} {\bibfnamefont {S.~K.}\ \bibnamefont {Lele}}, \ and\
  \bibinfo {author} {\bibfnamefont {A.~W.}\ \bibnamefont {Cook}},\ }\bibfield
  {title} {\enquote {\bibinfo {title} {Nonlinear effects in the combined
  {R}ayleigh-{T}aylor/{K}elvin-{H}elmholtz instability},}\ }\href@noop {}
  {\bibfield  {journal} {\bibinfo  {journal} {Phys. Fluids}\ }\textbf {\bibinfo
  {volume} {23}} (\bibinfo {year} {2011})}\BibitemShut {NoStop}%
\bibitem [{\citenamefont {Chaurasia}\ and\ \citenamefont
  {Thompson}(2011)}]{chaurasia2011CUP}%
  \BibitemOpen
  \bibfield  {author} {\bibinfo {author} {\bibfnamefont {H.~K.}\ \bibnamefont
  {Chaurasia}}\ and\ \bibinfo {author} {\bibfnamefont {M.~C.}\ \bibnamefont
  {Thompson}},\ }\bibfield  {title} {\enquote {\bibinfo {title}
  {Three-dimensional instabilities in the boundary-layer flow over a long
  rectangular plate},}\ }\href@noop {} {\bibfield  {journal} {\bibinfo
  {journal} {J. Fluid Mech.}\ }\textbf {\bibinfo {volume} {681}},\ \bibinfo
  {pages} {411--433} (\bibinfo {year} {2011})}\BibitemShut {NoStop}%
\bibitem [{\citenamefont {Lund}\ and\ \citenamefont
  {Werne}(2004)}]{lund2004IEEE}%
  \BibitemOpen
  \bibfield  {author} {\bibinfo {author} {\bibfnamefont {T.~S.}\ \bibnamefont
  {Lund}}\ and\ \bibinfo {author} {\bibfnamefont {J.~A.}\ \bibnamefont
  {Werne}},\ }\bibfield  {title} {\enquote {\bibinfo {title} {Numerical
  simulation of {K}elvin-{H}elmholtz instability via direct-and large-eddy
  simulation},}\ }in\ \href@noop {} {\emph {\bibinfo {booktitle} {2004 Users
  Group Conference (DOD\_UGC'04)}}}\ (\bibinfo {organization} {IEEE},\ \bibinfo
  {year} {2004})\ pp.\ \bibinfo {pages} {89--95}\BibitemShut {NoStop}%
\bibitem [{\citenamefont {Ganesh}(2010)}]{ganesh2010PRL}%
  \BibitemOpen
  \bibfield  {author} {\bibinfo {author} {\bibfnamefont {R.}~\bibnamefont
  {Ganesh}},\ }\bibfield  {title} {\enquote {\bibinfo {title} {Kelvin
  {H}elmholtz instability in strongly coupled yukawa liquids},}\ }\href@noop {}
  {\bibfield  {journal} {\bibinfo  {journal} {Phys. Rev. Lett.}\ }\textbf
  {\bibinfo {volume} {104}},\ \bibinfo {pages} {215003} (\bibinfo {year}
  {2010})}\BibitemShut {NoStop}%
\bibitem [{\citenamefont {Shadloo}\ and\ \citenamefont
  {Yildiz}(2011)}]{shadloo2011SMSA}%
  \BibitemOpen
  \bibfield  {author} {\bibinfo {author} {\bibfnamefont {M.~S.}\ \bibnamefont
  {Shadloo}}\ and\ \bibinfo {author} {\bibfnamefont {M.}~\bibnamefont
  {Yildiz}},\ }\bibfield  {title} {\enquote {\bibinfo {title} {Numerical
  modeling of {K}elvin--{H}elmholtz instability using smoothed particle
  hydrodynamics},}\ }\href@noop {} {\bibfield  {journal} {\bibinfo  {journal}
  {Int. J. Numer. Methods Eng.}\ }\textbf {\bibinfo {volume} {87}},\ \bibinfo
  {pages} {988--1006} (\bibinfo {year} {2011})}\BibitemShut {NoStop}%
\bibitem [{\citenamefont {Zellinger}\ \emph {et~al.}(2012)\citenamefont
  {Zellinger}, \citenamefont {M{\"o}stl}, \citenamefont {Erkaev},\ and\
  \citenamefont {Biernat}}]{zellinger2012POP}%
  \BibitemOpen
  \bibfield  {author} {\bibinfo {author} {\bibfnamefont {M.}~\bibnamefont
  {Zellinger}}, \bibinfo {author} {\bibfnamefont {U.}~\bibnamefont
  {M{\"o}stl}}, \bibinfo {author} {\bibfnamefont {N.}~\bibnamefont {Erkaev}}, \
  and\ \bibinfo {author} {\bibfnamefont {H.}~\bibnamefont {Biernat}},\
  }\bibfield  {title} {\enquote {\bibinfo {title} {2.5 {D} magnetohydrodynamic
  simulation of the {K}elvin-{H}elmholtz instability around
  {V}enus—{C}omparison of the influence of gravity and density increase},}\
  }\href@noop {} {\bibfield  {journal} {\bibinfo  {journal} {Phys. Plasma}\
  }\textbf {\bibinfo {volume} {19}} (\bibinfo {year} {2012})}\BibitemShut
  {NoStop}%
\bibitem [{\citenamefont {Rahmani}, \citenamefont {Lawrence},\ and\
  \citenamefont {Seymour}(2014)}]{rahmani2014JOFM}%
  \BibitemOpen
  \bibfield  {author} {\bibinfo {author} {\bibfnamefont {M.}~\bibnamefont
  {Rahmani}}, \bibinfo {author} {\bibfnamefont {G.}~\bibnamefont {Lawrence}}, \
  and\ \bibinfo {author} {\bibfnamefont {B.}~\bibnamefont {Seymour}},\
  }\bibfield  {title} {\enquote {\bibinfo {title} {The effect of reynolds
  number on mixing in {K}elvin--{H}elmholtz billows},}\ }\href@noop {}
  {\bibfield  {journal} {\bibinfo  {journal} {J. Fluid Mech.}\ }\textbf
  {\bibinfo {volume} {759}},\ \bibinfo {pages} {612--641} (\bibinfo {year}
  {2014})}\BibitemShut {NoStop}%
\bibitem [{\citenamefont {Liu}\ \emph {et~al.}(2018)\citenamefont {Liu},
  \citenamefont {Chen}, \citenamefont {Zhang},\ and\ \citenamefont
  {Lin}}]{liu2018POF}%
  \BibitemOpen
  \bibfield  {author} {\bibinfo {author} {\bibfnamefont {Y.}~\bibnamefont
  {Liu}}, \bibinfo {author} {\bibfnamefont {Z.}~\bibnamefont {Chen}}, \bibinfo
  {author} {\bibfnamefont {H.}~\bibnamefont {Zhang}}, \ and\ \bibinfo {author}
  {\bibfnamefont {Z.}~\bibnamefont {Lin}},\ }\bibfield  {title} {\enquote
  {\bibinfo {title} {Physical effects of magnetic fields on the
  {K}elvin-{H}elmholtz instability in a free shear layer},}\ }\href@noop {}
  {\bibfield  {journal} {\bibinfo  {journal} {Phys. Fluids}\ }\textbf {\bibinfo
  {volume} {30}} (\bibinfo {year} {2018})}\BibitemShut {NoStop}%
\bibitem [{\citenamefont {Fritts}\ \emph {et~al.}(2024)\citenamefont {Fritts},
  \citenamefont {Wang}, \citenamefont {Lund},\ and\ \citenamefont
  {Geller}}]{fritts2024JOTAS}%
  \BibitemOpen
  \bibfield  {author} {\bibinfo {author} {\bibfnamefont {D.~C.}\ \bibnamefont
  {Fritts}}, \bibinfo {author} {\bibfnamefont {L.}~\bibnamefont {Wang}},
  \bibinfo {author} {\bibfnamefont {T.}~\bibnamefont {Lund}}, \ and\ \bibinfo
  {author} {\bibfnamefont {M.~A.}\ \bibnamefont {Geller}},\ }\bibfield  {title}
  {\enquote {\bibinfo {title} {Kelvin {H}elmholtz instability “tube” and
  “knot” dynamics, {P}art {III}: Extension of elevated turbulence and
  energy dissipation into increasingly viscous flows},}\ }\href@noop {}
  {\bibfield  {journal} {\bibinfo  {journal} {J. Atmos. Sci.}\ } (\bibinfo
  {year} {2024})}\BibitemShut {NoStop}%
\bibitem [{\citenamefont {Succi}\ and\ \citenamefont
  {Succi}(2018)}]{succi2018lattice}%
  \BibitemOpen
  \bibfield  {author} {\bibinfo {author} {\bibfnamefont {S.}~\bibnamefont
  {Succi}}\ and\ \bibinfo {author} {\bibfnamefont {S.}~\bibnamefont {Succi}},\
  }\href@noop {} {\emph {\bibinfo {title} {The Lattice {B}oltzmann equation:
  for complex states of flowing matter}}}\ (\bibinfo  {publisher} {Oxford
  university press},\ \bibinfo {year} {2018})\BibitemShut {NoStop}%
\bibitem [{\citenamefont {Guo}\ and\ \citenamefont
  {Shu}(2013)}]{guo2013lattice}%
  \BibitemOpen
  \bibfield  {author} {\bibinfo {author} {\bibfnamefont {Z.}~\bibnamefont
  {Guo}}\ and\ \bibinfo {author} {\bibfnamefont {C.}~\bibnamefont {Shu}},\
  }\href@noop {} {\emph {\bibinfo {title} {Lattice {B}oltzmann method and its
  application in engineering}}},\ Vol.~\bibinfo {volume} {3}\ (\bibinfo
  {publisher} {World Scientific},\ \bibinfo {year} {2013})\BibitemShut
  {NoStop}%
\bibitem [{\citenamefont {Shan}, \citenamefont {Yuan},\ and\ \citenamefont
  {Chen}(2006)}]{Shan-JFM-2006}%
  \BibitemOpen
  \bibfield  {author} {\bibinfo {author} {\bibfnamefont {X.}~\bibnamefont
  {Shan}}, \bibinfo {author} {\bibfnamefont {X.-F.}\ \bibnamefont {Yuan}}, \
  and\ \bibinfo {author} {\bibfnamefont {H.}~\bibnamefont {Chen}},\ }\bibfield
  {title} {\enquote {\bibinfo {title} {Kinetic theory representation of
  hydrodynamics: a way beyond the navier--stokes equation},}\ }\href@noop {}
  {\bibfield  {journal} {\bibinfo  {journal} {J. Fluid Mech.}\ }\textbf
  {\bibinfo {volume} {550}},\ \bibinfo {pages} {413--441} (\bibinfo {year}
  {2006})}\BibitemShut {NoStop}%
\bibitem [{\citenamefont {Gonnella}, \citenamefont {Lamura},\ and\
  \citenamefont {Sofonea}(2007)}]{GLS-model-PRE2007}%
  \BibitemOpen
  \bibfield  {author} {\bibinfo {author} {\bibfnamefont {G.}~\bibnamefont
  {Gonnella}}, \bibinfo {author} {\bibfnamefont {A.}~\bibnamefont {Lamura}}, \
  and\ \bibinfo {author} {\bibfnamefont {V.}~\bibnamefont {Sofonea}},\
  }\bibfield  {title} {\enquote {\bibinfo {title} {Lattice {B}oltzmann
  simulation of thermal nonideal fluids},}\ }\href@noop {} {\bibfield
  {journal} {\bibinfo  {journal} {Phys. Rev. E}\ }\textbf {\bibinfo {volume}
  {76}},\ \bibinfo {pages} {036703} (\bibinfo {year} {2007})}\BibitemShut
  {NoStop}%
\bibitem [{\citenamefont {Fei}\ \emph {et~al.}(2023)\citenamefont {Fei},
  \citenamefont {Qin}, \citenamefont {Wang}, \citenamefont {Huang},
  \citenamefont {Wen}, \citenamefont {Zhao}, \citenamefont {Luo}, \citenamefont
  {Derome},\ and\ \citenamefont {Carmeliet}}]{fei2023coupled}%
  \BibitemOpen
  \bibfield  {author} {\bibinfo {author} {\bibfnamefont {L.}~\bibnamefont
  {Fei}}, \bibinfo {author} {\bibfnamefont {F.}~\bibnamefont {Qin}}, \bibinfo
  {author} {\bibfnamefont {G.}~\bibnamefont {Wang}}, \bibinfo {author}
  {\bibfnamefont {J.}~\bibnamefont {Huang}}, \bibinfo {author} {\bibfnamefont
  {B.}~\bibnamefont {Wen}}, \bibinfo {author} {\bibfnamefont {J.}~\bibnamefont
  {Zhao}}, \bibinfo {author} {\bibfnamefont {K.~H.}\ \bibnamefont {Luo}},
  \bibinfo {author} {\bibfnamefont {D.}~\bibnamefont {Derome}}, \ and\ \bibinfo
  {author} {\bibfnamefont {J.}~\bibnamefont {Carmeliet}},\ }\bibfield  {title}
  {\enquote {\bibinfo {title} {Coupled lattice {B}oltzmann method--discrete
  element method model for gas--liquid--solid interaction problems},}\
  }\href@noop {} {\bibfield  {journal} {\bibinfo  {journal} {J. Fluid Mech.}\
  }\textbf {\bibinfo {volume} {975}},\ \bibinfo {pages} {A20} (\bibinfo {year}
  {2023})}\BibitemShut {NoStop}%
\bibitem [{\citenamefont {Yang}\ \emph {et~al.}(2016)\citenamefont {Yang},
  \citenamefont {Shu}, \citenamefont {Wu},\ and\ \citenamefont
  {Wang}}]{yang2016JOCP}%
  \BibitemOpen
  \bibfield  {author} {\bibinfo {author} {\bibfnamefont {L.}~\bibnamefont
  {Yang}}, \bibinfo {author} {\bibfnamefont {C.}~\bibnamefont {Shu}}, \bibinfo
  {author} {\bibfnamefont {J.}~\bibnamefont {Wu}}, \ and\ \bibinfo {author}
  {\bibfnamefont {Y.}~\bibnamefont {Wang}},\ }\bibfield  {title} {\enquote
  {\bibinfo {title} {Numerical simulation of flows from free molecular regime
  to continuum regime by a dvm with streaming and collision processes},}\
  }\href@noop {} {\bibfield  {journal} {\bibinfo  {journal} {J. Comput. Phys.}\
  }\textbf {\bibinfo {volume} {306}},\ \bibinfo {pages} {291--310} (\bibinfo
  {year} {2016})}\BibitemShut {NoStop}%
\bibitem [{\citenamefont {Yang}\ \emph
  {et~al.}(2018{\natexlab{b}})\citenamefont {Yang}, \citenamefont {Shu},
  \citenamefont {Yang}, \citenamefont {Chen},\ and\ \citenamefont
  {Dong}}]{Shu-POF-2018}%
  \BibitemOpen
  \bibfield  {author} {\bibinfo {author} {\bibfnamefont {L.}~\bibnamefont
  {Yang}}, \bibinfo {author} {\bibfnamefont {C.}~\bibnamefont {Shu}}, \bibinfo
  {author} {\bibfnamefont {W.}~\bibnamefont {Yang}}, \bibinfo {author}
  {\bibfnamefont {Z.}~\bibnamefont {Chen}}, \ and\ \bibinfo {author}
  {\bibfnamefont {H.}~\bibnamefont {Dong}},\ }\bibfield  {title} {\enquote
  {\bibinfo {title} {An improved discrete velocity method (dvm) for efficient
  simulation of flows in all flow regimes},}\ }\href@noop {} {\bibfield
  {journal} {\bibinfo  {journal} {Phys. Fluids}\ }\textbf {\bibinfo {volume}
  {30}} (\bibinfo {year} {2018}{\natexlab{b}})}\BibitemShut {NoStop}%
\bibitem [{\citenamefont {Li}\ \emph {et~al.}(2015)\citenamefont {Li},
  \citenamefont {Peng}, \citenamefont {Zhang},\ and\ \citenamefont
  {Yang}}]{lzh-2015}%
  \BibitemOpen
  \bibfield  {author} {\bibinfo {author} {\bibfnamefont {Z.~H.}\ \bibnamefont
  {Li}}, \bibinfo {author} {\bibfnamefont {A.~P.}\ \bibnamefont {Peng}},
  \bibinfo {author} {\bibfnamefont {H.~X.}\ \bibnamefont {Zhang}}, \ and\
  \bibinfo {author} {\bibfnamefont {J.~Y.}\ \bibnamefont {Yang}},\ }\bibfield
  {title} {\enquote {\bibinfo {title} {Rarefied gas flow simulations using
  high-order gas-kinetic unified algorithms for {B}oltzmann model equations},}\
  }\href@noop {} {\bibfield  {journal} {\bibinfo  {journal} {Prog. Aerosp.
  Sci.}\ }\textbf {\bibinfo {volume} {74}},\ \bibinfo {pages} {81--113}
  (\bibinfo {year} {2015})}\BibitemShut {NoStop}%
\bibitem [{\citenamefont {Xu}\ and\ \citenamefont {Huang}(2010)}]{xu2010JOCP}%
  \BibitemOpen
  \bibfield  {author} {\bibinfo {author} {\bibfnamefont {K.}~\bibnamefont
  {Xu}}\ and\ \bibinfo {author} {\bibfnamefont {J.}~\bibnamefont {Huang}},\
  }\bibfield  {title} {\enquote {\bibinfo {title} {A unified gas-kinetic scheme
  for continuum and rarefied flows},}\ }\href@noop {} {\bibfield  {journal}
  {\bibinfo  {journal} {J. Comput. Phys.}\ }\textbf {\bibinfo {volume} {229}},\
  \bibinfo {pages} {7747--7764} (\bibinfo {year} {2010})}\BibitemShut {NoStop}%
\bibitem [{\citenamefont {Guo}, \citenamefont {Wang},\ and\ \citenamefont
  {Xu}(2015)}]{guo2015PRE}%
  \BibitemOpen
  \bibfield  {author} {\bibinfo {author} {\bibfnamefont {Z.}~\bibnamefont
  {Guo}}, \bibinfo {author} {\bibfnamefont {R.}~\bibnamefont {Wang}}, \ and\
  \bibinfo {author} {\bibfnamefont {K.}~\bibnamefont {Xu}},\ }\bibfield
  {title} {\enquote {\bibinfo {title} {Discrete unified gas kinetic scheme for
  all knudsen number flows. {II}. thermal compressible case},}\ }\href@noop {}
  {\bibfield  {journal} {\bibinfo  {journal} {Phys. Rev. E}\ }\textbf {\bibinfo
  {volume} {91}},\ \bibinfo {pages} {033313} (\bibinfo {year}
  {2015})}\BibitemShut {NoStop}%
\bibitem [{\citenamefont {Xu}\ and\ \citenamefont {Zhang}(2019)}]{Xu2022BSTP}%
  \BibitemOpen
  \bibfield  {author} {\bibinfo {author} {\bibfnamefont {A.~G.}\ \bibnamefont
  {Xu}}\ and\ \bibinfo {author} {\bibfnamefont {Y.~D.}\ \bibnamefont {Zhang}},\
  }\href@noop {} {\emph {\bibinfo {title} {Complex Media Kinetics}}}\ (\bibinfo
   {publisher} {Beijing Sci. Tech. Press},\ \bibinfo {year} {2019})\BibitemShut
  {NoStop}%
\bibitem [{\citenamefont {Xu}, \citenamefont {Zhang},\ and\ \citenamefont
  {Gan}(2024)}]{xu2024FOP}%
  \BibitemOpen
  \bibfield  {author} {\bibinfo {author} {\bibfnamefont {A.}~\bibnamefont
  {Xu}}, \bibinfo {author} {\bibfnamefont {D.}~\bibnamefont {Zhang}}, \ and\
  \bibinfo {author} {\bibfnamefont {Y.}~\bibnamefont {Gan}},\ }\bibfield
  {title} {\enquote {\bibinfo {title} {Advances in the kinetics of heat and
  mass transfer in near-continuous complex flows},}\ }\href@noop {} {\bibfield
  {journal} {\bibinfo  {journal} {Front. Phys.}\ }\textbf {\bibinfo {volume}
  {19}},\ \bibinfo {pages} {42500} (\bibinfo {year} {2024})}\BibitemShut
  {NoStop}%
\bibitem [{\citenamefont {Gan}\ \emph {et~al.}(2022)\citenamefont {Gan},
  \citenamefont {Xu}, \citenamefont {Lai}, \citenamefont {Li}, \citenamefont
  {Sun},\ and\ \citenamefont {Succi}}]{Gan2022}%
  \BibitemOpen
  \bibfield  {author} {\bibinfo {author} {\bibfnamefont {Y.}~\bibnamefont
  {Gan}}, \bibinfo {author} {\bibfnamefont {A.}~\bibnamefont {Xu}}, \bibinfo
  {author} {\bibfnamefont {H.}~\bibnamefont {Lai}}, \bibinfo {author}
  {\bibfnamefont {W.}~\bibnamefont {Li}}, \bibinfo {author} {\bibfnamefont
  {G.}~\bibnamefont {Sun}}, \ and\ \bibinfo {author} {\bibfnamefont
  {S.}~\bibnamefont {Succi}},\ }\bibfield  {title} {\enquote {\bibinfo {title}
  {{Discrete {B}oltzmann multi-scale modelling of non-equilibrium multiphase
  flows}},}\ }\href@noop {} {\bibfield  {journal} {\bibinfo  {journal} {{J.
  Fluid Mech.}}\ }\textbf {\bibinfo {volume} {951}},\ \bibinfo {pages} {A8}
  (\bibinfo {year} {2022})}\BibitemShut {NoStop}%
\bibitem [{\citenamefont {Xu}\ \emph {et~al.}(2012)\citenamefont {Xu},
  \citenamefont {Zhang}, \citenamefont {Gan}, \citenamefont {Chen},\ and\
  \citenamefont {Yu}}]{xu2012FOP}%
  \BibitemOpen
  \bibfield  {author} {\bibinfo {author} {\bibfnamefont {A.}~\bibnamefont
  {Xu}}, \bibinfo {author} {\bibfnamefont {G.}~\bibnamefont {Zhang}}, \bibinfo
  {author} {\bibfnamefont {Y.}~\bibnamefont {Gan}}, \bibinfo {author}
  {\bibfnamefont {F.}~\bibnamefont {Chen}}, \ and\ \bibinfo {author}
  {\bibfnamefont {X.}~\bibnamefont {Yu}},\ }\bibfield  {title} {\enquote
  {\bibinfo {title} {Lattice {B}oltzmann modeling and simulation of
  compressible flows},}\ }\href@noop {} {\bibfield  {journal} {\bibinfo
  {journal} {Front. Phys.}\ }\textbf {\bibinfo {volume} {7}},\ \bibinfo {pages}
  {582--600} (\bibinfo {year} {2012})}\BibitemShut {NoStop}%
\bibitem [{\citenamefont {Gan}\ \emph {et~al.}(2015)\citenamefont {Gan},
  \citenamefont {Xu}, \citenamefont {Zhang},\ and\ \citenamefont
  {Succi}}]{Gan2015}%
  \BibitemOpen
  \bibfield  {author} {\bibinfo {author} {\bibfnamefont {Y.}~\bibnamefont
  {Gan}}, \bibinfo {author} {\bibfnamefont {A.}~\bibnamefont {Xu}}, \bibinfo
  {author} {\bibfnamefont {G.}~\bibnamefont {Zhang}}, \ and\ \bibinfo {author}
  {\bibfnamefont {S.}~\bibnamefont {Succi}},\ }\bibfield  {title} {\enquote
  {\bibinfo {title} {{Discrete {B}oltzmann modeling of multiphase flows:
  Hydrodynamic and thermodynamic non-equilibrium effects}},}\ }\href@noop {}
  {\bibfield  {journal} {\bibinfo  {journal} {{Soft Matter}}\ }\textbf
  {\bibinfo {volume} {11}},\ \bibinfo {pages} {5336--5345} (\bibinfo {year}
  {2015})}\BibitemShut {NoStop}%
\bibitem [{\citenamefont {Lai}\ \emph {et~al.}(2016)\citenamefont {Lai},
  \citenamefont {Xu}, \citenamefont {Zhang}, \citenamefont {Gan}, \citenamefont
  {Ying},\ and\ \citenamefont {Succi}}]{Lai2016PRE}%
  \BibitemOpen
  \bibfield  {author} {\bibinfo {author} {\bibfnamefont {H.}~\bibnamefont
  {Lai}}, \bibinfo {author} {\bibfnamefont {A.}~\bibnamefont {Xu}}, \bibinfo
  {author} {\bibfnamefont {G.}~\bibnamefont {Zhang}}, \bibinfo {author}
  {\bibfnamefont {Y.}~\bibnamefont {Gan}}, \bibinfo {author} {\bibfnamefont
  {Y.}~\bibnamefont {Ying}}, \ and\ \bibinfo {author} {\bibfnamefont
  {S.}~\bibnamefont {Succi}},\ }\bibfield  {title} {\enquote {\bibinfo {title}
  {{Nonequilibrium thermohydrodynamic effects on the {R}ayleigh-{T}aylor
  instability in compressible flows}},}\ }\href@noop {} {\bibfield  {journal}
  {\bibinfo  {journal} {{Phys. Rev. E}}\ }\textbf {\bibinfo {volume} {94}},\
  \bibinfo {pages} {023106} (\bibinfo {year} {2016})}\BibitemShut {NoStop}%
\bibitem [{\citenamefont {Chen}\ \emph
  {et~al.}(2025{\natexlab{a}})\citenamefont {Chen}, \citenamefont {Xu},
  \citenamefont {Zhang}, \citenamefont {Chen},\ and\ \citenamefont
  {Chen}}]{JieChen}%
  \BibitemOpen
  \bibfield  {author} {\bibinfo {author} {\bibfnamefont {J.}~\bibnamefont
  {Chen}}, \bibinfo {author} {\bibfnamefont {A.}~\bibnamefont {Xu}}, \bibinfo
  {author} {\bibfnamefont {Y.}~\bibnamefont {Zhang}}, \bibinfo {author}
  {\bibfnamefont {D.}~\bibnamefont {Chen}}, \ and\ \bibinfo {author}
  {\bibfnamefont {Z.}~\bibnamefont {Chen}},\ }\bibfield  {title} {\enquote
  {\bibinfo {title} {Kinetics of {R}ayleigh-{T}aylor instability in van der
  {W}aals fluid: the influence of compressibility},}\ }\href@noop {} {\bibfield
   {journal} {\bibinfo  {journal} {Front. Phys.}\ }\textbf {\bibinfo {volume}
  {20}},\ \bibinfo {eid} {11201} (\bibinfo {year}
  {2025}{\natexlab{a}})}\BibitemShut {NoStop}%
\bibitem [{\citenamefont {Chen}\ \emph {et~al.}(2024)\citenamefont {Chen},
  \citenamefont {Xu}, \citenamefont {Song}, \citenamefont {Gan}, \citenamefont
  {Zhang},\ and\ \citenamefont {Guan}}]{chen2024surface}%
  \BibitemOpen
  \bibfield  {author} {\bibinfo {author} {\bibfnamefont {F.}~\bibnamefont
  {Chen}}, \bibinfo {author} {\bibfnamefont {A.}~\bibnamefont {Xu}}, \bibinfo
  {author} {\bibfnamefont {J.}~\bibnamefont {Song}}, \bibinfo {author}
  {\bibfnamefont {Y.}~\bibnamefont {Gan}}, \bibinfo {author} {\bibfnamefont
  {Y.}~\bibnamefont {Zhang}}, \ and\ \bibinfo {author} {\bibfnamefont
  {N.}~\bibnamefont {Guan}},\ }\bibfield  {title} {\enquote {\bibinfo {title}
  {Surface tension effects on {R}ayleigh-{T}aylor instability in nonideal
  fluids: A multiple-relaxation-time discrete {B}oltzmann study},}\ }\href@noop
  {} {\bibfield  {journal} {\bibinfo  {journal} {Sci. China Phys., Mech.
  Astron.}\ }\textbf {\bibinfo {volume} {67}},\ \bibinfo {pages} {124611}
  (\bibinfo {year} {2024})}\BibitemShut {NoStop}%
\bibitem [{\citenamefont {Xu}, \citenamefont {Lin},\ and\ \citenamefont
  {Lai}(2025)}]{xu2025influence}%
  \BibitemOpen
  \bibfield  {author} {\bibinfo {author} {\bibfnamefont {H.}~\bibnamefont
  {Xu}}, \bibinfo {author} {\bibfnamefont {C.}~\bibnamefont {Lin}}, \ and\
  \bibinfo {author} {\bibfnamefont {H.}~\bibnamefont {Lai}},\ }\bibfield
  {title} {\enquote {\bibinfo {title} {Influence of phase difference and
  amplitude ratio on {K}elvin--{H}elmholtz instability with dual-mode interface
  perturbations},}\ }\href@noop {} {\bibfield  {journal} {\bibinfo  {journal}
  {Phys. Fluids}\ }\textbf {\bibinfo {volume} {37}} (\bibinfo {year}
  {2025})}\BibitemShut {NoStop}%
\bibitem [{\citenamefont {Zhang}\ \emph {et~al.}(2016)\citenamefont {Zhang},
  \citenamefont {Xu}, \citenamefont {Zhang}, \citenamefont {Zhu},\ and\
  \citenamefont {Lin}}]{zhang2016kinetic}%
  \BibitemOpen
  \bibfield  {author} {\bibinfo {author} {\bibfnamefont {Y.}~\bibnamefont
  {Zhang}}, \bibinfo {author} {\bibfnamefont {A.}~\bibnamefont {Xu}}, \bibinfo
  {author} {\bibfnamefont {G.}~\bibnamefont {Zhang}}, \bibinfo {author}
  {\bibfnamefont {C.}~\bibnamefont {Zhu}}, \ and\ \bibinfo {author}
  {\bibfnamefont {C.}~\bibnamefont {Lin}},\ }\bibfield  {title} {\enquote
  {\bibinfo {title} {Kinetic modeling of detonation and effects of negative
  temperature coefficient},}\ }\href@noop {} {\bibfield  {journal} {\bibinfo
  {journal} {Combust. Flame}\ }\textbf {\bibinfo {volume} {173}},\ \bibinfo
  {pages} {483--492} (\bibinfo {year} {2016})}\BibitemShut {NoStop}%
\bibitem [{\citenamefont {Ji}, \citenamefont {Lin},\ and\ \citenamefont
  {Luo}(2022)}]{ji2022JOCP}%
  \BibitemOpen
  \bibfield  {author} {\bibinfo {author} {\bibfnamefont {Y.}~\bibnamefont
  {Ji}}, \bibinfo {author} {\bibfnamefont {C.}~\bibnamefont {Lin}}, \ and\
  \bibinfo {author} {\bibfnamefont {K.~H.}\ \bibnamefont {Luo}},\ }\bibfield
  {title} {\enquote {\bibinfo {title} {A three-dimensional discrete {B}oltzmann
  model for steady and unsteady detonation},}\ }\href@noop {} {\bibfield
  {journal} {\bibinfo  {journal} {J. Comput. Phys.}\ }\textbf {\bibinfo
  {volume} {455}},\ \bibinfo {pages} {111002} (\bibinfo {year}
  {2022})}\BibitemShut {NoStop}%
\bibitem [{\citenamefont {Bao}\ \emph {et~al.}(2022)\citenamefont {Bao},
  \citenamefont {Qiu}, \citenamefont {Zhou}, \citenamefont {Zhou},
  \citenamefont {Weng}, \citenamefont {Lin},\ and\ \citenamefont
  {You}}]{Qiu-2022-POF}%
  \BibitemOpen
  \bibfield  {author} {\bibinfo {author} {\bibfnamefont {Y.}~\bibnamefont
  {Bao}}, \bibinfo {author} {\bibfnamefont {R.}~\bibnamefont {Qiu}}, \bibinfo
  {author} {\bibfnamefont {K.}~\bibnamefont {Zhou}}, \bibinfo {author}
  {\bibfnamefont {T.}~\bibnamefont {Zhou}}, \bibinfo {author} {\bibfnamefont
  {Y.}~\bibnamefont {Weng}}, \bibinfo {author} {\bibfnamefont {K.}~\bibnamefont
  {Lin}}, \ and\ \bibinfo {author} {\bibfnamefont {Y.}~\bibnamefont {You}},\
  }\bibfield  {title} {\enquote {\bibinfo {title} {Study of shock wave/boundary
  layer interaction from the perspective of nonequilibrium effects},}\
  }\href@noop {} {\bibfield  {journal} {\bibinfo  {journal} {Phys. Fluids}\
  }\textbf {\bibinfo {volume} {34}},\ \bibinfo {pages} {046109} (\bibinfo
  {year} {2022})}\BibitemShut {NoStop}%
\bibitem [{\citenamefont {Qiu}\ \emph {et~al.}(2020)\citenamefont {Qiu},
  \citenamefont {Bao}, \citenamefont {Zhou}, \citenamefont {Che}, \citenamefont
  {Chen},\ and\ \citenamefont {You}}]{Qiu2020AP}%
  \BibitemOpen
  \bibfield  {author} {\bibinfo {author} {\bibfnamefont {R.}~\bibnamefont
  {Qiu}}, \bibinfo {author} {\bibfnamefont {Y.}~\bibnamefont {Bao}}, \bibinfo
  {author} {\bibfnamefont {T.}~\bibnamefont {Zhou}}, \bibinfo {author}
  {\bibfnamefont {H.}~\bibnamefont {Che}}, \bibinfo {author} {\bibfnamefont
  {R.}~\bibnamefont {Chen}}, \ and\ \bibinfo {author} {\bibfnamefont
  {Y.}~\bibnamefont {You}},\ }\bibfield  {title} {\enquote {\bibinfo {title}
  {{Study of regular reflection shock waves using a mesoscopic kinetic
  approach: Curvature pattern and effects of viscosity}},}\ }\href@noop {}
  {\bibfield  {journal} {\bibinfo  {journal} {{Phys. Fluids}}\ }\textbf
  {\bibinfo {volume} {32}},\ \bibinfo {pages} {106106} (\bibinfo {year}
  {2020})}\BibitemShut {NoStop}%
\bibitem [{\citenamefont {Qiu}\ \emph {et~al.}(2021)\citenamefont {Qiu},
  \citenamefont {Zhou}, \citenamefont {Bao}, \citenamefont {Zhou},
  \citenamefont {Che},\ and\ \citenamefont {You}}]{Qiu2021APS}%
  \BibitemOpen
  \bibfield  {author} {\bibinfo {author} {\bibfnamefont {R.}~\bibnamefont
  {Qiu}}, \bibinfo {author} {\bibfnamefont {T.}~\bibnamefont {Zhou}}, \bibinfo
  {author} {\bibfnamefont {Y.}~\bibnamefont {Bao}}, \bibinfo {author}
  {\bibfnamefont {K.}~\bibnamefont {Zhou}}, \bibinfo {author} {\bibfnamefont
  {H.}~\bibnamefont {Che}}, \ and\ \bibinfo {author} {\bibfnamefont
  {Y.}~\bibnamefont {You}},\ }\bibfield  {title} {\enquote {\bibinfo {title}
  {Mesoscopic kinetic approach for studying nonequilibrium hydrodynamic and
  thermodynamic effects of shock wave, contact discontinuity, and rarefaction
  wave in the unsteady shock tube},}\ }\href@noop {} {\bibfield  {journal}
  {\bibinfo  {journal} {{Phys. Rev. E}}\ }\textbf {\bibinfo {volume} {103}},\
  \bibinfo {pages} {053113} (\bibinfo {year} {2021})}\BibitemShut {NoStop}%
\bibitem [{\citenamefont {Zhang}\ \emph {et~al.}(2019)\citenamefont {Zhang},
  \citenamefont {Xu}, \citenamefont {Zhang}, \citenamefont {Chen},\ and\
  \citenamefont {Wang}}]{zhang2019discrete}%
  \BibitemOpen
  \bibfield  {author} {\bibinfo {author} {\bibfnamefont {Y.}~\bibnamefont
  {Zhang}}, \bibinfo {author} {\bibfnamefont {A.}~\bibnamefont {Xu}}, \bibinfo
  {author} {\bibfnamefont {G.}~\bibnamefont {Zhang}}, \bibinfo {author}
  {\bibfnamefont {Z.}~\bibnamefont {Chen}}, \ and\ \bibinfo {author}
  {\bibfnamefont {P.}~\bibnamefont {Wang}},\ }\bibfield  {title} {\enquote
  {\bibinfo {title} {Discrete {B}oltzmann method for non-equilibrium flows:
  Based on {S}hakhov model},}\ }\href@noop {} {\bibfield  {journal} {\bibinfo
  {journal} {Comput. Phys. Commun.}\ }\textbf {\bibinfo {volume} {238}},\
  \bibinfo {pages} {50--65} (\bibinfo {year} {2019})}\BibitemShut {NoStop}%
\bibitem [{\citenamefont {Zhang}\ \emph {et~al.}(2023)\citenamefont {Zhang},
  \citenamefont {Wu}, \citenamefont {Nie}, \citenamefont {Xu}, \citenamefont
  {Chen},\ and\ \citenamefont {Wei}}]{zhang2023lagrangian}%
  \BibitemOpen
  \bibfield  {author} {\bibinfo {author} {\bibfnamefont {Y.}~\bibnamefont
  {Zhang}}, \bibinfo {author} {\bibfnamefont {X.}~\bibnamefont {Wu}}, \bibinfo
  {author} {\bibfnamefont {B.}~\bibnamefont {Nie}}, \bibinfo {author}
  {\bibfnamefont {A.}~\bibnamefont {Xu}}, \bibinfo {author} {\bibfnamefont
  {F.}~\bibnamefont {Chen}}, \ and\ \bibinfo {author} {\bibfnamefont
  {R.}~\bibnamefont {Wei}},\ }\bibfield  {title} {\enquote {\bibinfo {title}
  {{Lagrangian steady-state discrete {B}oltzmann model for non-equilibrium
  flows at micro--nanoscale}},}\ }\href@noop {} {\bibfield  {journal} {\bibinfo
   {journal} {{Phys. Fluids}}\ }\textbf {\bibinfo {volume} {35}} (\bibinfo
  {year} {2023})}\BibitemShut {NoStop}%
\bibitem [{\citenamefont {Liu}\ \emph {et~al.}(2023)\citenamefont {Liu},
  \citenamefont {Song}, \citenamefont {Xu}, \citenamefont {Zhang},\ and\
  \citenamefont {Xie}}]{liu2023POT}%
  \BibitemOpen
  \bibfield  {author} {\bibinfo {author} {\bibfnamefont {Z.}~\bibnamefont
  {Liu}}, \bibinfo {author} {\bibfnamefont {J.}~\bibnamefont {Song}}, \bibinfo
  {author} {\bibfnamefont {A.}~\bibnamefont {Xu}}, \bibinfo {author}
  {\bibfnamefont {Y.}~\bibnamefont {Zhang}}, \ and\ \bibinfo {author}
  {\bibfnamefont {K.}~\bibnamefont {Xie}},\ }\bibfield  {title} {\enquote
  {\bibinfo {title} {Discrete {B}oltzmann modeling of plasma shock wave},}\
  }\href@noop {} {\bibfield  {journal} {\bibinfo  {journal} {Proc. Inst. Mech.
  Eng., Part C: J. Mech. Eng. Sci.}\ }\textbf {\bibinfo {volume} {237}},\
  \bibinfo {pages} {2532--2548} (\bibinfo {year} {2023})}\BibitemShut {NoStop}%
\bibitem [{\citenamefont {Song}\ \emph {et~al.}(2024)\citenamefont {Song},
  \citenamefont {Xu}, \citenamefont {Miao}, \citenamefont {Chen}, \citenamefont
  {Liu}, \citenamefont {Wang}, \citenamefont {Wang},\ and\ \citenamefont
  {Hou}}]{Song2024POF}%
  \BibitemOpen
  \bibfield  {author} {\bibinfo {author} {\bibfnamefont {J.}~\bibnamefont
  {Song}}, \bibinfo {author} {\bibfnamefont {A.}~\bibnamefont {Xu}}, \bibinfo
  {author} {\bibfnamefont {L.}~\bibnamefont {Miao}}, \bibinfo {author}
  {\bibfnamefont {F.}~\bibnamefont {Chen}}, \bibinfo {author} {\bibfnamefont
  {Z.}~\bibnamefont {Liu}}, \bibinfo {author} {\bibfnamefont {L.}~\bibnamefont
  {Wang}}, \bibinfo {author} {\bibfnamefont {N.}~\bibnamefont {Wang}}, \ and\
  \bibinfo {author} {\bibfnamefont {X.}~\bibnamefont {Hou}},\ }\bibfield
  {title} {\enquote {\bibinfo {title} {Plasma kinetics: Discrete {B}oltzmann
  modeling and {R}ichtmyer--{M}eshkov instability},}\ }\href@noop {} {\bibfield
   {journal} {\bibinfo  {journal} {Phys. Fluids}\ }\textbf {\bibinfo {volume}
  {36}} (\bibinfo {year} {2024})}\BibitemShut {NoStop}%
\bibitem [{\citenamefont {Sun}\ \emph {et~al.}(2024)\citenamefont {Sun},
  \citenamefont {Gan}, \citenamefont {Xu},\ and\ \citenamefont
  {Shi}}]{Sun2024POF}%
  \BibitemOpen
  \bibfield  {author} {\bibinfo {author} {\bibfnamefont {G.}~\bibnamefont
  {Sun}}, \bibinfo {author} {\bibfnamefont {Y.}~\bibnamefont {Gan}}, \bibinfo
  {author} {\bibfnamefont {A.}~\bibnamefont {Xu}}, \ and\ \bibinfo {author}
  {\bibfnamefont {Q.}~\bibnamefont {Shi}},\ }\bibfield  {title} {\enquote
  {\bibinfo {title} {Droplet coalescence kinetics: {T}hermodynamic
  non-equilibrium effects and entropy production mechanism},}\ }\href@noop {}
  {\bibfield  {journal} {\bibinfo  {journal} {Phys. Fluids}\ }\textbf {\bibinfo
  {volume} {36}} (\bibinfo {year} {2024})}\BibitemShut {NoStop}%
\bibitem [{\citenamefont {Lin}\ \emph {et~al.}(2016)\citenamefont {Lin},
  \citenamefont {Xu}, \citenamefont {Zhang},\ and\ \citenamefont
  {Li}}]{lin2016CAF}%
  \BibitemOpen
  \bibfield  {author} {\bibinfo {author} {\bibfnamefont {C.}~\bibnamefont
  {Lin}}, \bibinfo {author} {\bibfnamefont {A.}~\bibnamefont {Xu}}, \bibinfo
  {author} {\bibfnamefont {G.}~\bibnamefont {Zhang}}, \ and\ \bibinfo {author}
  {\bibfnamefont {Y.}~\bibnamefont {Li}},\ }\bibfield  {title} {\enquote
  {\bibinfo {title} {Double-distribution-function discrete {B}oltzmann model
  for combustion},}\ }\href@noop {} {\bibfield  {journal} {\bibinfo  {journal}
  {Combust. Flame}\ }\textbf {\bibinfo {volume} {164}},\ \bibinfo {pages}
  {137--151} (\bibinfo {year} {2016})}\BibitemShut {NoStop}%
\bibitem [{\citenamefont {Gan}\ \emph {et~al.}(2019)\citenamefont {Gan},
  \citenamefont {Xu}, \citenamefont {Zhang}, \citenamefont {Lin}, \citenamefont
  {Lai},\ and\ \citenamefont {Liu}}]{gan2019FOP}%
  \BibitemOpen
  \bibfield  {author} {\bibinfo {author} {\bibfnamefont {Y.}~\bibnamefont
  {Gan}}, \bibinfo {author} {\bibfnamefont {A.}~\bibnamefont {Xu}}, \bibinfo
  {author} {\bibfnamefont {G.}~\bibnamefont {Zhang}}, \bibinfo {author}
  {\bibfnamefont {C.}~\bibnamefont {Lin}}, \bibinfo {author} {\bibfnamefont
  {H.}~\bibnamefont {Lai}}, \ and\ \bibinfo {author} {\bibfnamefont
  {Z.}~\bibnamefont {Liu}},\ }\bibfield  {title} {\enquote {\bibinfo {title}
  {Nonequilibrium and morphological characterizations of {K}elvin--{H}elmholtz
  instability in compressible flows},}\ }\href@noop {} {\bibfield  {journal}
  {\bibinfo  {journal} {Front. Phys.}\ }\textbf {\bibinfo {volume} {14}},\
  \bibinfo {pages} {1--17} (\bibinfo {year} {2019})}\BibitemShut {NoStop}%
\bibitem [{\citenamefont {Lin}\ \emph {et~al.}(2019)\citenamefont {Lin},
  \citenamefont {Luo}, \citenamefont {Gan},\ and\ \citenamefont
  {Liu}}]{lin2019CITP}%
  \BibitemOpen
  \bibfield  {author} {\bibinfo {author} {\bibfnamefont {C.}~\bibnamefont
  {Lin}}, \bibinfo {author} {\bibfnamefont {K.~H.}\ \bibnamefont {Luo}},
  \bibinfo {author} {\bibfnamefont {Y.}~\bibnamefont {Gan}}, \ and\ \bibinfo
  {author} {\bibfnamefont {Z.}~\bibnamefont {Liu}},\ }\bibfield  {title}
  {\enquote {\bibinfo {title} {Kinetic simulation of nonequilibrium
  {K}elvin-{H}elmholtz instability},}\ }\href@noop {} {\bibfield  {journal}
  {\bibinfo  {journal} {Commun. Theor. Phys.}\ }\textbf {\bibinfo {volume}
  {71}},\ \bibinfo {pages} {132} (\bibinfo {year} {2019})}\BibitemShut
  {NoStop}%
\bibitem [{\citenamefont {Lin}\ \emph {et~al.}(2021)\citenamefont {Lin},
  \citenamefont {Luo}, \citenamefont {Xu}, \citenamefont {Gan},\ and\
  \citenamefont {Lai}}]{lin2021multiple}%
  \BibitemOpen
  \bibfield  {author} {\bibinfo {author} {\bibfnamefont {C.}~\bibnamefont
  {Lin}}, \bibinfo {author} {\bibfnamefont {K.~H.}\ \bibnamefont {Luo}},
  \bibinfo {author} {\bibfnamefont {A.}~\bibnamefont {Xu}}, \bibinfo {author}
  {\bibfnamefont {Y.}~\bibnamefont {Gan}}, \ and\ \bibinfo {author}
  {\bibfnamefont {H.}~\bibnamefont {Lai}},\ }\bibfield  {title} {\enquote
  {\bibinfo {title} {Multiple-relaxation-time discrete {B}oltzmann modeling of
  multicomponent mixture with nonequilibrium effects},}\ }\href@noop {}
  {\bibfield  {journal} {\bibinfo  {journal} {Phys. Rev. E}\ }\textbf {\bibinfo
  {volume} {103}},\ \bibinfo {pages} {013305} (\bibinfo {year}
  {2021})}\BibitemShut {NoStop}%
\bibitem [{\citenamefont {Li}\ and\ \citenamefont {Lin}(2024)}]{li2024POF}%
  \BibitemOpen
  \bibfield  {author} {\bibinfo {author} {\bibfnamefont {Y.}~\bibnamefont
  {Li}}\ and\ \bibinfo {author} {\bibfnamefont {C.}~\bibnamefont {Lin}},\
  }\bibfield  {title} {\enquote {\bibinfo {title} {Kinetic investigation of
  {K}elvin--{H}elmholtz instability with nonequilibrium effects in a force
  field},}\ }\href@noop {} {\bibfield  {journal} {\bibinfo  {journal} {Phys.
  Fluids}\ }\textbf {\bibinfo {volume} {36}} (\bibinfo {year}
  {2024})}\BibitemShut {NoStop}%
\bibitem [{\citenamefont {Wang}, \citenamefont {Ye},\ and\ \citenamefont
  {Li}(2010)}]{wang2010POP}%
  \BibitemOpen
  \bibfield  {author} {\bibinfo {author} {\bibfnamefont {L.}~\bibnamefont
  {Wang}}, \bibinfo {author} {\bibfnamefont {W.}~\bibnamefont {Ye}}, \ and\
  \bibinfo {author} {\bibfnamefont {Y.}~\bibnamefont {Li}},\ }\bibfield
  {title} {\enquote {\bibinfo {title} {Combined effect of the density and
  velocity gradients in the combination of {K}elvin--{H}elmholtz and
  {R}ayleigh--{T}aylor instabilities},}\ }\href@noop {} {\bibfield  {journal}
  {\bibinfo  {journal} {Phys. Plasma}\ }\textbf {\bibinfo {volume} {17}}
  (\bibinfo {year} {2010})}\BibitemShut {NoStop}%
\bibitem [{\citenamefont {Li}\ \emph {et~al.}(2022)\citenamefont {Li},
  \citenamefont {Lai}, \citenamefont {Lin},\ and\ \citenamefont
  {Li}}]{li2022FOP}%
  \BibitemOpen
  \bibfield  {author} {\bibinfo {author} {\bibfnamefont {Y.}~\bibnamefont
  {Li}}, \bibinfo {author} {\bibfnamefont {H.}~\bibnamefont {Lai}}, \bibinfo
  {author} {\bibfnamefont {C.}~\bibnamefont {Lin}}, \ and\ \bibinfo {author}
  {\bibfnamefont {D.}~\bibnamefont {Li}},\ }\bibfield  {title} {\enquote
  {\bibinfo {title} {Influence of the tangential velocity on the compressible
  {K}elvin-{H}elmholtz instability with nonequilibrium effects},}\ }\href@noop
  {} {\bibfield  {journal} {\bibinfo  {journal} {Front. Phys.}\ }\textbf
  {\bibinfo {volume} {17}},\ \bibinfo {pages} {63500} (\bibinfo {year}
  {2022})}\BibitemShut {NoStop}%
\bibitem [{\citenamefont {Soler}\ and\ \citenamefont
  {Ballester}(2022)}]{soler2022FIA}%
  \BibitemOpen
  \bibfield  {author} {\bibinfo {author} {\bibfnamefont {R.}~\bibnamefont
  {Soler}}\ and\ \bibinfo {author} {\bibfnamefont {J.~L.}\ \bibnamefont
  {Ballester}},\ }\bibfield  {title} {\enquote {\bibinfo {title} {Theory of
  fluid instabilities in partially ionized plasmas: an overview},}\ }\href@noop
  {} {\bibfield  {journal} {\bibinfo  {journal} {Front. Astron. Space Sci.}\
  }\textbf {\bibinfo {volume} {9}},\ \bibinfo {pages} {789083} (\bibinfo {year}
  {2022})}\BibitemShut {NoStop}%
\bibitem [{\citenamefont {Burrows}(2000)}]{burrows2000NAT}%
  \BibitemOpen
  \bibfield  {author} {\bibinfo {author} {\bibfnamefont {A.}~\bibnamefont
  {Burrows}},\ }\bibfield  {title} {\enquote {\bibinfo {title} {Supernova
  explosions in the universe},}\ }\href@noop {} {\bibfield  {journal} {\bibinfo
   {journal} {Nature}\ }\textbf {\bibinfo {volume} {403}},\ \bibinfo {pages}
  {727--733} (\bibinfo {year} {2000})}\BibitemShut {NoStop}%
\bibitem [{\citenamefont {Drake}(2010)}]{drake2010PT}%
  \BibitemOpen
  \bibfield  {author} {\bibinfo {author} {\bibfnamefont {R.~P.}\ \bibnamefont
  {Drake}},\ }\bibfield  {title} {\enquote {\bibinfo {title}
  {High-energy-density physics},}\ }\href@noop {} {\bibfield  {journal}
  {\bibinfo  {journal} {Phys. Today}\ }\textbf {\bibinfo {volume} {63}},\
  \bibinfo {pages} {28--33} (\bibinfo {year} {2010})}\BibitemShut {NoStop}%
\bibitem [{\citenamefont {Shi}\ \emph {et~al.}(2023)\citenamefont {Shi},
  \citenamefont {Zhang}, \citenamefont {Zhao}, \citenamefont {Chen},\ and\
  \citenamefont {Zheng}}]{shi2023POP}%
  \BibitemOpen
  \bibfield  {author} {\bibinfo {author} {\bibfnamefont {Q.}~\bibnamefont
  {Shi}}, \bibinfo {author} {\bibfnamefont {H.}~\bibnamefont {Zhang}}, \bibinfo
  {author} {\bibfnamefont {Z.}~\bibnamefont {Zhao}}, \bibinfo {author}
  {\bibfnamefont {Z.}~\bibnamefont {Chen}}, \ and\ \bibinfo {author}
  {\bibfnamefont {C.}~\bibnamefont {Zheng}},\ }\bibfield  {title} {\enquote
  {\bibinfo {title} {Effect of the transverse magnetic field on the
  {K}elvin--{H}elmholtz instability of the supersonic mixing layer},}\
  }\href@noop {} {\bibfield  {journal} {\bibinfo  {journal} {Phys. Plasma}\
  }\textbf {\bibinfo {volume} {30}} (\bibinfo {year} {2023})}\BibitemShut
  {NoStop}%
\bibitem [{\citenamefont {Gan}\ \emph {et~al.}(2018)\citenamefont {Gan},
  \citenamefont {Xu}, \citenamefont {Zhang}, \citenamefont {Zhang},\ and\
  \citenamefont {Succi}}]{gan2018PRE}%
  \BibitemOpen
  \bibfield  {author} {\bibinfo {author} {\bibfnamefont {Y.}~\bibnamefont
  {Gan}}, \bibinfo {author} {\bibfnamefont {A.}~\bibnamefont {Xu}}, \bibinfo
  {author} {\bibfnamefont {G.}~\bibnamefont {Zhang}}, \bibinfo {author}
  {\bibfnamefont {Y.}~\bibnamefont {Zhang}}, \ and\ \bibinfo {author}
  {\bibfnamefont {S.}~\bibnamefont {Succi}},\ }\bibfield  {title} {\enquote
  {\bibinfo {title} {Discrete {B}oltzmann trans-scale modeling of high-speed
  compressible flows},}\ }\href@noop {} {\bibfield  {journal} {\bibinfo
  {journal} {Phys. Rev. E}\ }\textbf {\bibinfo {volume} {97}},\ \bibinfo
  {pages} {053312} (\bibinfo {year} {2018})}\BibitemShut {NoStop}%
\bibitem [{\citenamefont {Gan}\ \emph {et~al.}(2013)\citenamefont {Gan},
  \citenamefont {Xu}, \citenamefont {Zhang},\ and\ \citenamefont
  {Yang}}]{gan2013EPL}%
  \BibitemOpen
  \bibfield  {author} {\bibinfo {author} {\bibfnamefont {Y.}~\bibnamefont
  {Gan}}, \bibinfo {author} {\bibfnamefont {A.}~\bibnamefont {Xu}}, \bibinfo
  {author} {\bibfnamefont {G.}~\bibnamefont {Zhang}}, \ and\ \bibinfo {author}
  {\bibfnamefont {Y.}~\bibnamefont {Yang}},\ }\bibfield  {title} {\enquote
  {\bibinfo {title} {Lattice {BGK} kinetic model for high-speed compressible
  flows: Hydrodynamic and nonequilibrium behaviors},}\ }\href@noop {}
  {\bibfield  {journal} {\bibinfo  {journal} {Europhys. Lett.}\ }\textbf
  {\bibinfo {volume} {103}},\ \bibinfo {pages} {24003} (\bibinfo {year}
  {2013})}\BibitemShut {NoStop}%
\bibitem [{\citenamefont {Chen}\ \emph
  {et~al.}(2025{\natexlab{b}})\citenamefont {Chen}, \citenamefont {Lin},
  \citenamefont {Li},\ and\ \citenamefont {Lai}}]{chen2025arXiv}%
  \BibitemOpen
  \bibfield  {author} {\bibinfo {author} {\bibfnamefont {S.}~\bibnamefont
  {Chen}}, \bibinfo {author} {\bibfnamefont {C.}~\bibnamefont {Lin}}, \bibinfo
  {author} {\bibfnamefont {D.}~\bibnamefont {Li}}, \ and\ \bibinfo {author}
  {\bibfnamefont {H.}~\bibnamefont {Lai}},\ }\bibfield  {title} {\enquote
  {\bibinfo {title} {Burnett-level discrete boltzmann modeling of compressible
  flows under force},}\ }\href@noop {} {\bibfield  {journal} {\bibinfo
  {journal} {arXiv preprint arXiv:2502.02241}\ } (\bibinfo {year}
  {2025}{\natexlab{b}})}\BibitemShut {NoStop}%
\bibitem [{\citenamefont {Jiang}\ and\ \citenamefont
  {Shu}(1996)}]{jiang1996JOCP}%
  \BibitemOpen
  \bibfield  {author} {\bibinfo {author} {\bibfnamefont {G.}~\bibnamefont
  {Jiang}}\ and\ \bibinfo {author} {\bibfnamefont {C.}~\bibnamefont {Shu}},\
  }\bibfield  {title} {\enquote {\bibinfo {title} {Efficient implementation of
  weighted {ENO} schemes},}\ }\href@noop {} {\bibfield  {journal} {\bibinfo
  {journal} {J. Comput. Phys.}\ }\textbf {\bibinfo {volume} {126}},\ \bibinfo
  {pages} {202--228} (\bibinfo {year} {1996})}\BibitemShut {NoStop}%
\bibitem [{\citenamefont {Ascher}, \citenamefont {Ruuth},\ and\ \citenamefont
  {Spiteri}(1997)}]{ascher1997ANM}%
  \BibitemOpen
  \bibfield  {author} {\bibinfo {author} {\bibfnamefont {U.~M.}\ \bibnamefont
  {Ascher}}, \bibinfo {author} {\bibfnamefont {S.~J.}\ \bibnamefont {Ruuth}}, \
  and\ \bibinfo {author} {\bibfnamefont {R.~J.}\ \bibnamefont {Spiteri}},\
  }\bibfield  {title} {\enquote {\bibinfo {title} {Implicit-explicit
  {R}unge-{K}utta methods for time-dependent partial differential equations},}\
  }\href@noop {} {\bibfield  {journal} {\bibinfo  {journal} {Appl. Numer.
  Math.}\ }\textbf {\bibinfo {volume} {25}},\ \bibinfo {pages} {151--167}
  (\bibinfo {year} {1997})}\BibitemShut {NoStop}%
\bibitem [{\citenamefont {Gan}\ \emph {et~al.}(2008)\citenamefont {Gan},
  \citenamefont {Xu}, \citenamefont {Zhang}, \citenamefont {Yu},\ and\
  \citenamefont {Li}}]{Gan2008PA}%
  \BibitemOpen
  \bibfield  {author} {\bibinfo {author} {\bibfnamefont {Y.}~\bibnamefont
  {Gan}}, \bibinfo {author} {\bibfnamefont {A.}~\bibnamefont {Xu}}, \bibinfo
  {author} {\bibfnamefont {G.}~\bibnamefont {Zhang}}, \bibinfo {author}
  {\bibfnamefont {X.}~\bibnamefont {Yu}}, \ and\ \bibinfo {author}
  {\bibfnamefont {Y.}~\bibnamefont {Li}},\ }\bibfield  {title} {\enquote
  {\bibinfo {title} {{Two-dimensional lattice Boltzmann model for compressible
  flows with high Mach number}},}\ }\href@noop {} {\bibfield  {journal}
  {\bibinfo  {journal} {Physica A}\ }\textbf {\bibinfo {volume} {387}},\
  \bibinfo {pages} {1721--1732} (\bibinfo {year} {2008})}\BibitemShut {NoStop}%
\bibitem [{\citenamefont {Keppens}\ and\ \citenamefont
  {T{\'o}th}(1999)}]{keppens1999POP}%
  \BibitemOpen
  \bibfield  {author} {\bibinfo {author} {\bibfnamefont {R.}~\bibnamefont
  {Keppens}}\ and\ \bibinfo {author} {\bibfnamefont {G.}~\bibnamefont
  {T{\'o}th}},\ }\bibfield  {title} {\enquote {\bibinfo {title} {Nonlinear
  dynamics of {K}elvin--{H}elmholtz unstable magnetized jets: Three-dimensional
  effects},}\ }\href@noop {} {\bibfield  {journal} {\bibinfo  {journal} {Phys.
  Plasma}\ }\textbf {\bibinfo {volume} {6}},\ \bibinfo {pages} {1461--1469}
  (\bibinfo {year} {1999})}\BibitemShut {NoStop}%
\bibitem [{\citenamefont {Wang}\ \emph {et~al.}(2009)\citenamefont {Wang},
  \citenamefont {Teng}, \citenamefont {Ye}, \citenamefont {Xue}, \citenamefont
  {Fan},\ and\ \citenamefont {Li}}]{li2009CITP}%
  \BibitemOpen
  \bibfield  {author} {\bibinfo {author} {\bibfnamefont {L.}~\bibnamefont
  {Wang}}, \bibinfo {author} {\bibfnamefont {A.}~\bibnamefont {Teng}}, \bibinfo
  {author} {\bibfnamefont {W.}~\bibnamefont {Ye}}, \bibinfo {author}
  {\bibfnamefont {C.}~\bibnamefont {Xue}}, \bibinfo {author} {\bibfnamefont
  {Z.}~\bibnamefont {Fan}}, \ and\ \bibinfo {author} {\bibfnamefont
  {Y.}~\bibnamefont {Li}},\ }\bibfield  {title} {\enquote {\bibinfo {title}
  {Phase effect on mode coupling in {K}elvin--{H}elmholtz instability for
  two-dimensional incompressible fluid},}\ }\href@noop {} {\bibfield  {journal}
  {\bibinfo  {journal} {Commun. Theor. Phys.}\ }\textbf {\bibinfo {volume}
  {52}},\ \bibinfo {pages} {694} (\bibinfo {year} {2009})}\BibitemShut
  {NoStop}%
\bibitem [{\citenamefont {Amerstorfer}\ \emph {et~al.}(2010)\citenamefont
  {Amerstorfer}, \citenamefont {Erkaev}, \citenamefont {Taubenschuss},\ and\
  \citenamefont {Biernat}}]{amerstorfer2010POP}%
  \BibitemOpen
  \bibfield  {author} {\bibinfo {author} {\bibfnamefont {U.}~\bibnamefont
  {Amerstorfer}}, \bibinfo {author} {\bibfnamefont {N.}~\bibnamefont {Erkaev}},
  \bibinfo {author} {\bibfnamefont {U.}~\bibnamefont {Taubenschuss}}, \ and\
  \bibinfo {author} {\bibfnamefont {H.}~\bibnamefont {Biernat}},\ }\bibfield
  {title} {\enquote {\bibinfo {title} {Influence of a density increase on the
  evolution of the {K}elvin--{H}elmholtz instability and vortices},}\
  }\href@noop {} {\bibfield  {journal} {\bibinfo  {journal} {Phys. Plasma}\
  }\textbf {\bibinfo {volume} {17}} (\bibinfo {year} {2010})}\BibitemShut
  {NoStop}%
\bibitem [{\citenamefont {Obergaulinger}, \citenamefont {Aloy},\ and\
  \citenamefont {M{\"u}ller}(2010)}]{obergaulinger2010AA}%
  \BibitemOpen
  \bibfield  {author} {\bibinfo {author} {\bibfnamefont {M.}~\bibnamefont
  {Obergaulinger}}, \bibinfo {author} {\bibfnamefont {M.}~\bibnamefont {Aloy}},
  \ and\ \bibinfo {author} {\bibfnamefont {E.}~\bibnamefont {M{\"u}ller}},\
  }\bibfield  {title} {\enquote {\bibinfo {title} {Local simulations of the
  magnetized {K}elvin-{H}elmholtz instability in neutron-star mergers},}\
  }\href@noop {} {\bibfield  {journal} {\bibinfo  {journal} {Astron.
  Astrophys.}\ }\textbf {\bibinfo {volume} {515}},\ \bibinfo {pages} {A30}
  (\bibinfo {year} {2010})}\BibitemShut {NoStop}%
\bibitem [{\citenamefont {Kim}, \citenamefont {Padrino},\ and\ \citenamefont
  {Joseph}(2011)}]{kim2011JOF}%
  \BibitemOpen
  \bibfield  {author} {\bibinfo {author} {\bibfnamefont {H.}~\bibnamefont
  {Kim}}, \bibinfo {author} {\bibfnamefont {J.}~\bibnamefont {Padrino}}, \ and\
  \bibinfo {author} {\bibfnamefont {D.}~\bibnamefont {Joseph}},\ }\bibfield
  {title} {\enquote {\bibinfo {title} {Viscous effects on {K}elvin--{H}elmholtz
  instability in a channel},}\ }\href@noop {} {\bibfield  {journal} {\bibinfo
  {journal} {J. Fluid Mech.}\ }\textbf {\bibinfo {volume} {680}},\ \bibinfo
  {pages} {398--416} (\bibinfo {year} {2011})}\BibitemShut {NoStop}%
\bibitem [{\citenamefont {Awasthi}, \citenamefont {Asthana},\ and\
  \citenamefont {Agrawal}(2014)}]{awasthi2014IJO}%
  \BibitemOpen
  \bibfield  {author} {\bibinfo {author} {\bibfnamefont {M.~K.}\ \bibnamefont
  {Awasthi}}, \bibinfo {author} {\bibfnamefont {R.}~\bibnamefont {Asthana}}, \
  and\ \bibinfo {author} {\bibfnamefont {G.}~\bibnamefont {Agrawal}},\
  }\bibfield  {title} {\enquote {\bibinfo {title} {Viscous correction for the
  viscous potential flow analysis of {K}elvin--{H}elmholtz instability of
  cylindrical flow with heat and mass transfer},}\ }\href@noop {} {\bibfield
  {journal} {\bibinfo  {journal} {Int. J. Heat Mass Transfer}\ }\textbf
  {\bibinfo {volume} {78}},\ \bibinfo {pages} {251--259} (\bibinfo {year}
  {2014})}\BibitemShut {NoStop}%
\bibitem [{\citenamefont {Gan}\ \emph {et~al.}(2025)\citenamefont {Gan},
  \citenamefont {Zhuang}, \citenamefont {Yang}, \citenamefont {Xu},
  \citenamefont {Zhang}, \citenamefont {Chen}, \citenamefont {Song},\ and\
  \citenamefont {Wu}}]{Gan2025}%
  \BibitemOpen
  \bibfield  {author} {\bibinfo {author} {\bibfnamefont {Y.}~\bibnamefont
  {Gan}}, \bibinfo {author} {\bibfnamefont {Z.}~\bibnamefont {Zhuang}},
  \bibinfo {author} {\bibfnamefont {B.}~\bibnamefont {Yang}}, \bibinfo {author}
  {\bibfnamefont {A.}~\bibnamefont {Xu}}, \bibinfo {author} {\bibfnamefont
  {D.}~\bibnamefont {Zhang}}, \bibinfo {author} {\bibfnamefont
  {F.}~\bibnamefont {Chen}}, \bibinfo {author} {\bibfnamefont {J.}~\bibnamefont
  {Song}}, \ and\ \bibinfo {author} {\bibfnamefont {Y.}~\bibnamefont {Wu}},\
  }\bibfield  {title} {\enquote {\bibinfo {title} {{Supersonic ow kinetics:
  Mesoscale structures, thermodynamic nonequilibrium effects and entropy
  production mechanisms}},}\ }\href@noop {} {\bibfield  {journal} {\bibinfo
  {journal} {arXiv preprint arXiv:2502.10832}\ } (\bibinfo {year}
  {2025})}\BibitemShut {NoStop}%
\end{thebibliography}%

\end{document}